\documentclass[manuscript]{aastex62} 
\usepackage[utf8]{inputenc}

\usepackage[version=4]{mhchem}
\usepackage[inline]{enumitem}
\usepackage{gensymb}
\usepackage{float}
\usepackage{capt-of}
\usepackage[T1]{fontenc}

\newcommand{\hms}[3]{#1\mathrm{h}#2\mathrm{m}#3\mathrm{s}}
\newcommand{\dms}[3]{#1\degr#2\arcmin#3\arcsec}

\received{September 14, 2018}
\revised{November 26, 2018}
\accepted{December 7, 2018}

\submitjournal{ApJ}

\shorttitle{\ce{C2H4O2} toward Sgr B2(N)}
\shortauthors{Xue et al.}

\begin{document}
\title{ALMA Observations of the Spatial Distribution of three \ce{C2H4O2} Isomers towards Sgr B2(N)}

\correspondingauthor{CI XUE}
\email{cx5up@virginia.edu}

\author{CI XUE}
\affiliation{Department of Chemistry, University of Virginia, Charlottesville, VA 22904, U.S.A}

\author{ANTHONY J. REMIJAN}
\affiliation{National Radio Astronomy Observatory, Charlottesville, VA 22903, U.S.A}

\author{ANDREW M. BURKHARDT}
\affiliation{Department of Astronomy, University of Virginia, Charlottesville, VA 22904, U.S.A}
\affiliation{Harvard-Smithsonian Center for Astrophysics, Cambridge, MA 02138, U.S.A}

\author{ERIC HERBST}
\affiliation{Department of Chemistry, University of Virginia, Charlottesville, VA 22904, U.S.A}
\affiliation{Department of Astronomy, University of Virginia, Charlottesville, VA 22904, U.S.A}

\begin{abstract}
The \ce{C2H4O2} isomers have been previously investigated primarily via disparate sets of observations involving single dish and array measurements. The only attempt at using a uniform set of observations was performed with the IRAM 30 m observation in 2013 \citep{bmmsc2013}. In this study, we present an intensive and rigorous spectral and morphological analysis of the \ce{C2H4O2} isomers towards Sgr B2(N) with interferometers, $ALMA$ Band 3 observations. We propose a quantitative selection method, which automates the determination of the most uncontaminated transitions and allows us to report the discovery of previously undetected transitions of the three isomers. With the least contaminated transitions, the high spatial-resolution millimeter (mm) maps of the \ce{C2H4O2} isomers reveal that \ce{HCOOCH3} and \ce{CH2OHCHO} each display two different velocity components, while only one velocity component of \ce{CH3COOH} is resolved. Moreover, the distribution of \ce{HCOOCH3} is extended and offset from the continuum emission, unlike \ce{CH2OHCHO} and \ce{CH3COOH}, for which the low-velocity component is found to be compact and concentrated toward the continuum emission peak of Sgr B2(N). The distinct morphologies of these \ce{C2H4O2} isomeric species indicate that \ce{HCOOCH3} have significant differences in chemical processes than \ce{CH2OHCHO} and \ce{CH3COOH}, which display similar spatial distributions.\\
\end{abstract}


\section{Introduction}
Isomers are families of molecules that share the same constituent atoms, but are uniquely arranged and, as such, have very different chemical properties. Several isomeric groups are well known to exist in the interstellar medium (ISM). Examples include \ce{HCN} and \ce{HNC} towards translucent clouds \citep{tur1991} and proroplanetary disks \citep{goqk2015}, as well as \ce{CH3CN} and \ce{CH3NC} \citep{rhlpj2005} and three isomers of \ce{C3H2O} \citep{lmsjbrr2015} toward Sagittarius B2 North (Sgr B2(N)). Because isomers have different chemical and physical properties, further insight can be gained into the astrochemical evolution of different physical environments of astronomical sources by making use of isomers.

One way this can be accomplished is with the Minimum Energy Principle \citep{lpec2009, lpec2010, lbm2011}. However, it has already been refuted with numerous examples of isomers \citep{2014Sci...345.1584B, kt2014}. Instead, it is widely accepted that the relative abundance of structural isomers in the ISM depends more on their formation mechanisms, i.e. kinetically controlled, than their relative energy differences, assuming they have same destruction routes, which has been further supported by \citet{lmsjbrr2015}. Thus, the study of isomers can help to constrain temperature- and density-dependent synthetic routes to complex interstellar molecules \citep{mgkm2007}, which may in turn help to explain spatial distributions.

This work focuses on three isomers of \ce{C2H4O2}: glycolaldehyde (\ce{CH2OHCHO}, GLA), acetic acid (\ce{CH3COOH}, AcA) and methyl formate (\ce{HCOOCH3}, MF), the chemical evolution of which has been widely investigated \citep{lghw2011, sbc2018, lif2015}. In astrochemical models, the most widely used mechanism to efficiently form the \ce{C2H4O2} isomeric family in hot cores is through radical-radical recombination on interstellar grain surfaces \citep{gwh2008, cfiv2016}. However, in addition to grain surface chemistry, it was also suggested that MF could be formed via gas-phase processes in cold regions \citep{lghw2011, bct2015}. An understanding of the astrochemical evolution of the \ce{C2H4O2} isomers can shed light on calibrating astrochemical modeling networks. Furthermore, there has been increasing interest in studying these isomers towards astronomical environments because of their importance in the formation of large biologically-relevant molecules \citep{w2000, lms2001, lgrs2002, pmmc2010}. 

The northern clump of Sgr B2(N) molecular cloud, which lies $\sim$120 pc from the Galactic center, is one of the most massive star-forming regions in the Galaxy. In 2011, observations with the SMA at a spatial resolution of $0.\arcsec4 \times 0.\arcsec24$ revealed two emission cores within the millimeter continuum map of Sgr B2(N) \citep{qsrcl2011}. One of these located at $\alpha_\text{J200}=\hms{17}{47}{19.889}$, $\delta_\text{J2000}=\dms{-28}{22}{18.22}$ was designated as N1 in \citet{bmgk2016} and the other located at $\alpha_\text{J2000}= \hms{17}{47}{19.885}$, $\delta_{J2000}=\dms{-28}{22}{13.29}$ was designated as N2 in \citet{bmgk2016}. The prominent core, N1, which contains the well known Sgr B2(N) Large Molecule Heimat (LMH) source \citep{skm1994}, is located $\sim 5 \arcsec$ south of N2. The two cores have different systematic velocities, $V_\text{lsr}, \sim 64\ \mathrm{km\ s^{-1}}$ for N1 and $\sim 73\ \mathrm{km\ s^{-1}}$ for N2 \citep{bmmsc2013}. Recently, the hot core structures in Sgr B2(N) was further characterized and up to 20 continuum sources were reported \citep{bbmgm2017, sss2017}. Here, we adopt the naming conventions LMH and the shorthand notation of the continuum emission cores of N1, N2 and N3. The coordinates of N3 is $\alpha_\text{J200}=\hms{17}{47}{19.248}$, $\delta_\text{J2000}=\dms{-28}{22}{14.91}$ as described in \citet{bbmgm2017}. See Section \ref{sec:result} for the corresponding positions.

Sgr B2(N) is one of the most prolific regions for detecting Large Astronomical Molecules (LAMs). Here, we move away from Complex Organic Molecules (COMs) \citep{hd2009} and use LAMs as a more accurate definition for astronomical molecules with more than 6 atoms on account of the fact that, in astronomical environment, many detected molecules are not complex by chemical standards and have no direct relationship to organic or prebiotic chemistry. Many LAMs were first detected toward this source, such as \ce{CH2CHCHO} and \ce{CH3CH2CHO} \citep{hjlrm2004}, \ce{NH2CH2CN} \citep{bmcmso2008}, \ce{CH3NCO} \citep{hiz2015} and \ce{CH3CHCH2O} \citep{mcl2016}. The first detection of all of the three \ce{C2H4O2} isomers were also reported towards Sgr B2. The discovery of MF was achieved with the Parkes 64 m telescope in 1975 \citep{bcggrw1975}, while the lower abundance isomers, AcA and GLA, were not detected until several decades later due to the lack of accurate laboratory data. \citet{msml1997} reported the first detection of 4 strong transitions of AcA towards the LMH region in Sgr B2(N) using the Berkeley Illinois Maryland Association (BIMA) Array and the Caltech Owens Valley Radio Observatory (OVRO) millimeter array, while the first detection of GLA was not reported until \citet{hlj2000} who used the NRAO 12m telescope.

The \ce{C2H4O2} isomers have been previously investigated primarily via disparate sets of observations involving single dish and array measurements. The MF and GLA abundance ratio,\ [\ce{HCOOCH3}]:[\ce{CH2OHCHO}], toward Sgr B2(N) was reported to be $\sim$ 6.7:1 as observed with the NRAO 12 m single-dish telescope \citep{hlj2000}. \citet{hvsjl2001} further studied MF and GLA with the BIMA array and found the abundance ratio to be $\sim$ 52:1 for [\ce{HCOOCH3}]:[\ce{CH2OHCHO}]. The difference between the ratios was interpreted to be due to the different beam sizes between the single dish and interferometric array observations and to the size of the source of the molecular emission towards the LMH region. The one attempt at using a uniform set of observations was performed with the IRAM 30 m \citep{bmmsc2013}. They deduced the abundance ratios among the three molecules [HCOOCH3]:[CH2OHCHO]:[CH3COOH] to be 242:1:5.8 with unconstrained temperatures and source sizes. Regarding the spatial distributions, although \citet{hvsjl2001} mapped both MF and GLA with the BIMA array, the images were only based on a single transition of GLA, $8(0,8)-7(1,7)$, and the blended transitions of MF: $7(5,3)-6(5,2)\ E$ and $7(5,2)-6(5,1)\ A$. The extended and cool distribution of GLA suggested with the BIMA array observations indicated a gas phase formation mechanism. MF is ubiquitous in molecular cores with various physical conditions, suggesting that its formation involves both gas-phase and grain-surface processes. AcA always possesses a warm, compact distribution, which favors grain surface formation mechanisms \citep{msml1997} followed presumably by thermal desorption.

Therefore, a rigorous comparison between the abundances and overall spatial distributions of these three isomers has not been achieved. High-sensitivity observations with better spatial and spectral resolution are necessary to determine accurate abundance ratios and molecular morphology in different regions towards Sgr B2(N). As such, a new careful comparison of these triplets towards the Sgr B2(N) region is needed. The Atacama Large Millimeter/submillimeter Array (ALMA) has the ability to carry out observations of molecules with low abundances and weak emission lines and to image LAMs at small spatial scales due to its high sensitivity and angular resolution \citep{ogfca2015}.

This work reports a self-consistent and rigorous investigation of three \ce{C2H4O2} isomeric species with interferometric observations towards Sgr B2(N). The observations are summarized in Section \ref{sec:alma}. Section \ref{sec:dataanlysis} describes the methods and results of the data analysis, including the selection processes for selecting the unblended transitions of each species. The spatial distributions of \ce{C2H4O2} isomers are presented in detail in Section \ref{sec:result}. The possible explanations for the difference of the spatial distribution of each molecular species is discussed in Section \ref{sec:diss}, and the conclusion is given in Section \ref{sec:sum}.

\section{OBSERVATIONS} \label{sec:alma}
The interferometric data used here were acquired from the ALMA Science Archive\footnote{\footnotesize{http://almascience.nrao.edu/aq/}} of the EMoCA survey (Exploring Molecular Complexity with ALMA) \citep{bmgk2016}. This survey included 5 spectral setups as shown in Table \ref{obs-par} of which S1-S4 were performed during Cycle 0 (2012) with ALMA project code 2011.0.00017.S and S5 during Cycle 1 (2014) with project code 2012.1.00012.S. Each setup contains 4 spectral windows within the overall frequency range from 84.1 to 114.4 GHz. The phase center of the observations was pointed toward Sgr B2 with the field center at $\alpha_\text{J2000}=\hms{17}{47}{19.87}$, $\delta_\text{J2000}=\dms{-28}{22}{16}$. The Half Power Beam Width (HPBW) of the primary beam measured from actual 12 m ALMA antennas was $69 \arcsec$ and $51 \arcsec$ at 84 and 114 GHz respectively \citep{raj2015}. The spectral resolution was 488.3 kHz, which corresponds to a velocity resolution of 1.29 to 1.70 $\mathrm{km\ s^{-1}}$ across the observing band. A detailed description of the observations and the data calibration was presented in \citet{bmgk2016}. 

The Common Astronomy Software Applications package (CASA) \citep{mws2007} was used for imaging and analysis. For each spectral window, the relatively line-free continuum channel ranges were selected with the techniques, as described in \citet{gswb2017}, which are based on the mean spectrum generated from the CASA image cube. Afterwards, the ranges were used to subtract the continuum emission in the UV-plane before imaging the spectral cubes. The imaging parameters of each spectral window are presented in Table 1. For all data sets, we used the Briggs scheme with a robust parameter of 0.5 and a cell size of $0.3 \arcsec$ for imaging. The resulting size of the synthesized beam (HPBW) of each data cube has a medium value $\sim$ $1.6 \arcsec$. The median of resultant typical rms noise levels is 4.4 mJy beam$ ^{-1}$.

\begin{deluxetable*}{ccCCCCCC}
    \tablewidth{0pt}
    \tablecaption{Summary of Analyzing Parameters \label{obs-par}}
    \tablehead{
        \colhead{Setup}	&\colhead{SPW}	&\colhead{Frequency Range}	&\colhead{Primary Beam}  &\multicolumn{2}{c}{Synthesized Beam}\tablenotemark{a}   &\colhead{Spectral}   &\colhead{Median rms}\tablenotemark{b}\\
	                    &               &                           &\colhead{$HPBW$}   &\colhead{$HPBW$}&\colhead{PA}      &\colhead{Resolution}  &\colhead{Per Channel}\\ 
	                &                   &\colhead{GHz}              &\colhead{$\arcsec$}&\colhead{$\arcsec \times \arcsec$}&\colhead{$\degr$} &\colhead{$\mathrm{km\ s^{-1}}$}     &\colhead{mJy beam$^{-1}$}\\
	}
    \startdata
        S1  &   0   &   84.091-85.966     &69   &2.14\times 1.56 &   -85.5   &1.70   &4.1\\
            &   1   &   85.904-87.779     &67   &2.06\times 1.54 &   -84.4   &1.69   &3.5\\
            &   2   &   96.154-98.029     &60   &1.85\times 1.41 &   -87.2   &1.51   &3.5\\
            &   3   &   97.904-99.779     &59   &1.81\times 1.39 &   -86.7   &1.48   &3.8\\
        S2  &   0   &   87.729-89.604     &66   &3.00\times 1.36 &   -82.8   &1.65   &4.5\\
            &   1   &   89.554-91.429     &64   &2.96\times 1.36 &   -83.4   &1.62   &4.5\\
            &   2   &   99.728-101.602    &58   &2.68\times 1.21 &   -83.7   &1.46   &4.4\\
            &   3   &   101.552-103.427   &57   &2.58\times 1.21 &   84.4    &1.43   &4.6\\
        S3  &   0   &   91.368-93.242     &63   &3.01\times 1.47 &   84.8    &1.59   &5.7\\
            &   1   &   93.193-95.067     &62   &2.94\times 1.48 &   84.7    &1.56   &5.6\\ 
            &   2   &   103.365-105.239   &56   &2.68\times 1.32 &   85.1    &1.40   &6.8\\
            &   3   &   105.189-107.064   &55   &2.59\times 1.31 &   83.9    &1.38   &7.1\\
        S4  &   0   &   95.021-96.896     &61   &1.82\times 1.40 &   -83.1   &1.53   &4.0\\   
            &   1   &   96.846-98.720     &60   &1.80\times 1.38 &   -82.1   &1.50   &3.1\\
            &   2   &   107.019-108.893   &54   &1.62\times 1.24 &   -84.0   &1.36   &3.8\\
            &   3   &   108.843-110.718   &53   &1.58\times 1.22 &   -82.3   &1.33   &3.8\\
        S5  &   0   &   98.672-100.546    &58   &1.79\times 1.46 &   -74.2   &1.47   &3.2\\
            &   1   &   100.496-102.370   &57   &1.76\times 1.43 &   -73.7   &1.44   &5.0\\
            &   2   &   110.669-112.543   &52   &1.61\times 1.30 &   -75.6   &1.31   &5.7\\
            &   3   &   112.494-114.368   &51   &1.56\times 1.29 &   -73.4   &1.29   &5.6
   \enddata
    \tablecomments{\tablenotetext{a}{The synthesized beam size varies depending on the $robust$ parameter. Here we used a $robust$ parameter of 0.5.} \tablenotetext{b}{The estimated rms noise level of each spectral window is the median of the noise levels measured in several channels with moderate emission, where the noise levels are each measured in the off-source region.}}
\end{deluxetable*}

\section{SPECTRAL ANALYSIS}\label{sec:dataanlysis}
\subsection{Transition Selection Processes} \label{sec:selection}
Although MF, GLA and AcA have been unambiguously detected toward Sgr B2, the identification of clear individual spectral features of GLA and AcA were not obvious because of the spectral confusion, which limited the number of uncontaminated lines. Here, we propose a more reliable way to uniquely identify the spectral features of the \ce{C2H4O2} isomers towards Sgr B2(N), particularly the spectral features of AcA and GLA. Through a more accurate continuum subtraction and our methods of line identification, we identified weaker, and previously undetected, transitions of this isomeric triplet from the EMoCA survey \citep{bmgk2016} and further selected the least contaminated transitions for imaging.

\citet{slhf2005} proposed five criteria, hereafter referred to as the Snyder Criteria, for line assignments, particularly toward interstellar clouds with high spectral line densities greater than 10 lines per 100 MHz, see also \citep{cvc2014, rsm2014, bdl2015, cmb2015}. The Snyder Criteria, which are utilized in our study, include: (i) accurate rest frequencies, (ii) beam dilution, (iii) frequency agreement, (iv) line intensity and (v) presence of transitions with observable intensity. Here we applied these criteria for identifying the mostly uncontaminated transitions of the \ce{C2H4O2} isomers. In the following, we discuss each criterion of the Snyder Criteria with our line identification process for the \ce{C2H4O2} isomers. Details about other applications and the scope of the criteria are presented in \citet{slhf2005}.

\begin{enumerate}[label=(\roman*)]
  \item \textit{Accurate Rest Frequencies}. The first criterion is the need for high degree of accuracy of the rest frequency of the target molecular transitions. The millimeter wave spectrum of GLA has been studied in the laboratory by \citet{blp2001}, \citet{wbd2005}, and \citet{cdw2010}; the spectrum of MF by \citet{oads1999}, \citet{oott2004}, and \citet{cwdk2007}, and the spectrum of AcA by \citet{iadp2001, ielsd2013}. Their transitions are included in the current public databases: the CDMS catalog \footnote{\footnotesize{https://www.astro.uni-koeln.de/cdms/catalog}} \citep{mssw2005}, the JPL catalog \footnote{\footnotesize{https://spec.jpl.nasa.gov}} \citep{ppcd1998} and the Spectral Line Atlas of Interstellar Molecules (SLAIM) database, which are available in the SPLATALOGUE spectroscopy database \footnote{\footnotesize{https://www.splatalogue.net}}. For the analysis of the present ALMA observations, the rest frequencies and other spectral line parameters of the three isomers were taken from the above databases. In the range of 84-114 GHz, the transition uncertainties of the three molecules were estimated to be less than 150 kHz, compared to a spectral resolution of 488.3 kHz. As a result, our study satisfied the criteria of the high accuracy in rest frequency.
 
  In this study, we consider the transitions for both the ground and first vibrationally excited states of MF but only the vibrational ground state of AcA and GLA. The high hot cores temperatures ($\sim$ 150-200 K, \citet{bmgk2016}) along with the high abundance of interstellar MF make the low-energy vibrationally excited states of MF ($v_\mathrm{t} = 1$ at $\mathrm{\sim 132 cm^{-1}}$ or 190 K) likely to be populated in hot cores \citep{skh2015}. In fact, transitions within $v_\mathrm{t} = 1$ of MF have been reported toward Orion KL \citep{kott2007} and W51 e2 \citep{dwc2008}, which are comparable hot cores to Sgr B2. \citet{bmgk2016} derived the rotational temperatures in Sgr B2(N) ranging from 112-278 K for different LAMs. Considering the excitation temperature and the previously determined column density of $\mathrm{N_{HCOOCH_3} \sim 1.2 \times 10^{18} cm^{-2}}$ \citep{bmgk2016}, both the ground and $v_\mathrm{t} = 1$ states are studied in this work. Even though the comparable first vibrationally excited state of AcA lies at about 170 $\mathrm{cm^{-1}}$ (or 245 K) above the ground state \citep{iadp2001} and that of GLA at 195 $\mathrm{cm^{-1}}$ (or 280 K) \citep{wbd2005}, their abundances are not high enough to be detectable toward Sgr B2(N) like MF \citep{hvsjl2001, rslmk2002}. Therefore, only the vibrational ground states for AcA and GLA are considered in this study.

  \item \textit{Beam Dilution}. The correction based on the beam dilution effects arises from the different synthesized beam sizes of the telescopes, which affect the relative intensities measured for spectral lines toward the same pointing position. Therefore, the relative observed intensities between any transitions need to be corrected for beam dilution effects when the observations are carried out with different telescopes \citep{rsm2014}. In this study, the observations were carried out in the same receiver band, ALMA Band 3. The minimum size of the synthesized beam of the EMoCA survey was reported as $1.58 \arcsec \times 1.22 \arcsec$ while the source diameters, the FWHM from two-dimentional Gaussian fits to the integrated intensity maps, of many LAMs, including MF, were reported to be not larger than $1.5 \arcsec$ \citep{bmgk2016}. Therefore beam dilution should not affect the corresponding line intensity difference of the three isomers if they have similar source sizes.  In fact, we find that the three isomers have different source sizes as shown in Section \ref{sec:result}. But, given the measurement accuracy and the resolution of the EMoCA survey, the angular extent of all emission above 3 $\sigma$ for the three isomers is larger than the synthesized beam of the observations. The beam dilution effect in our analysis is thus negligible. 
\end{enumerate}

To examine whether our line assignments satisfy the other three Snyder criteria, we applied a single-excitation temperature simulation followed with a quantitative analysis approach, which automatically reduces the number of emission line candidates by filtering out the obviously blended lines. The remaining less blended transitions go through a more rigorous analysis including spatial distribution imaging.

In this study, the observed spectra presented are extracted from the largest synthesized beam with a size of 3.01" $\times$ 1.47" centered at LMH with the centroid position at $\alpha_\text{J200}=\hms{17}{47}{19.930}$, $\delta_\text{J2000}=\dms{-28}{22}{18.200}$. The LMH region is dense with $n_{\ce{H2}} (\text{Sgr B2(N)}) > 10^{9}\ \mathrm{cm}^{-3}$ \citep{sss2017}, and hence LTE is assumed to be reached.Although there is still some likelihood of non-thermal excitation, the molecular emission toward LMH should be well-described by spectra simulated with a single-excitation temperature ($T_\text{ex}$) \citep{gswb2017}. The single-excitation model follows the convention of \citet{hjlr2004} with corrections for optical depth as discussed in \citet{tur1991} and \citet{ms2015}. The spectra are simulated from the catalog data available in the public databases, which were generated from the laboratory data.

By comparing the simulated spectra with the observed spectra, we fit the observed spectrum of each molecule to constrain (1) the spectral line width ($\Delta V$), (2) the source velocity ($V_\text{lsr}$), (3) the molecular column density ($N_T$), and (4) the excitation temperature ($T_\text{ex}$) which contributes to the relative intensities among transitions of a certain species. 

To determine the best fit to the observed spectrum, a variety of simulations with different values of these key quantities was constructed. We left both $N_T$ and $T_\text{ex}$ as free parameters to be adjusted and tried to obtain the best-fit simulation. By setting the $\Delta V$ to 6.5 $\mathrm{km\ s^{-1}}$, $V_\text{lsr}$ to 64 $\mathrm{km\ s^{-1}}$, $T_\text{ex}$ to 190 K, we find that though subjective the single-excitation temperature simulations yield the most consistent fit to the observed spectra. The $V_\text{lsr}$ and $\Delta V$ toward LMH were constrained by previous studies to the values of 64 $\mathrm{km\ s^{-1}}$ for $V_\text{lsr}$ and $\sim$ 7 $\mathrm{km\ s^{-1}}$ for $\Delta V$ \citep{bmcmso2008}. The parameter $T_\text{ex}$ used here is consistent with the value of 200 K for $T_\text{rot}$ used in \citet{msml1997} for AcA and MF. It should be noted that these spectra are used for the initial analysis for identifying the potential uncontaminated transitions only and the values of $N_T$ and $T_\text{ex}$ will be refined later by $\chi^2$ fitting using only the uncontaminated transitions as discussed in Section \ref{sec:columndensity}.

In order to quantify the degree of consistency between the simulated and observed spectra, we further defined two factors to describe this degree of consistency of line profiles, which quantify two of the Snyder Criteria: (iii) frequency and (iv) line intensity agreement. 

\begin{figure}[ht]
    \centering
    \epsscale{1.2}
    \plotone{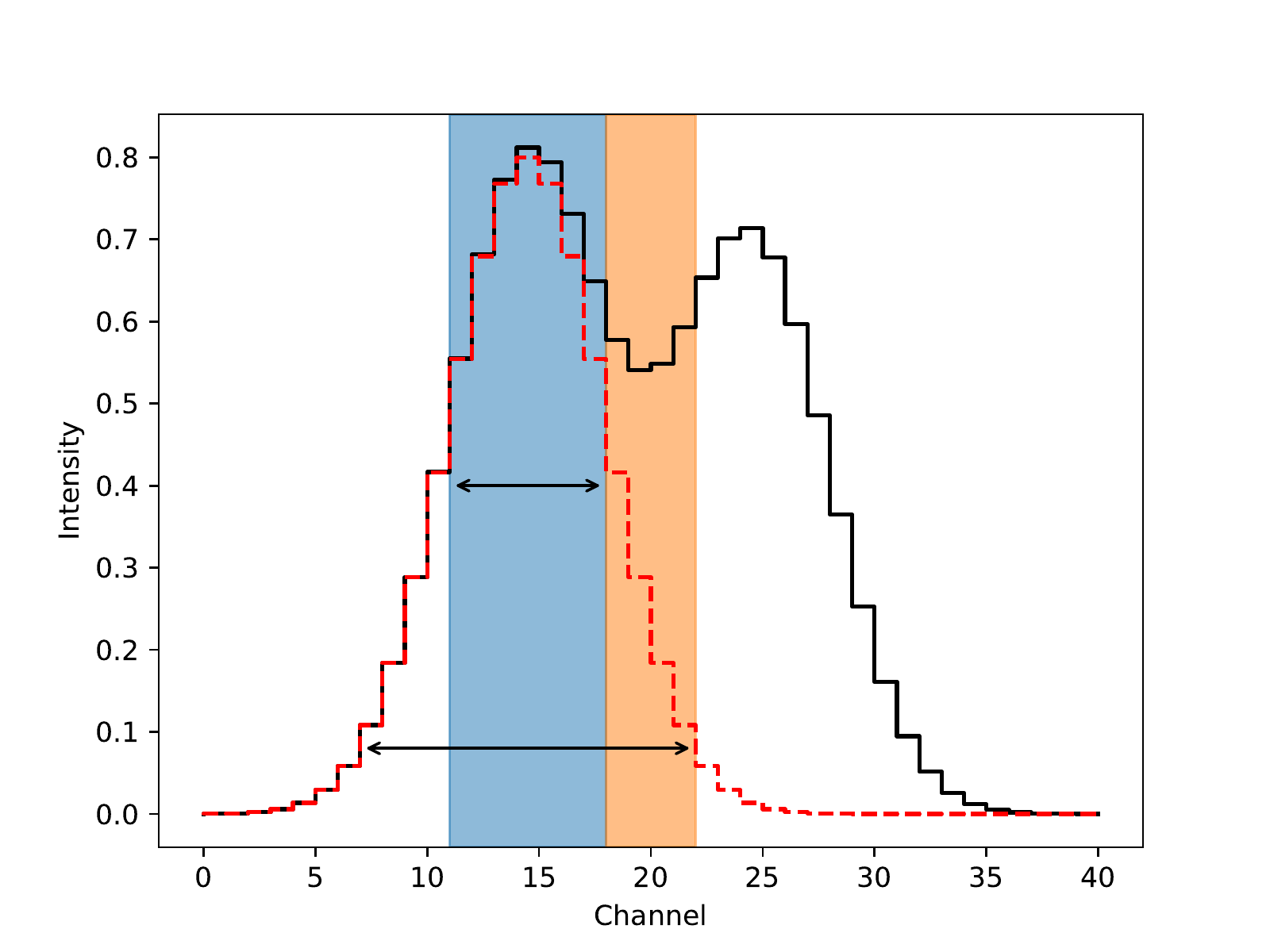}
    \caption{\footnotesize{Illustration of the effect of different spectral range for comparing line profiles. Simulated spectrum and observed spectrum are shown in red dashed line and black solid line respectively.}}
    \label{fig:compared_range}
\end{figure}

The spectral range for comparing line profiles from observation and simulation is crucial for calculating these consistency factors. As illustrated in Figure \ref{fig:compared_range}, within the small spectral range shown in blue, the simulated line shows a high degree of agreement in line intensity with the observed line. As a matter of fact there are two observed lines blended with each other. But, if a larger spectral range is used for comparing, the blended part colored in orange would be taken into account and thus reduce the degree of consistency. Therefore, using a large spectral range yields a value that describes the situation more accurately. In addition, as required by the Snyder criterion for frequency agreement, a resolved transition should be separated from its adjacent lines by more than its FWHM. So the spectral range for comparing line profiles should at least include the contiguous channels of which the simulated intensities are larger than 50\% of the peak intensity. As it is critical to have a clean transition for imaging, in our study, we adopt an even wider spectral range where only the channels with intensities larger than 10\% of the maximum are considered.

\begin{enumerate}[label=(\roman*), resume]
  \item \textit{Frequency agreement}. For characterizing the proximity in central frequencies, we define the $P$ (Product) factor, which is based on the product of the observed and simulated line profiles, in the following manner:
  \begin{equation}
    P\text{ factor} = \frac{\sum_{\nu_1}^{\nu_2}{I_\text{obs} \times I_\text{sim}}}{\sum_{\nu_1}^{\nu_2}{I_\text{obs}^\text{sorted} \times I_\text{sim}^\text{sorted}}}
    \label{eqn:pfactor} 
  \end{equation}
  
  In Equation \eqref{eqn:pfactor}, $I_\text{obs}$ and $I_\text{sim}$ are the observed and simulated intensity as functions of frequency respectively. Figure \ref{fig:pfactor}(a) shows a fictitious simulated spectrum in red and a fictitious observed spectrum in black. The product of the profiles in each channel is first calculated and plotted in Figure \ref{fig:pfactor}(b). For example, in Figure \ref{fig:pfactor}(a), the observed value in channel 10 is 0.835 and the simulated value is 0.643. Therefore, the product is 0.537 as shown in channel 10 of Figure \ref{fig:pfactor}(b). The simulated and observed spectrum are then sorted in intensity bins from lowest intensity (left) to highest intensity (right) in Figure \ref{fig:pfactor}(c) and denoted by $I_\text{sim}^\text{sorted}$ and $I_\text{obs}^\text{sorted}$ respectively. Once again, similar to Figure \ref{fig:pfactor}(b), we derived the product of the sorted spectra for each bin as shown in Figure \ref{fig:pfactor}(d). Finally, the P factor is calculated by dividing the area in Figure \ref{fig:pfactor}(b) by the area in Figure \ref{fig:pfactor}(d). The denominator represents the maximum achievable value of the numerator and normalizes the P factor to 100\%. In this example, since the central frequencies of the spectra are inconsistent, we obtain a rather low $P$ factor, 82\%.
  
  If the central frequencies of the compared lines are same, the bin of the simulated and observed intensity in each channel would be the same. So, after sorting, only the order would change but the product of intensity would remain the same. The P factor thus would reach 100\%. Figure \ref{fig:pexample} shows a couple of examples of the spectra with corresponding 5\%, 25\%, 50\% and 100\% P-values.
  
  In our analysis, the summation operations in Equation \eqref{eqn:pfactor} sum over the spectral range in which the simulated intensities larger than 10\% of its maximum as mentioned above. The threshold $P$ factor for assigning a line depends on the situation. In the study of \ce{C2H4O2} toward Sgr B2(N), we have a high confidence of the accuracy of rest frequencies for the three molecules and the source velocity of LMH, which are the major factors affecting the frequency agreement. As a result, we are rigorous with the $P$ factor and assign a transition with a $P$ factor at least larger than 90\% to be an unblended transition candidate.
  
  \begin{figure}[ht]
    \centering
    \epsscale{1.2}
    \plotone{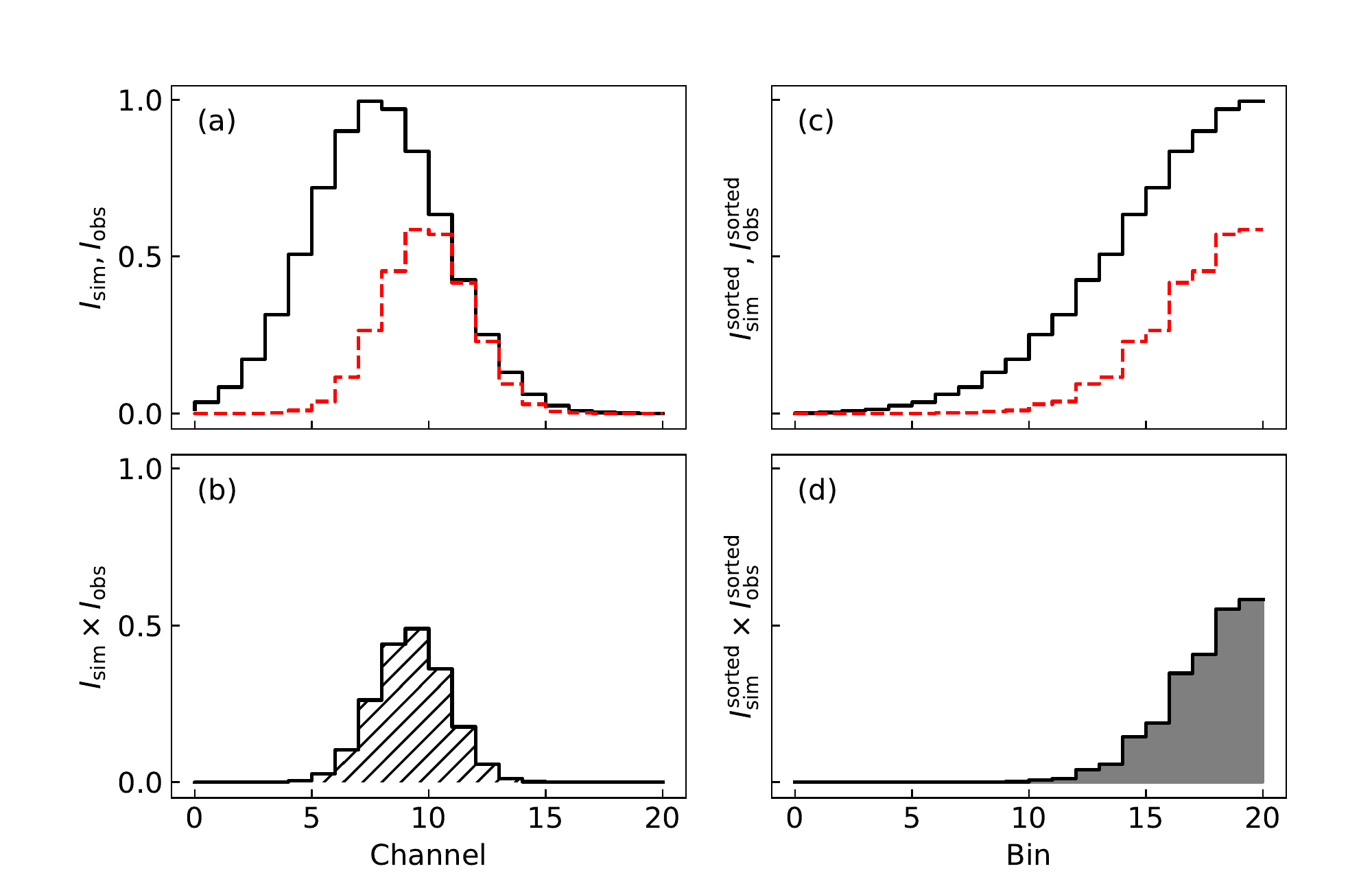}
    \caption{\footnotesize{Illustration of the calculation of the $P$ factor. In panel (a), the black solid lines represent the fictitious observed data while the red dashed lines represent the fictitious simulated data. Panel (b) represents the product of the compared spectra, i.e. the numerator in equation \eqref{eqn:pfactor}. Panel (c) shows the spectra sorted in their intensities. The products of the sorted spectra are plotted in panel (d) of which the areas represents the denominator in Eq. \eqref{eqn:pfactor}. The $P$ factor is 82\% in this case.}}
    \label{fig:pfactor}
  \end{figure}
  
   \begin{figure*}[ht]
    \centering
    \epsscale{1.2}
    \plotone{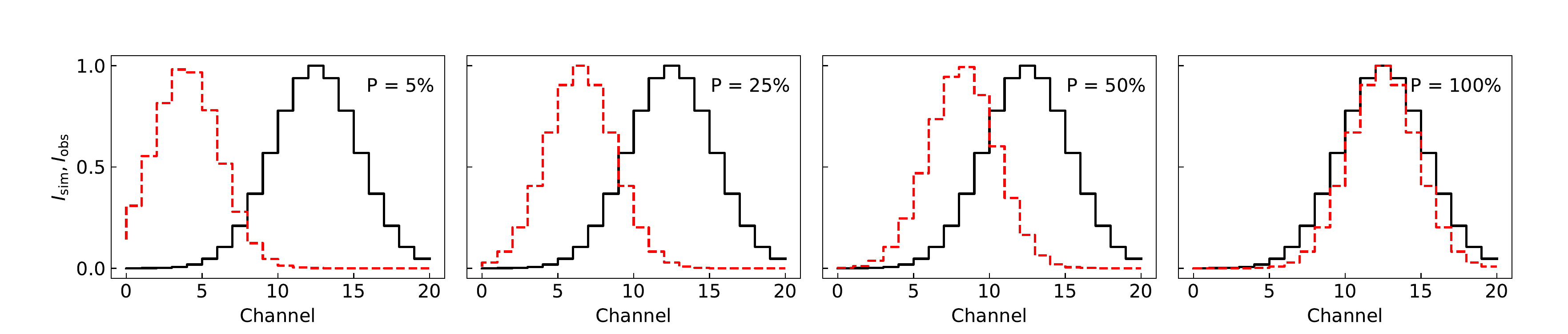}
    \caption{\footnotesize{Examples of the compared spectra with $P$ factors of 5\%, 25\%, 50\% and 100\% which are plotted in solid and dashed lines respectively.}}
    \label{fig:pexample}
  \end{figure*}
  
  \item \textit{Line Intensity}. The difference of the line intensities is measured by the $D$ (Difference) factor. It is the ratio between the difference of the integrated intensities of the compared spectra and the maximum of the two subtracted from 1:
  \begin{equation}
    D\text{ factor} = 1-\frac{\left|\sum_{\nu_1}^{\nu_2}{I_\text{obs}} - \sum_{\nu_1}^{\nu_2}{I_\text{sim}}\right|}{\max{\left(\sum_{\nu_1}^{\nu_2}{I_\text{obs}}, \sum_{\nu_1}^{\nu_2}{I_\text{sim}}\right)}}
    \label{eqn:dfactor}
  \end{equation}
  
  Similar to Equation \eqref{eqn:pfactor}, in Equation \eqref{eqn:dfactor}, $I_\text{obs}$ and $I_\text{sim}$ are the observed and simulated intensity, respectively, and the $D$ factor is calculated using the same spectral range of the $P$ factor. Figure \ref{fig:dfactor} shows fictitious observed and simulated lines with different intensities. The hashed area in Figure \ref{fig:dfactor}(a), which represents the numerator in Equation \eqref{eqn:dfactor}, is the difference in integrated line intensities while the filled area in Figure \ref{fig:dfactor}(b), which represents the denominator is the maximum of the two integrated intensities. In this example, the $D$ factor is 39\%.
  
  When there is no difference in the intensities, the numerator is 0. As a result, the $D$ factor is normalized to 100\%. In the worst case, where there is no emission in one of the spectra, the numerator would be same as the denominator and the $D$ factor equals 0\%. Figure \ref{fig:dexample} shows a couple examples of the spectra with a 5\%, 25\%, 50\% and 100\% D-value correspondingly.
  
  The threshold of the $D$ factor depends on the accuracy of the input temperature, line width and column density for simulating spectra to the actual values. However, there are still uncertainties in these input parameters, so that a relatively high tolerance of the difference of line intensities between the simulated and observed spectrum is needed. To decide the threshold of the $D$ factor, many simulations were run over a wide range of temperature and column density, with significant deviation in line intensity occurring when the $D$ factor dropped below 60\%. Therefore, in this study, we set a threshold of 60\% for the $D$ factor.
  
  \begin{figure}[ht]
    \centering
    \epsscale{1.2}
    \plotone{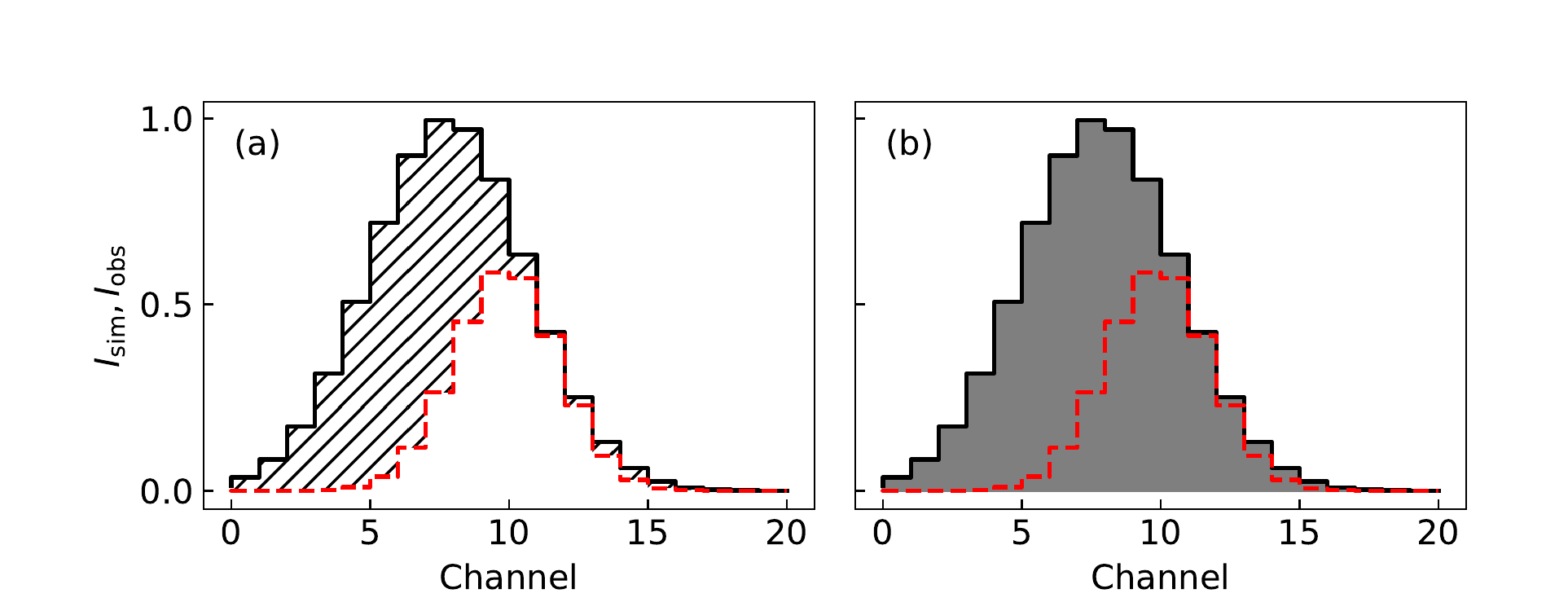}
    \caption{\footnotesize{Illustration of the calculation of the $D$ factor. The fictitious compared spectra are plotted in solid and dashed lines respectively. The hatched area in panel (a) represents the numerator in equation \eqref{eqn:dfactor} while the filled area in panel (b) represents the denominator. The $D$ factor is 39\% in this case.}}
    \label{fig:dfactor}
  \end{figure}

  \begin{figure*}[ht]
    \centering
    \epsscale{1.2}
    \plotone{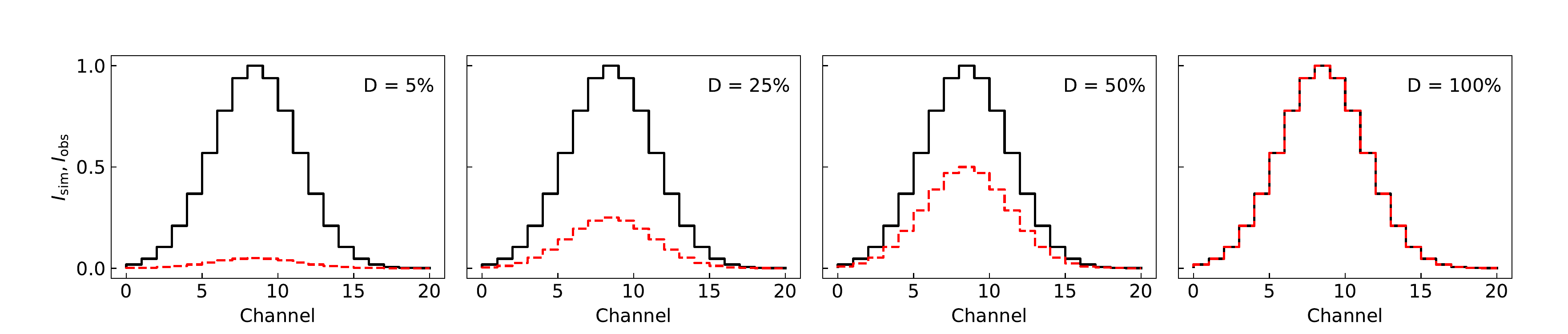}
    \caption{\footnotesize{Examples of the compared spectra with $D$ factors of 5\%, 25\%, 50\% and 100\% which are plotted in solid and dashed lines respectively. The compared spectra are given the same central frequencies.}}
    \label{fig:dexample}
  \end{figure*}
\end{enumerate}

\begin{enumerate}[label=(\roman*), resume]
  \item \textit{Presence of Transitions with Observable Intensity}. The fifth and last Snyder criterion suggests that all transitions with intensity predictions leading to detectable signal levels must be present. We further confirmed the candidate transitions by checking their corresponding connected transitions, i.e. the transitions connected by favorable transition probabilities \citep{slhf2005}. Similarly to the connected transitions, the A-E pair transitions for MF and AcA have similar line strengths, energy levels, and transition probabilities, and therefore each pair of transitions should have roughly the same intensity and line shape. Both the $A$ and $E$ state transitions must be detected if any of them has a detectable signal level. In this work, all the transitions of the three isomers with observable intensities are present and none of the strong lines is missing within the observed spectra, although a large number of these lines are blended with neighboring features which leads to small $P$ and $D$ factors for these lines. 
\end{enumerate} 

We are aware that our selection method suffers from several limitations: (1) There is the possibility of having absorption in the observed spectra. In that case, the methods of making use of the above consistency factors will be broken and the resulting factors would have abnormal values such as values larger than 100\% or negative values. Therefore, we exclude the transitions when there is obvious absorption into our calculations of P and D values. (2) The transition maps of unblended transitions might suffer from the contamination of spatial distributions, which could not be recognized by analyzing the spectrum toward one region. To avoid this problem, we have additionally checked every interferometric transition map of the less contaminated transitions found by our methods and marked the spatially polluted transitions, as discussed in Section \ref{sec:transition_maps}. (3) Unfortunately, due to the large line width and the high density of emission lines at mm wavelengths, lines could be blended very frequently with adjacent emission lines of the same molecule and appear as single-line spectral features. These transitions still have high values of $P$ and $D$ and would be identified as clean transitions. However, since their line profiles overlap with one another, they do not qualify for imaging. Thus, we carefully inspect the spectra of the transitions with the two consistency factors over our thresholds and highlight these kinds of "grouped" transitions, as discussed in Section \ref{sec:transition}-\ce{HCOOCH3}. These limitations reveal the complexity and difficulty of identifying and mapping a molecule in a source with high spectral line densities.

In spite of its shortcomings, as the focus of the study is on the rotational emission transitions of the \ce{C2H4O2} isomers, the transition selection processes in this study is in accordance with the rigorous Snyder criteria. This method also facilitated the detection of weaker and undiscovered transitions of our targets and enabled automation of the analysis of a survey with a large bandwidth by quantifying the consistency of the observed and simulated spectra. The details of the transition selection for the three isomers are summarized below.

\subsection{Transitions of \ce{C2H4O2}} \label{sec:transition}
For assigning the least contaminated transitions, we set thresholds for both $P$ and $D$ factors respectively to select the corresponding transition candidates first. Theoretically, the two factors for a clean transition, which has no difference in either center frequency or line intensity between observed and simulated lines, are both 100\%. In this study, we assign a transition with a $P$ factor at least larger than 90\% and a $D$ factor larger than 60\% to be an unblended transition candidate. After that, we further check if there is serious contamination from other transitions of the same molecule and if the spatial distribution is consistent with other transitions. The details of the selection of the most uncontaminated transitions of MF, GLA and AcA are now detailed.

\paragraph{\ce{HCOOCH3}} Within the observed spectral range, 84091 to 114368 MHz, there are 1516 transitions of MF recorded in the JPL catalog, including both the vibrationally ground state and first torsionally excited state. In the LTE simulated spectra of MF with an excitation temperature of $T_\text{ex}=190\ \mathrm{K}$, 197 transitions were found to be above a signal-to-noise ($S/N$) level of 3 and are potentially detectable.

\begin{figure*}[htp!]
\epsscale{1.2}
\plotone{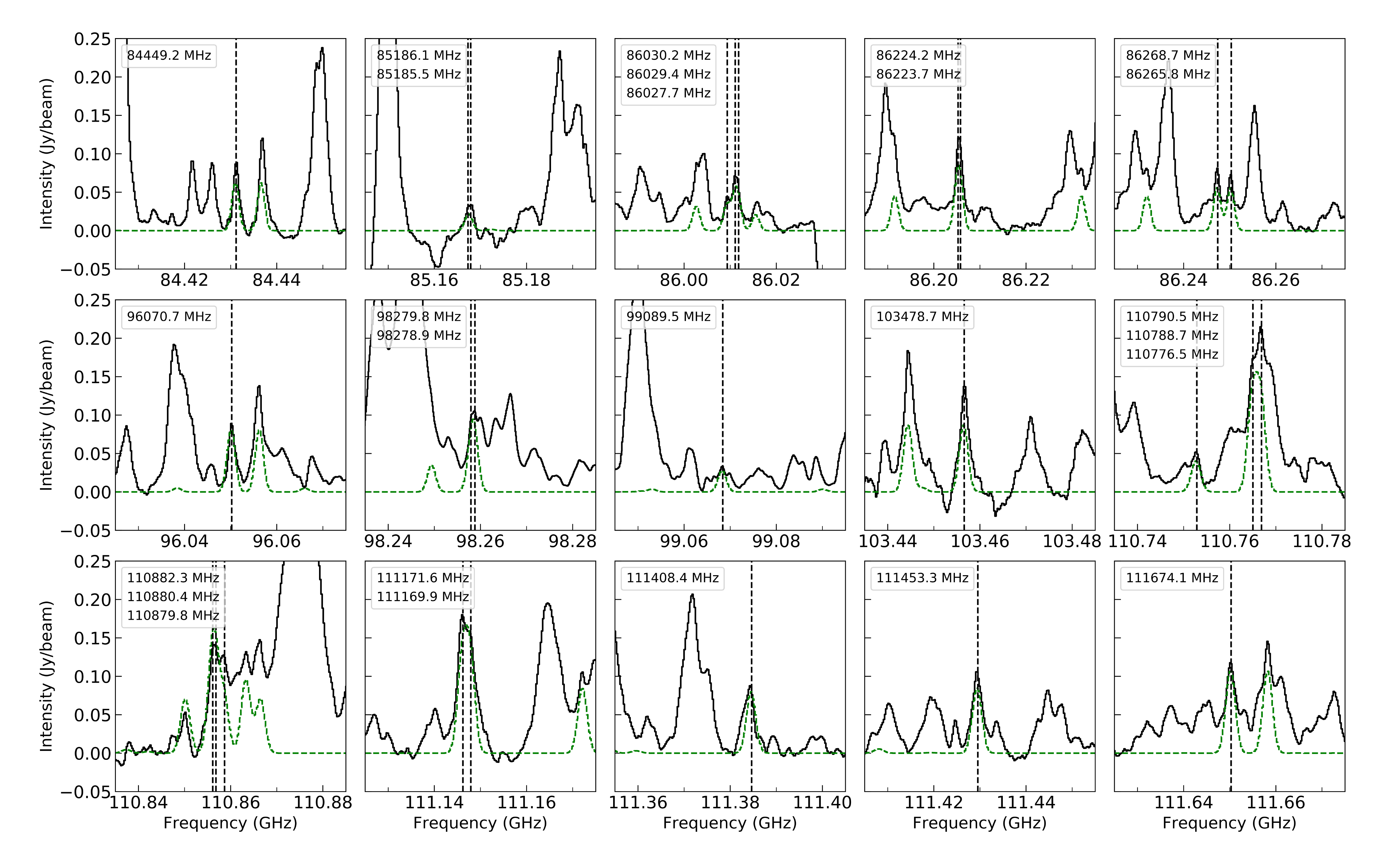}
\caption{\footnotesize{Observed spectra of the 26 transitions of MF with a $P$ factor $> 95\%$ and a $D$ factor $> 60\%$ toward LMH are plotted in black over the best-fit LTE simulated spectra of MF in green. The observed spectra are extracted from a uniform synthesized beam of $3.01\arcsec \times 1.47\arcsec$. The observed frequencies in GHz are marked below each panel. The intensities are indicated in $\mathrm{Jy beam^{-1}}$. }\label{MF-spec}}
\end{figure*}

Through calculating the correlation factors, there are 26 MF transitions, which have $P$ factors larger than 95\% and $D$ factors larger than 60\% and satisfy our criteria. These 26 transitions of MF are shown in Figure \ref{MF-spec} and given in Table \ref{MF-table} with the following parameters: rest frequency(uncertainty) in MHz, rotational quantum number, upper-state energy ($E_u$) in Kelvin, intensities expressed as the product of the relevant component of the dipole moment squared times transition strength ($S_{ij}\mu^{2}$) in Debye$^{2}$ \citep{ts1975}, FWHM in MHz, critical density ($n_\mathrm{crit}$) in $cm^{-3}$ and the two resulting correlation factors. 

Many of A-E pairs and connected transitions of MF usually have little difference in frequencies compared to the observed line width, some of which are almost completely blended with each other and appear as a single line spectral feature. Therefore, as mentioned before, the strict use of our methodology for comparing spectral lines fails to identify these wide spectral features. As expected, there are several sets of transitions blended together among the 26 MF transitions. Therefore, we only imaged the transition maps of the remaining 10 transitions of MF which are the least contaminated, including 2 transitions in the $v_\mathrm{t} = 1$ state, $9(1,8)-8(1,7)\ A\ v=1$ and $8(3,5)-7(3,4)\ A\ v=1$. 

MF has been widely studied in the ISM. Previous studies that targeted Sgr B2(N) reported several transitions of MF, within the frequency coverage of the EMoCA survey \citep{hvsjl2001, rslmk2002, rsfls2003, rmr2006}. We critically examined seven transitions of MF from \citet{hvsjl2001}, two from \citet{rslmk2002}, six from \citet{rsfls2003}, and seven from \citet{rmr2006}, and found that most of these previously reported transitions are contaminated by adjacent lines in the observed spectra toward LMH and have relatively low $P$ and $D$ values. Only the $7(3,5)-6(3,4)$ A and E lines \citep{hvsjl2001}, the $9(4,5)-8(4,4)\ E$ transition \citep{rmr2006}, the $9(4,5)-8(4,4)\ A$ transition \citep{rsfls2003}, and the $9(1,8)-8(1,7)\ E$ transition \citep{rmr2006} qualify for imaging.

\startlongtable
\begin{deluxetable*}{llrrrrrccl}
    \tabletypesize{\scriptsize}
    \tablecaption{The Most Uncontaminated Transitions of MF \label{MF-table}}
    \tablehead{
        \\
        \colhead{Rest Frequency}	& \colhead{Transition}	& \colhead{$E_u$}	    & \colhead{$\log_{10}{\frac{A_{ul}}{\mathrm{s^{-1}}}}$}	&
        \colhead{$S_{ij}\mu^{2}$}	& \colhead{FWHM}    & \colhead{$n_\mathrm{crit}$}  & \colhead{P Factor}	& \colhead{D Factor}	& \colhead{Comments} \\
        \colhead{(MHz)}	            & ~                 	& \colhead{(K)} 	    & ~                             &
        \colhead{(Debye$^2$)}    	& \colhead{(MHz)}   & \colhead{($cm^{-3}$)} & (\%) 	                & (\%)                  & ~
    }
    \startdata
        84449.169(10)   &   $7(2,6)-6(2,5)\ E$                  &   19.00   &   -5.0992 &17.0   &1.533 &1992.9  &97.6   &79.9   &l, h\\
        85185.466(10)   &   $7(5,3)-6(5,2)\ A\ v=1$             &   220.89  &   -5.3602 &9.1    &      &        &97.4   &63.4   &\\    
        85186.063(10)   &   $7(5,2)-6(5,1)\ A\ v=1$             &   220.89  &   -5.3602 &9.1    &      &        & "     & "     &\\
        86027.723(10)\tablenotemark{a}  & $7(5,3)-6(5,2)\ E$    &   33.12   &   -5.3461 &9.1    &      &        &99.6   &70.8	&\\
		86029.442(10)\tablenotemark{a}  & $7(5,3)-6(5,2)\ A$    &   33.11   &   -5.3459 &9.1    &      &        & "     & "     &\\
		86030.186(10)\tablenotemark{a}  & $7(5,2)-6(5,1)\ A$    &   33.11   &   -5.3459 &9.1    &      &        & "     & "     &\\
		86223.655(10)   &   $7(4,3)-6(4,2)\ E$                  &   27.17   &   -5.2074 &12.5   &      &        &99.3   &65.2   &\\
		86224.160(10)   &   $7(4,4)-6(4,3)\ E$                  &   27.16   &   -5.2075 &12.5   &      &        & "     & "     &\\
		86265.796(10)\tablenotemark{a}  &   $7(3,5)-6(3,4)\ A$  &   22.51   &   -5.1209 &15.2   &1.739 &1707.4  &99.7   &63.2   &h\\
		86268.739(10)\tablenotemark{a}  &   $7(3,5)-6(3,4)\ E$  &   22.52   &   -5.1248 &15.1   &1.800 &1635.3  &99.7   &69.3   &l\\
        96070.725(10)   &   $8(2,7)-7(2,6)\ E$                  &   23.61   &   -4.9191 &19.8   &1.940 &2712.5  &97.7   &97.6	&l, h\\
        98278.921(10)   &   $8(6,3)-7(6,2)\ E$                  &   45.13   &   -5.2179 &9.3    &      &        &96.4   &65.3   &\\  
        98279.762(10)   &   $8(6,3)-7(6,2)\ A$                  &   45.13   &   -5.2179 &9.3    &      &        & "     & "     &\\
        99089.518(10)   &   $8(3,5)-7(3,4)\ A\ v=1$             &   215.08  &   -4.9159 &18.2   &2.095 &2609.9  &98.7   &64.8   &l\\
        103478.663(10)  &   $8(2,6)-7(2,5)\ A$                  &   24.63   &   -4.8188 &20.0   &2.326 &3070.6  &98.9   &63.4   &l, h\\
		110776.499(10)  &   $9(1,8)-8(1,7)\ A\ v=1$             &   215.75  &   -4.7151 &23.1   &2.320 &4183.4  &96.9   &68.0	&l, h\\
		110788.664(10)  &   $10(1,10)-9(1,9)\ E$		        &   30.27	&   -4.7051	&26.2	&      &        &96.6	&62.5	&\\
    	110790.526(10)  &   $10(1,10)-9(1,9)\ A$                &   30.26	&   -4.7049	&26.2	&      &        & "     & "     &\\
		110879.766(10)  &   $9(3,7)-8(3,6)\ E$                  &   32.58   &   -4.7510 &21.2   &      &        &97.1	&89.6	&\\
		110880.447(10)  &   $9(5,5)-8(5,4)\ A$                  &   43.16   &   -4.8592 &16.6   &      &        & "     & "     &\\
		110882.331(10)  &   $9(5,5)-8(5,4)\ E$                  &   43.16   &   -4.8594 &16.6   &      &        & "     & "     &\\
		111169.903(10)  &   $10(0,10)-9(0,9)\ E$                &   30.25   &   -4.7002 &26.2   &      &        &99.6	&82.7	&\\
		111171.634(10)  &   $10(0,10)-9(0,9)\ A$                &   30.23   &   -4.7002 &26.2   &      &        & "     & "     &\\
    	111408.412(10)\tablenotemark{b}  &   $9(4,5)-8(4,4)\ E$ &   37.26   &   -4.8123 &18.2   &2.032 &3841.2  &94.8	&86.1	&l, h\\
    	111453.300(10)\tablenotemark{c}  &   $9(4,5)-8(4,4)\ A$                  &   37.24   &   -4.7878 &19.2   &2.681 &3081.6  &99.9	&75.0	&l, h\\
    	111674.131(10)\tablenotemark{b}  &   $9(1,8)-8(1,7)\ E$                  &   28.14   &   -4.7037 &23.2   &2.988 &3362.0  &99.6	&78.7	&l, h
    \enddata
    \tablecomments{\footnotesize{Pertinent parameters of the transitions assigned to MF with $P > 95\%$ and $D > 60\%$ taken from JPL catalog. Here we consider the transitions for both the vibrational ground state and first excited state. $Q_r = 12.45\times T_r^{1.5}$ \citet{rsfls2003}. P and D factors are calculated based on the observed spectra toward LMH and the simulated spectra assuming an excited temperature of 190 K. "l" and "h" mark the transitions for which the low- and high-velocity components are used for generating the chemical map of MF. The transitions with ditto marks in the column of P and D factors are blended with the transitions in the previous row. \tablenotetext{a}{Lines previously reported in \citet{hvsjl2001}.} \tablenotetext{b}{Lines previously reported in \citet{rmr2006}.}\tablenotetext{c}{Lines previously reported in \citet{rsfls2003}.}}}
\end{deluxetable*}

\begin{figure*}[htp!]
\epsscale{1.2}
\plotone{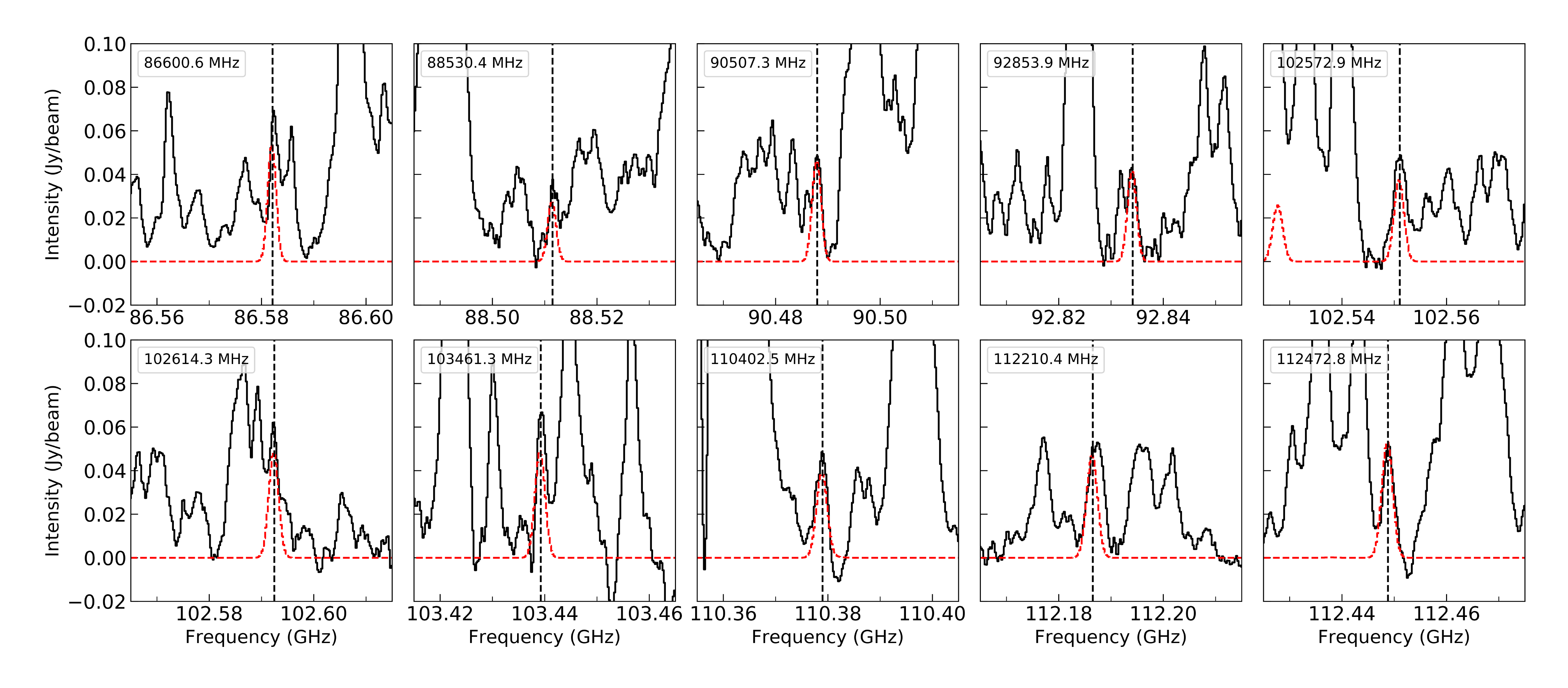}
\caption{\footnotesize{Same as Figure \ref{MF-spec} for the 10 transitions of GLA. The best-fit LTE simulated spectra of GLA are shown in red.}\label{GLA-spec}}
\end{figure*}

\paragraph{\ce{CH2OHCHO}} In the 84091-114368 MHz frequency range of the EMoCA survey, 402 ground-state transitions of GLA are listed in the CDMS catalog. We focused on the 83 transitions of GLA with $S/N > 3$ under the simulation condition $T_\text{ex}=190\ \mathrm{K}$, covering $J$ values up to 18. We found 10 such lines with a $P$ factor $>90\%$ and a $D$ factor $>60\%$ which were assigned to be the transitions of GLA. Table \ref{GLA-table} summarizes the 10 most uncontaminated transitions with the pertinent molecular parameters and the $P$ and $D$ factors. The spectral lines used in the detection of GLA toward LMH are shown in Figure \ref{GLA-spec}. 

As in the case of MF, we studied the three lines of GLA reported by \citet{hlj2000} and the 12 lines from \citet{hawpz2006}. But none of them has consistency factors over our threshold. The observed line intensities of these 11 lines are much higher than the simulated values; this problem is interpreted as serious blending with other molecules. In contrast, the previously unreported 10 transitions, following our criteria, are the most uncontaminated lines identified with GLA at this frequency range toward Sgr B2(N).

\begin{deluxetable*}{llrrrrrccl}
    \tablewidth{0pt}
    \tablecaption{The Most Uncontaminated Transitions of GLA\label{GLA-table}}
    \tablehead{
        \\
        \colhead{Rest Frequency}	& \colhead{Transition}	& \colhead{$E_u$}	    & \colhead{$\log_{10}{\frac{A_{ul}}{\mathrm{s^{-1}}}}$}	&
        \colhead{$S_{ij}\mu^{2}$}	& \colhead{FWHM}    & \colhead{$n_\mathrm{crit}$}  & \colhead{P Factor}	& \colhead{D Factor}	& \colhead{Comments} \\
        \colhead{(MHz)}	            & ~                 	& \colhead{(K)} 	    & ~                             &
        \colhead{(Debye$^2$)}    	& \colhead{(MHz)}   & \colhead{($cm^{-3}$)} & (\%) 	                & (\%)                  & ~
    }
    \startdata
        86600.5997(67)  &   $17(5,12)-17(4,13)$ &101.14 &-4.8624    &   63.6    &1.684  &3210.0 &91.0   &86.6   &\\	
		88530.4001(46)  &   $8(4,5)-8(3,6)$     &29.73  &-4.9874    &   21.7    &2.125  &1950.2 &94.9   &68.9   &l, h\\
		90507.2610(133) &   $17(3,14)-17(2,15)$ &93.77  &-4.9064    &   50.3    &2.012  &2536.7 &98.1   &80.4   &l\\
		92853.9487(49)  &   $12(4,9)-12(3,10)$  &53.23  &-4.8762    &   35.7    &2.231  &2517.1 &97.2   &79.2	&l\\
		102572.9391(82) &   $14(3,12)-14(2,13)$ &63.60  &-4.8764    &   30.7    &2.960  &2094.5 &99.0   &62.2   &l, h\\
		102614.3429(137)&   $18(3,15)-18(2,16)$ &104.10 &-4.7784    &   49.0    &2.373  &3274.6 &97.9   &64.4   &l\\
		103461.3106(66) &   $15(4,12)-15(3,13)$ &76.75  &-4.7501    &   42.8    &2.102  &3979.1 &95.0   &77.7	&h\\
		110402.5367(87) &   $15(3,13)-15(2,14)$	&71.79  &-4.8048    &   31.0	&2.549  &3087.6 &99.4   &80.3	&h\\
		112210.4422(52) &   $12(5,8)-12(4,9)$   &58.62  &-4.6471    &   34.3    &3.638  &3161.2 &92.2   &70.7	&l, h\\
		112472.7946(50) &   $14(5,10)-14(4,11)$ &73.76  &-4.6197    &   42.0    &2.524  &4862.7 &98.6	&98.4   &l\\ 
    \enddata
    \tablecomments{\footnotesize{Pertinent parameters of the transitions assigned to GLA with $P > 90\%$ and $D > 60\%$ are taken from CDMS catalog. Here we only consider the transitions for vibrational ground state. $Q_r = 6.868\times T_r^{1.5}$ is derived from the partition function at different temperatures given in CDMS. P and D Factors are calculated based on the simulated spectra assuming the excited temperature as 190 K. "l" and "h" mark the transitions of which the low- and high-velocity components are used for generating the chemical map of GLA respectively.}}
\end{deluxetable*}

\paragraph{\ce{CH3COOH}} There are 573 vibrational ground-state transitions of AcA from the SLAIM database in the 84091-114368 MHz spectral range; 136 of AcA transitions have $S/N$ larger than 3 in the simulated spectra with $T_\text{ex}=190\ \mathrm{K}$ including the best known six four-fold degenerate transitions, 90203.35, 90246.26, 100855.43, 100897.46, 11507.27, and 111548.53 MHz lines \citep{msml1997, rsfls2003}. However, the emission lines of AcA suffer seriously from contamination from other species. The observed intensity of some of the well-known four-fold degenerate transitions of AcA, for example $9(*,9)-8(*,8)\ A$ at 100897.46 MHz where asterisks can be 0 or -1, are much higher than their simulated intensity and the observed intensity of their corresponding E pairs in spite of their similar upper-level energy. 

\begin{figure*}[htp!]
\epsscale{1.2}
\plotone{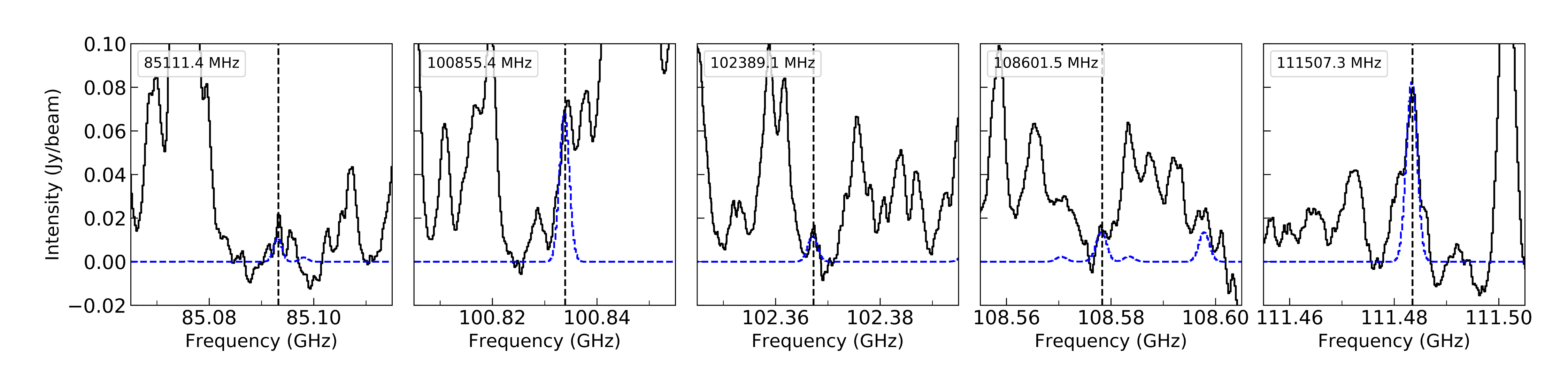}
\caption{\footnotesize{Same as Figure \ref{MF-spec} for the five transitions of AcA. The best-fit LTE simulated spectra of AcA are shown in blue.}\label{AcA-spec}}
\end{figure*}

There are only three transitions of AcA over the threshold $P$ factor $>90\%$ and $D$ factor $>60\%$ with $S/N > 3$ as shown in Table \ref{AcA-table}. If we go to weaker lines, there are two more transitions of AcA satisfying the requirement of the consistency factors but with lower $S/N$. They are the $23(15,9)-23(14,10)$ transition with $S/N \sim 2.7$ and the $24(-13,11)-24(-12,12)$ transition with $S/N \sim 2.6$. The relatively high values of the consistency factors and the similar spatial distribution of these two transitions with that of the three stronger lines make their assignments reliable. Therefore, we image the transition maps of the five transitions shown in Table \ref{AcA-table}. Figure \ref{AcA-spec} shows the observed and simulated spectra of these transitions toward the LMH region. 

\begin{deluxetable*}{llrrrrrccl}
    \tablewidth{0pt}
    \tablecaption{The Most Uncontaminated Transitions of AcA \label{AcA-table}}
    \tablehead{
        \\
        \colhead{Rest Frequency}	& \colhead{Transition}	& \colhead{$E_u$}	    & \colhead{$\log_{10}{\frac{A_{ul}}{\mathrm{s^{-1}}}}$}	&
        \colhead{$S_{ij}\mu^{2}$}	& \colhead{FWHM}    & \colhead{$n_\mathrm{crit}$}  & \colhead{P Factor}	& \colhead{D Factor}	& \colhead{Comments} \\
        \colhead{(MHz)}	            & ~                 	& \colhead{(K)} 	    & ~                             &
        \colhead{(Debye$^2$)}    	& \colhead{(MHz)}   & \colhead{($cm^{-3}$)} & (\%) 	                & (\%)                  & ~
    }
    \startdata
        85111.3900(30)                  &   $23(15,9)-23(14,10)$                &251.54 &-5.3344    & 30.3  &1.201  &1491.4 & 95.9  & 65.7  &l\\  
		100855.4300(10)\tablenotemark{a}&   $9(*,9)-8(*,8)\ E$\tablenotemark{a} &25.67  &-4.5116    & 49.0	&3.450  &4092.4 & 94.7	& 70.7	&l\\
		102389.0800(10)	                &   $24(-13,11)-24(-12,12)$             &266.21 &-5.1281    & 29.2	&1.544  &2244.7 & 96.7	& 86.0	&l\\
		108601.4600(10)                	&   $21(-10,11)-21(-9,12)$              &199.07 &-5.1076    & 22.5	&2.008  &1919.4 & 90.1	& 82.6	&\\
		111507.2800(10)\tablenotemark{b}&   $10(*,10)-9(*,9)\ E$\tablenotemark{a}&31.02 &-4.3757    & 54.8	&3.037  &7029.4 & 99.4	& 86.7	&\\
    \enddata
    \tablecomments{\footnotesize{Pertinent parameters of the transitions assigned to AcA with $P > 90\%$ and $D > 60\%$ are taken from SLAIM catalog. Here we only consider the transitions for vibrational ground state. $Q_r = 14.10\times T_r^{1.5}$ \citet{rsfls2003}. P and D Factors are calculated based on the simulated spectra assuming the excited temperature as 190 K. "l" marks the transitions that are most uncontaminated and of which the low-velocity component is used for generating the chemical map of AcA respectively. Each of the fourfold degenerate transitions consists of two a-type and two b-type degenerate transitions, and is shown with the K quantum numbers as substituted by an asterisk. \tablenotetext{a}{Fourfold transition detected by \citet{msml1997}; asterisk could be 0 or -1.} \tablenotetext{b}{Fourfold transition detected by \citet{rsfls2003}; asterisk could be 0 or 1.}}}
\end{deluxetable*}

Pertinent line parameters of all the detectable transitions of the isomeric species of \ce{C2H4O2} observed in this survey with signal to noise ratio ($S/N$) larger than $\sim 3$ are summarized in Appendix Tables \ref{MF-appendix-table}, \ref{GLA-appendix-table} and \ref{AcA-appendix-table} respectively. 

\subsection{Column Density}\label{sec:columndensity}
To more accurately determine the excitation temperature, we perform a $\chi^2$ fitting over these less contaminated transitions to statistically determine $\mathrm{T_{ex}}$ toward the LMH region. The formula for $\chi^2$ is defined as:
\begin{equation}
    \chi^2 = \sum_{i=1}^{n} \frac{(I_{\text{obs},i} - I_{\text{sim},i})^2}{\sigma^2},
\label{eqn:chi} 
\end{equation}\ where $I_{\text{sim}}$ and $I_{\text{obs}}$ represent the simulated peak intensity and the observed values at the corresponding frequency respectively, and $\sigma$ is the rms noise level of the observed spectrum.  

For MF, a $T_\text{ex}$ of 201(12) K and a $N_{T,MF}$ of $1.18(0.05) \times 10^{17}\ \mathrm{cm}^{-2}$ minimized the $\chi^2$ to 314.6 with the number of degrees of freedom of 15 and yield the best fit. The $\chi^2$ fitting over these 10 GLA transitions suggested that a $T_\text{ex}$ of 146(25) K and a $N_{T,GLA}$ of $1.57(0.12) \times 10^{16}\ \mathrm{cm}^{-2}$, which minimized the $\chi^2$ to 14.7 with the number of degrees of freedom of 8, best fit the observation data. A result of $T_\text{ex} = 296(59) \mathrm{K}$ and $N_{T,AcA} = 4.63(0.92) \times 10^{16}\ \mathrm{cm}^{-2}$ is obtained by $\chi^2$ fitting using the five AcA transitions. The minimum value of $\chi^2$ is 3.3 with a degree of freedom of 3.

Based on the ALMA Band 3 observations, the best fit to the observed spectra toward LMH in this study suggests abundance ratios of MF, AcA, and GLA [\ce{HCOOCH3}]:[\ce{CH3COOH}]:[\ce{CH2OHCHO}] of $\sim$ 7.5:2.9:1. Although this result is in agreement with the suggested ratio of [\ce{HCOOCH3}]:[\ce{CH2OHCHO}] $\sim$ 6.7 observed with the OVRO array \citep{hlj2000}, we find an enhancement of the abundance of AcA with respect to those reported by \citet{rslmk2002}. In \citet{rslmk2002}, the observations conducted with the OVRO array reported an MF to AcA abundance ratio, [\ce{HCOOCH3}]:[\ce{CH3COOH}], of (16 - 32):1 toward the LMH region. Based on a uniform analysis of observations with the IRAM 30m telescope, \citet{bmmsc2013} suggest that $T_\text{ex}$ = 80 K and $N_{T,MF} = 4.37 \times 10^{17}\ \mathrm{cm}^{-2}$ with a source size of 4.0$\arcsec$ for MF; $T_\text{ex}$ = 80 K and $N_{T,GLA} = 1.8 \times 10^{15}\ \mathrm{cm}^{-2}$ with a source size of 10.0$\arcsec$ for GLA; and $T_\text{ex}$ = 100 K and $N_{T,AcA} = 1.05 \times 10^{16}\ \mathrm{cm}^{-2}$ with a source size of 4.0$\arcsec$ for AcA, based on which the abundance ratios among the three molecules are [\ce{HCOOCH3}]:[\ce{CH3COOH}]:[\ce{CH2OHCHO}] $\sim$ 242:5.8:1. The inconsistency between the two values for [\ce{HCOOCH3}]:[\ce{CH3COOH}]:[\ce{CH2OHCHO}] can be justified in part by the different observations and data processing methods. The 30m IRAM observation has beam sizes larger than 11$\arcsec$ and, therefore, cannot constrain the source sizes of the three molecules. In addition, the rotational diagram method that \citet{bmmsc2013} used to determine column density requires a large number of the less contaminated transitions covering a large range of energy levels. But the detected AcA transitions in \citet{bmmsc2013} have a narrow range of energy levels, $E_l$ of 16-25 K, and there are only 5 detected GLA transitions, which could introduce large uncertainties to the results. In contrast, in our study, we are more rigorous with line contamination and use $\chi^2$ fitting to constrain the excitation temperature. In addition, the small synthesized beam sizes of the ALMA observations resolve the source sizes of the three molecules well.

\section{RESULTS} \label{sec:result}
\begin{figure}[ht]
    \centering
    \epsscale{1.2}
    \plotone{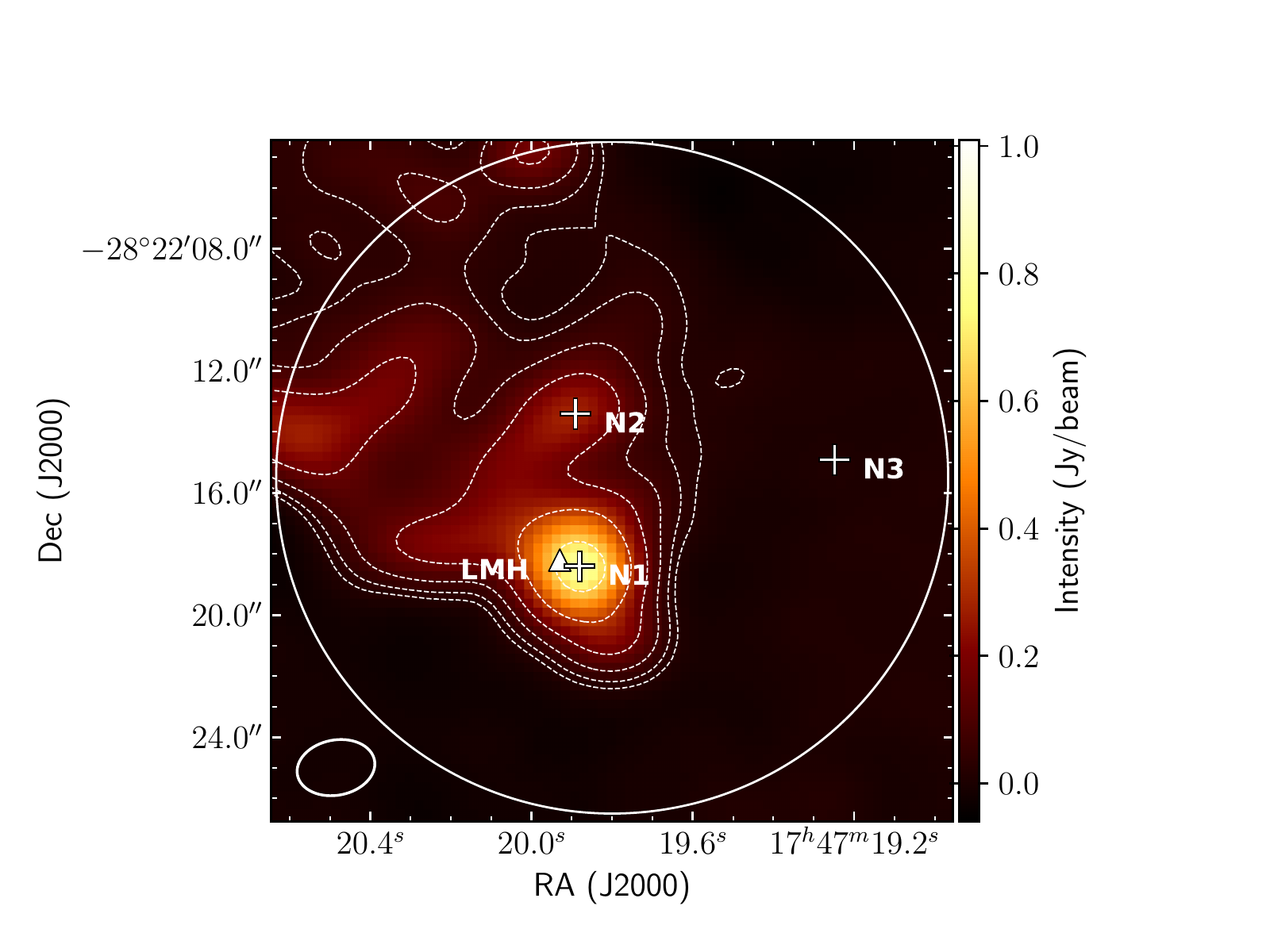}
    \caption{\footnotesize{Continuum map of the Sgr B2(N) region obtained with EMoCA survey at 85.9 GHz \citep{bmgk2016, bbmgm2017}. Contour levels start at five times the rms noise level (4.1 mJy/beam) and double in value up to the peak value. The solid circle shows the aperture ($11\arcsec \times 11\arcsec$) from which the spectra was extracted from to examine the velocity components of the three isomers. The triangle marker shows the LMH region.}}
    \label{fig:cont}
\end{figure}

In order to determine the velocity components of the three isomers, we examined the spectral transitions of the three isomers as extracted from an aperture with a diameter of $11 \arcsec$ and covering most of the continuum emission region. Figure \ref{fig:cont} shows the thermal dust continuum map of the Sgr B2(N) region as obtained with ALMA Band 3 observations at 85.9 GHz along with the LMH, N1, N2 and N3 emission cores \citep{bmgk2016, bbmgm2017} which will be referenced below. The average systemic $V_{lsr}$ of the N1 core is $\sim 64\ \mathrm{km\ s^{-1}}$ while that for N2 and N3 is $V_{lsr} \sim 73$.

In the most unblended spectral transitions within an aperture comparable to the size of the major continuum emission region, MF has two emission features at two different systemic velocities, $V_{lsr} \sim 64$ and $73\ \mathrm{km\ s^{-1}}$. The velocity constitution of MF is consistent with the findings in \citet{bmgk2016} and \citet{bbmgm2017}. In contrast to \citet{bmmsc2013}, who did not detect the high-velocity component of GLA toward Sgr B2(N) with the IRAM 30m line survey, two velocity components of GLA are also identified in our study. This difference from our result could be explained by the low sensitivity of the IRAM telescope due to beam dilution and smaller total collecting area. ALMA's high angular resolution and sensitivity give us a better understanding of the velocity components of these LAMs. In contrast to MF and GLA, however, AcA had only a single velocity component near $64\ \mathrm{km\ s^{-1}}$ resolved. There is no AcA emission above $1 \sigma$ around the N2 or N3 regions for any AcA transitions.  

\subsection{Transition Maps}\label{sec:transition_maps}
We carried out an intensive study on the spatial mapping of the \ce{C2H4O2} isomeric family with the most uncontaminated transitions that were discussed in Sec \ref{sec:transition}. We generated peak intensity images (i.e. moment-8 map) for the least contaminated transitions of the \ce{C2H4O2} isomeric species and refer to them as transition maps. For MF and GLA, we imaged each of the transitions near the two systemic velocity components separately. For the low-velocity component, the integrated velocity range is 59-68 $\mathrm{km\ s^{-1}}$ while for the high-velocity component, the range is 70-75 $\mathrm{km\ s^{-1}}$. For AcA, only the one velocity range, 59-67 $\mathrm{km\ s^{-1}}$, was imaged.

\begin{turnpage}
\begin{figure*}
    \includegraphics[width=1\textheight]{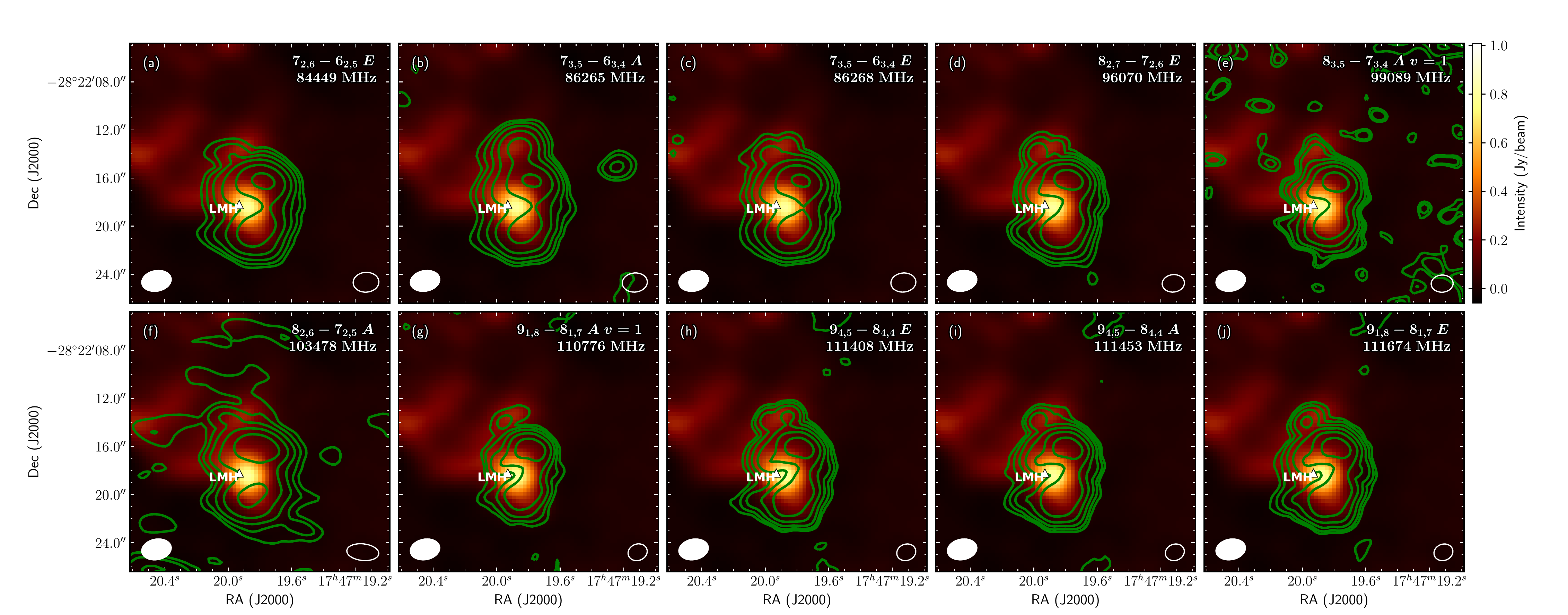}
    \caption{Peak intensity images of the low-velocity component ($64\ \mathrm{km\ s^{-1}}$) of the 10 MF transitions overlaid on the continuum emission at 85.9 GHz (background color image). The integrated velocity range of the low-velocity component ranges from 59 to 68 $\mathrm{km\ s^{-1}}$. The MF contours start at $3\sigma$ and double in value up to the peak value. Most of transitions of MF show consistent morphology. Spatial contamination source: panel (b) is contaminated with the high-velocity component of the MF $7(3,5)-6(3,4)\ E\ $ transition at 86265 MHz.} \label{MF-64kms-mom8}
\end{figure*}
\end{turnpage}

\begin{turnpage}
\begin{figure*}
    \includegraphics[width=1\textheight]{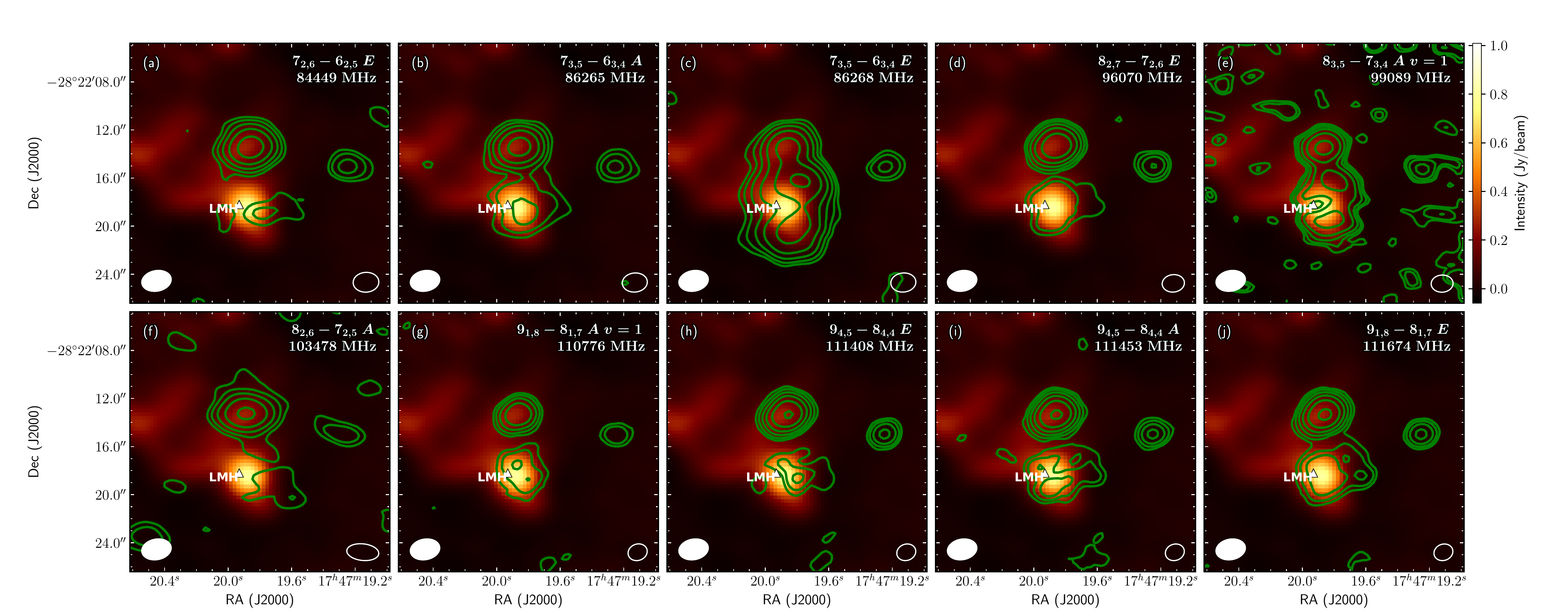}
    \caption{Peak intensity images of the high-velocity component ($73\ \mathrm{km\ s^{-1}}$) of the 10 MF transitions overlaid on the continuum emission at 85.9 GHz (background color image). The integrated velocity range of the high-velocity component ranges from 70 to 75 $\mathrm{km\ s^{-1}}$. The MF contours start at $3\sigma$ and double in value up to the peak value. We resolve the high-velocity component as the second velocity-component of MF. Spatial contamination sources: panel (c): the low-velocity component of the MF $7(3,5)-6(3,4)\ A\ $ transition at 86265 MHz; panel (e): the thioformaldehyde (\ce{H2^{13}CS}) $3(2,1)-2(2,0)$ transition at 99086 MHz.}\label{MF-73kms-mom8}
\end{figure*}
\end{turnpage}

\paragraph{\ce{HCOOCH3}} For MF, we imaged the emission from the 10 most uncontaminated transitions with $P$ factor $>95$ and $D$ factor $>60$, over a range of upper-state energies from 23 K to 221 K. The peak intensity images of the low-velocity component and high-velocity component are shown in Figure \ref{MF-64kms-mom8} and \ref{MF-73kms-mom8} respectively. In Figure \ref{MF-64kms-mom8} and \ref{MF-73kms-mom8}, the background color images are the 85.9 GHz continuum of the Sgr B2(N) region. The contours indicate the location of the MF emission of each transition. As the most uncontaminated transitions are selected based on the spectrum toward LMH only, some of these MF lines still suffer from spatial contamination, as summarized in the captions of Figure \ref{MF-64kms-mom8} and \ref{MF-73kms-mom8}.

Despite the spatial contamination, based on the peak intensity images of MF, it is clear that different transition maps of MF show nearly identical spatial distributions to each other at each characteristic velocity over a range of excitation levels. We identified two cores of emission of the low-velocity component and two of the high-velocity component among all the transition maps in \ref{MF-64kms-mom8} and \ref{MF-73kms-mom8}. The low-velocity component has a extended distribution which is offset from the continuum emission peak. The distribution of the emission cores of MF is described more completely in Section \ref{chemmap} in a quantitative way using its chemical map, which is, in short, a weighted composite transition map.

As outlined in \citet{bdl2015}, the spatial distributions from interferometric observations were suggested to be similar among transitions with similar excitation conditions. However, we find that the morphology of MF is consistent across a wide range of upper-state energy levels; i.e., the intensity maps obtained from $v = 0$ and $v = 1$ of MF are similar.

\begin{turnpage}
\begin{figure*}[htp!]
    \includegraphics[width=1\textheight]{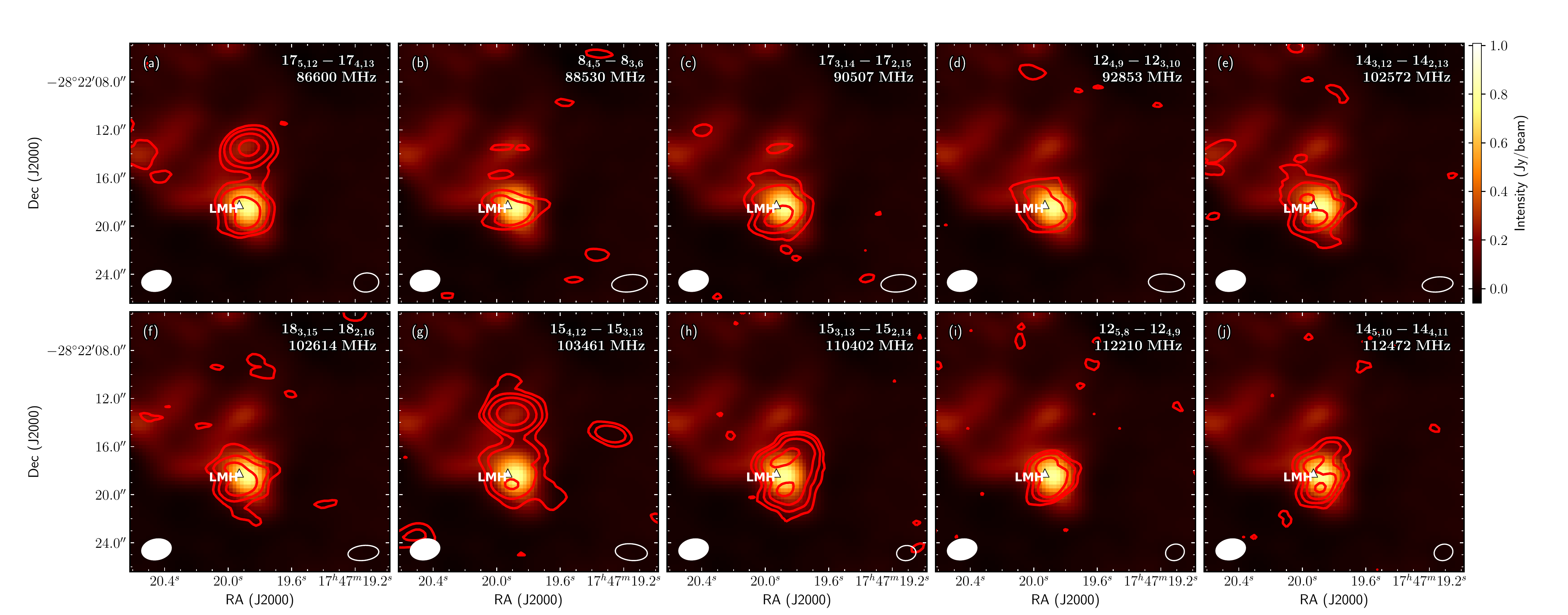}
    \caption{Peak intensity images of the low-velocity component ($64\ \mathrm{km\ s^{-1}}$) of the 10 GLA transitions overlaid on the continuum emission at 85.9 GHz (background color image). The integrated velocity range of the high-velocity component ranges from 59 to 68 $\mathrm{km\ s^{-1}}$. The GLA contours start at $3\sigma$ and double in value up to the peak value. Despite the transitions suffering from spatial contamination, GLA shows consistent distribution among different transitions. Spatial contamination sources: panel (a): the high-velocity components of the acetone (\ce{(CH3)2CO}) 19(10,10)-19(9,11) AA transition at 86604.63 MHz, and the gauche-ethanol (\ce{g-CH3CH2OH}) $5(3,3)-4(3,2)$ transition at 86604.34 MHz; panel (g): the high-velocity component of MF $8(2,6)-7(2,5)\ E$ at 103466 MHz; panel (h): the \ce{(CH3)2CO} 6(5,2)-5(4,1) EE transition at 110401 MHz.} \label{GLA-64kms-mom8}
\end{figure*}
\end{turnpage}

\begin{turnpage}
\begin{figure*}[htp!]
    \includegraphics[width=1\textheight]{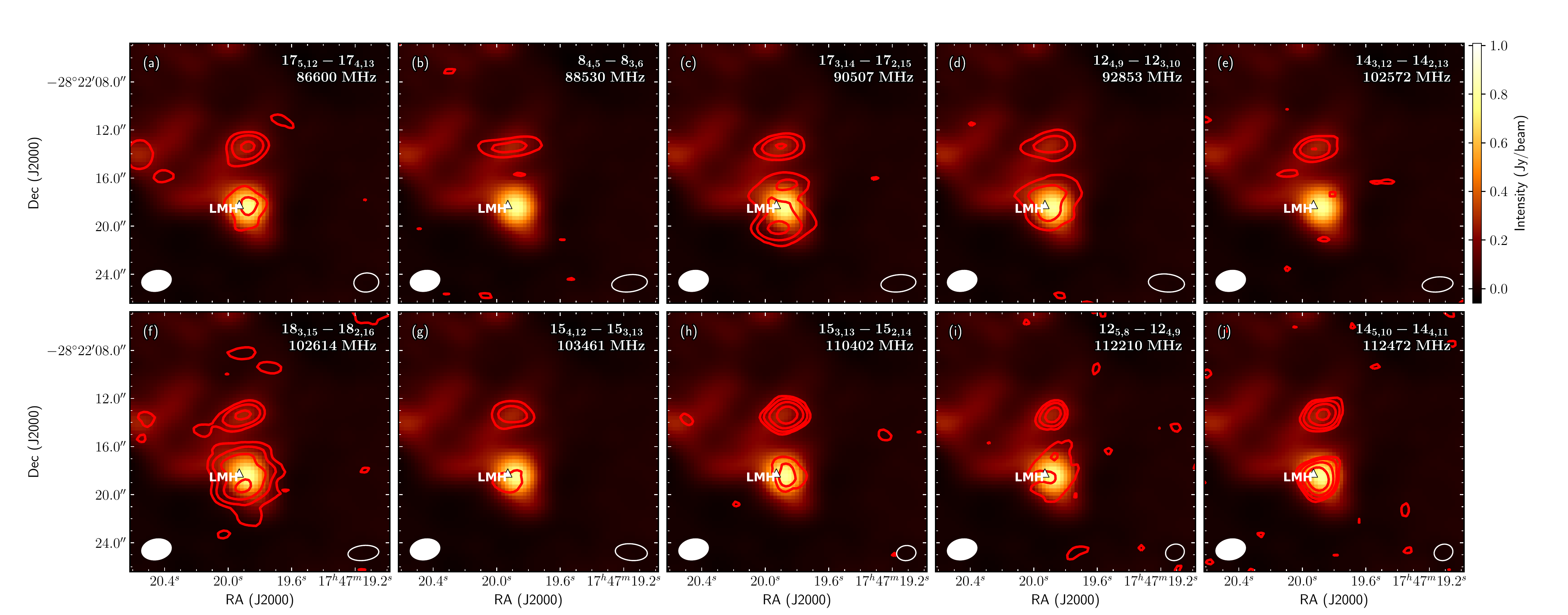}
    \caption{Peak intensity images of the high-velocity component ($73\ \mathrm{km\ s^{-1}}$) of the 10 GLA transitions overlaid on the continuum emission at 85.9 GHz (background color image). The integrated velocity range of the high-velocity component ranges from 70 to 75 $\mathrm{km\ s^{-1}}$. The GLA contours start at $3\sigma$ and double in value up to the peak value. We resolve the high-velocity component near the N2 region as the second velocity-component of GLA. Spatial contamination sources: panel (a): a weak and unidentified emission feature; panel (c): a strong emission from \ce{g-CH3CH2OH} $6(2,5)-5(1,5)$ at 90503 MHz; panel (d): unidentified; panel (f): unidentified; panel (j): the low-velocity component of the strong emission feature from the formic acid (\ce{HCOOH}) $5(3,2)-4(3,1)$ transition at 112467 MHz.}\label{GLA-73kms-mom8}
\end{figure*}
\end{turnpage}

\paragraph{\ce{CH2OHCHO}} We generated the interferometric images of the 10 least blended GLA transitions with upper-state energy levels ranging from 22 K to 64 K for both the low-velocity and high-velocity components, as shown in Figures \ref{GLA-64kms-mom8} and \ref{GLA-73kms-mom8}, respectively. In contrast to MF, there is substantial contamination among the spatial distributions of GLA. While the LMH region line profiles are relatively clean, the N2 region suffers much more significantly from this. 

Despite the contamination mentioned above, the remaining peak intensity images of GLA are consistent with each other, including the low-velocity components of seven transitions which are marked with "l" in Table \ref{GLA-table} and the high-velocity components of four transitions marked with "h" in Table \ref{GLA-table} respectively. From the uncontaminated transitions, as evident in Figures \ref{GLA-64kms-mom8} and \ref{GLA-73kms-mom8}, we resolved one emission core from the low-velocity component of GLA with a source size smaller than 7 $\arcsec$ and one from the high-velocity component with a source size $\sim$ 4 $\arcsec$. The emission regions of GLA are more compact that those of MF. A detailed comparison of the distributions of the isomers is discussed in Section \ref{chemmap}.

The result of a compact emission region of GLA is in contradiction with previous research on the spatial distribution of GLA. \citet{hvsjl2001} mapped the $8(0,8)-7(1,7)$ transition of GLA at 82470.67 MHz with the BIMA array and suggested an extended distribution of GLA toward Sgr B2(N) with a source size of around 60 $\arcsec$. The difference found between the distributions from our study may be due to the contamination of the transition detected by \citet{hvsjl2001}. In their study, only one transition was mapped. Therefore, they could not distinguish if the transition was contaminated by comparing maps from different transitions which is what we do in the present study. In addition, the broad line width of the GLA $8(0,8)-7(1,7)$ transition imaged by \citet{hvsjl2001}, 24.3 $\mathrm{km\ s^{-1}}$, compared with 8.2 $\mathrm{km\ s^{-1}}$ for the MF line in their observations is also consistent with the existence of contamination. Thus, the conclusions reached by \citet{hvsjl2001} are most likely inaccurate given the fact that it is probably based on a blended transition of GLA.

\begin{turnpage}
\begin{figure*}[htp!]
    \includegraphics[width=1\textheight]{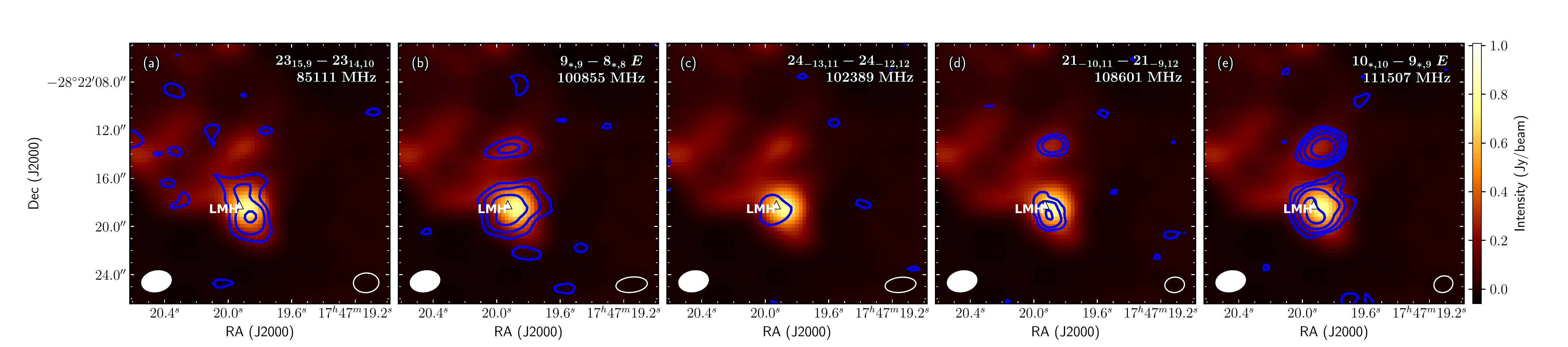}
    \caption{Peak intensity images of the five AcA transitions overlaid on the continuum emission at 85.9 GHz (background color image). The integrated velocity range of the high-velocity component ranges from 59 to 67 $\mathrm{km\ s^{-1}}$. The AcA contours start at $3\sigma$ and double in value up to the peak value. We only resolved one velocity component of AcA (($\sim 64\ \mathrm{km\ s^{-1}}$)) and have limited number of the transition maps of AcA. Spatial contamination sources: panel (b) is contaminated by a weak emission feature toward the N2 region contributed by the high-velocity component of the GLA transition $24(7,18)-23(8,15)$ at 100859 MHz; panel (d) is probably contaminated by the high-velocity component of a transition with a rest frequency of $\sim$ 108606 MHz, of which the low-velocity component was also observed toward LMH as a strong emission line. However, this transition is not found in any public database and, as such, remains unidentified; panel (e) is contaminated by the high-velocity component of the overlapping transitions of trans-ethanol (\ce{t-CH3CH2OH})-$12(7,6)-13(6,7)$ and $12(7,5)-13(6,8)$)-at 111510 MHz in the north region.}\label{AcA-mom8}
\end{figure*}
\end{turnpage}

\paragraph{\ce{CH3COOH}}
Figure \ref{AcA-mom8} shows the peak intensity images of AcA for the five transitions listed in Table \ref{AcA-table}. The upper-state energy levels range from 26 to 266 K. Because there is no emission with $S/N >1$ in the $\sim 73\ \mathrm{km\ s^{-1}}$ regions, we only resolve one low-velocity component of AcA ($64\ \mathrm{km\ s^{-1}}$). In addition to the limited number of clean emission transitions, AcA also suffers from severe spatial contamination. The least contaminated three transitions are marked with "l" in Table \ref{AcA-table}.

Given these limitations, the morphology of AcA cannot be accurately determined based on the limited number and the spatial contamination of the transition maps. But we can still see that the emission peaks are co-spatial among all the transition maps. Besides, \citet{msml1997} mapped Sgr B2(N) with the OVRO array in two transitions of AcA at 90246 MHz and 100855 MHz and \citet{rslmk2002} did the same at 111507 MHz and 111547 MHz. Although the two transitions, 90246 MHz and 111547 MHz, are are known to be contaminated toward LMH and the distribution of the transition at 111507 MHz is spatially contaminated toward the N2 region, the AcA emission peak resolved with the previous observations is consistent with our study. The peak positions are all at $\alpha_\text{J200}=\hms{17}{47}{19.9(0.1)}$, $\delta_\text{J2000}=\dms{-28}{22}{19(1)}$ within 1 $\arcsec$ which confirms the assertion that the AcA lines are coming from a common source \citep{rslmk2002}.

\subsection{Chemical Maps of \ce{C2H4O2}} \label{chemmap}
Because the uncontaminated transitions covering an extensive range of upper-state energies show similar morphology, the effect of excitation conditions is ruled out and the morphology of these transition maps is determined by the true distribution of molecules. For this reason, with an increased number of unblended transitions with similar morphologies, we can combine the transition maps to achieve mm chemical maps for the \ce{C2H4O2} isomers. A chemical map is obtained by stacking the peak intensity images of transitions which are mostly free of contamination. In the stacking process, we used the inverse-variance weighting for each transition, where the variance is the square of the noise of each data cube. Compared with a single transition map, the chemical map allows us to infer molecular distributions at a high $S/N$. It is worthwhile noting that the ultimate units of the chemical map are non-physical and thus nothing should be inferred from the value of intensity directly.

 \begin{figure*}[ht]
    \centering
    \epsscale{1.2}
        \plotone{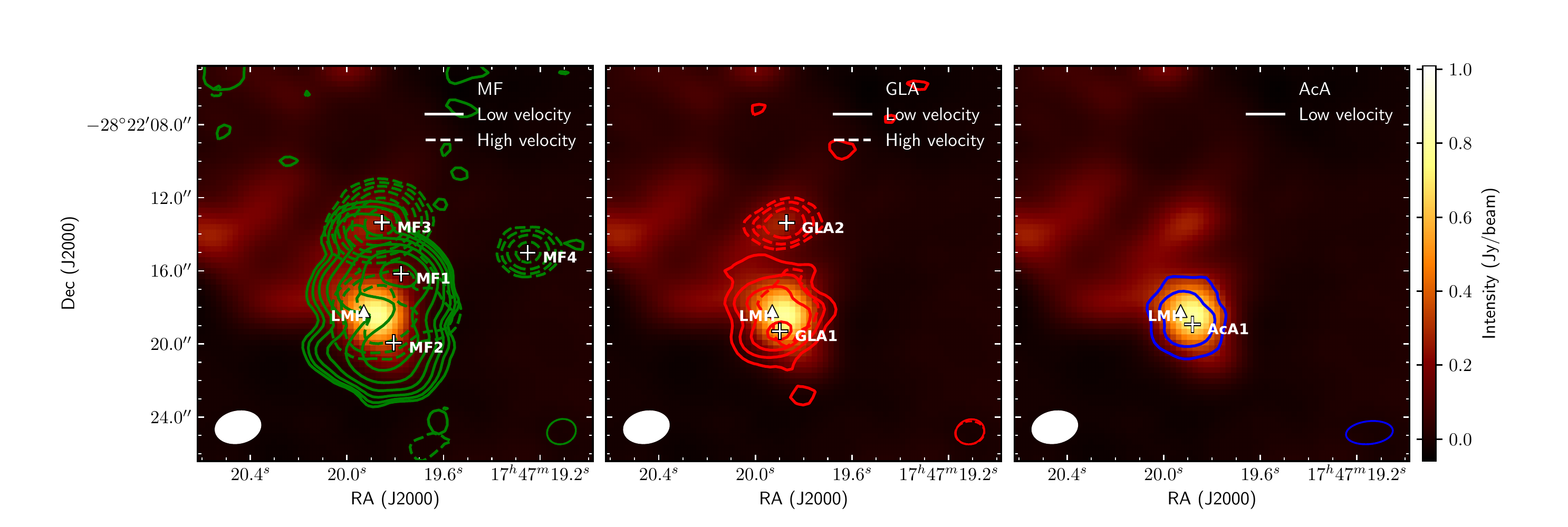}
        \caption{\footnotesize{Emission contours from the chemical maps of MF (left), GLA (middle) and AcA (right) overlaid on the continuum emission at 85.9 GHz (background color image) toward Sgr B2(N). The solid lines represent the low-velocity component, while the dashed lines correspond to the high-velocity component. The contour levels start at $5\sigma$ and double in value up to the peak value, with $\sigma \sim 1.7\ \mathrm{mJy\ beam^{-1}}$ for MF, $\sim 2.0\ \mathrm{mJy\ beam^{-1}}$ for GLA and $\sim 2.5\ \mathrm{mJy\ beam^{-1}}$ for AcA. The emission cores of each molecule are marked by a cross and labeled $\mathrm{MF_{NUMBER}}$, $\mathrm{GLA_{NUMBER}}$ and $\mathrm{AcA_{NUMBER}}$ correspondingly.}
        \label{fig:chemmap}}
\end{figure*}

For MF, we stacked the peak intensity images of the nine least contaminated transitions marked with "l" in Table \ref{MF-table} to obtain the low-velocity component and the eight transitions marked with "h" in Table \ref{MF-table} for the high-velocity component of the chemical map of MF. Figure \ref{fig:chemmap} (left panel) shows the chemical maps of MF in green with solid lines representing the low-velocity component and dashed lines the high-velocity component over the 85.9 GHz continuum emission. Compared with transition maps, we can better quantify the distribution of the emission cores with the chemical map. In the chemical map of MF, four distinct emission cores of MF are resolved. The low-velocity component contains two emission cores marked as MF1 and MF2, which are connected with each other over a distance of $3.8 \arcsec$ and located within the cloud. The other two emission cores, MF3 and MF4, with high velocity, are separated by around $8.1 \arcsec$. The coordinates of the four cores are summarized in Table \ref{core-coord}. The overall distribution of MF is offset from the continuum emission and more extended than that of GLA and AcA. The two emission peaks of MF with low velocity are offset from LMH by $\sim 2.9 \arcsec$ for MF1 and $\sim 2.4 \arcsec$ for MF2 respectively, while MF3 and MF4 are spatially coincident with the N2 core and N3 core defined in \cite{bbmgm2017}, respectively.  

The chemical map of GLA is represented in Figure \ref{fig:chemmap} (middle panel), which is stacked with the transition maps of the seven transitions marked with "l" in Table \ref{GLA-table} for the low-velocity component and the four transitions marked "with h" in Table \ref{GLA-table} for the high-velocity component. With the chemical map of GLA, we can better examine the spatial distribution features of GLA that are not clearly resolved in its transition maps. GLA also has distinct high- and low- velocity components. For the low-velocity component, we locate the major emission core (GLA1) at $\alpha_\text{J200}=\hms{17}{47}{19.9}$, $\delta_\text{J2000}=\dms{-28}{22}{19.3}$ and resolve an extended feature to the north-east. An emission core with high velocity (GLA2) is resolved and co-spatial with the MF3 core. GLA2 is $\sim 5.9 \arcsec$ north of GLA1. Compared with MF, the distribution of GLA is more compact and spatially consistent with the continuum emission. The GLA1 core is offset from the LMH peaks by only $\sim 1.2 \arcsec$ while the GLA2 core is co-spatial with the N2 continuum emission core \citep{bbmgm2017}.

Each of the transition maps of AcA are quite different from each another, perhaps due to the low $S/N$ of the most uncontaminated lines. The difference, as well as the limited number of the unblended transitions, 5 in total, make the morphology of AcA difficult to characterise. But, we can gain insight as to its distribution by looking at its chemical map with increased $S/N$. In the chemical map of AcA as shown in Figure \ref{fig:chemmap} (right), AcA only emits from the region within the low-velocity cloud and has one emission core (AcA1). We do not resolve any high-velocity component for AcA. In addition, the emission region of AcA appeared to be, interestingly, coincident with the distribution of the low-velocity component of GLA, and the spatial separation of AcA1 and GLA1 is smaller than $0.5 \arcsec$. As with GLA, we find the emission region of AcA to be compact and spatial coincident with the continuum emission. AcA1 is offset from the LMH peaks by only $\sim 1.0 \arcsec$. 

\begin{deluxetable}{cccc}
    \tablewidth{0pt}
    \tablecaption{Positions of the LMH Region and the Molecular Emission Peaks of the \ce{C2H4O2} isomers \label{core-coord}}
    \tablehead{
        \\
        \colhead{Source}	&\multicolumn{2}{c}{Coordinate} &\colhead{$V_\text{lsr}$}\\
        ~   &\colhead{$17^h47^m$}   &\colhead{$-28 \degree 22\arcmin$}  &$\mathrm{km\ s^{-1}}$
    }
    \startdata
        LMH(centroid)   &   19.930$^s$  &18.200$\arcsec$    &64 \\
        MF1 &               19.776$^s$  &16.158$\arcsec$    &64 \\
        MF2 &               19.806$^s$  &19.934$\arcsec$    &64 \\
        MF3 &               19.855$^s$  &13.357$\arcsec$    &73 \\
        MF4 &               19.252$^s$  &15.002$\arcsec$    &73 \\
        GLA1&               19.899$^s$  &19.294$\arcsec$    &64 \\
        GLA2&               19.871$^s$  &13.379$\arcsec$    &73 \\
        AcA1&               19.881$^s$  &18.926$\arcsec$    &64 \\
    \enddata
\end{deluxetable}

\section{DISCUSSION}\label{sec:diss}
Our observations indicate that the three isomers of \ce{C2H4O2} have distinct morphologies, as shown in Figure \ref{fig:chemmap}. The question remains as to why the three isomers have such different distributions compared to each other despite their identical atomic composition. Here, we discuss the possible mechanisms that could contribute to this difference.

Although different observations could bias the result of imaging and contribute to spatial difference, this is not the case in our study. Because we self-consistently analyzed these three isomers in the same observations, we can rule out any differences caused by observations with different facilities, techniques, or frequency regimes. Instead, there are both physical and chemical causes, which could be interwoven, to explain the spatial difference.

Different physical conditions and therefore different excitation conditions inside the source might lead to the diversity in morphology. While our best fit to the observed data suggests the $T_\text{ex}$ of MF toward LMH to be $\sim 200\ \text{K}$, \citet{bbmgm2017} derived a similar $T_\text{ex}$, $\sim 150\ \text{K}$, toward both N2 and N3, the two high-velocity cores. Because the transition maps of each molecule are spatially similar across a large range of upper-state energy levels, we can largely rule out the excitation temperatures of the multiple components of Sgr B2(N) as the source of the distinct distributions. In addition, the three species are similar in terms of both photo-absorption cross-sections and ionization potentials \citep{pmmc2010}. Therefore, selective photo-destruction fails to account for the differentiation in these morphologies. Regarding cloud density, \citet{sss2017} characterized the small scale structures of Sgr B2(N), which include the cores associated with the seven molecular emission peaks of the three isomers, with various hydrogen densities in a range of $10^7-10^9\ \mathrm{cm^{-3}}$. To estimate $n_\mathrm{crit}$, we adopted a typical \ce{H2} collisional cross section $\sim 10^{-15}\ \mathrm{cm^2}$ \citep{cr2016}. Compared with $n_\mathrm{crit}$ of the \ce{C2H4O2} transitions listed in Table \ref{MF-table}, \ref{GLA-table} and \ref{AcA-table}, we indeed find $n_\text{\ce{H2}} > n_\mathrm{crit} \sim10^3 \mathrm{cm^{-3}}$ for the detected lines so that the transitions are probably all thermalized. We also noticed that absorption by the extended material in the cold envelope can cause the offset of MF from the continuum emission peak and should not be neglected. Finally, evolving source properties with time may contribute to the differences in their spatial distributions. A more thorough analysis of other physical properties of the sources would provide further insights into addressing this point.

The morphological differences among the \ce{C2H4O2} isomeric family could be explained in part by the difference in their production mechanisms and the large energy barriers for the isomerization among them. The grain surface formation routes originally suggested by \citet{gh2006} and \citet{gwh2008} are accepted as the most likely mechanisms to efficiently form LAMs during the warm-up phase of the hot cores. Because thermal desorption is efficient in a warm region, a compact distribution of warm molecules toward hot cores is usually associated with a grain-surface formation scheme followed by thermal desorption, although, as discussed below, gas-phase synthesis cannot be ruled out. In this scenario, the \ce{HCO} radical becomes mobile on grain surfaces and then reacts with the \ce{CH3O} radical to form \ce{HCOOCH3} or with the \ce{CH2OH} radical to form \ce{CH2OHCHO}. Recent laboratory studies show that H-atom addition and abstraction reactions in ice mantles can lead to more reactive intermediates, which enhance the probability to form MF and GLA through radical-radical recombination \citep{cfiv2016}. Although the grain-surface formation routes of MF and GLA are similar, as both involve \ce{HCO}, the observed abundances of MF and GLA may be influenced by the abundance difference of the \ce{CH3O} and \ce{CH2OH} radicals on grain surfaces, which, as of yet, have not been measured. Although direct measurements of the abundances of \ce{CH3O} and \ce{CH2OH} on grain by IR absorption lines will be difficult to achieve, indirect determinations could be based on possible future observation of gaseous \ce{CH2OH} and chemical models. 

 \begin{figure*}[ht]
    \centering
    \epsscale{1.2}
        \plotone{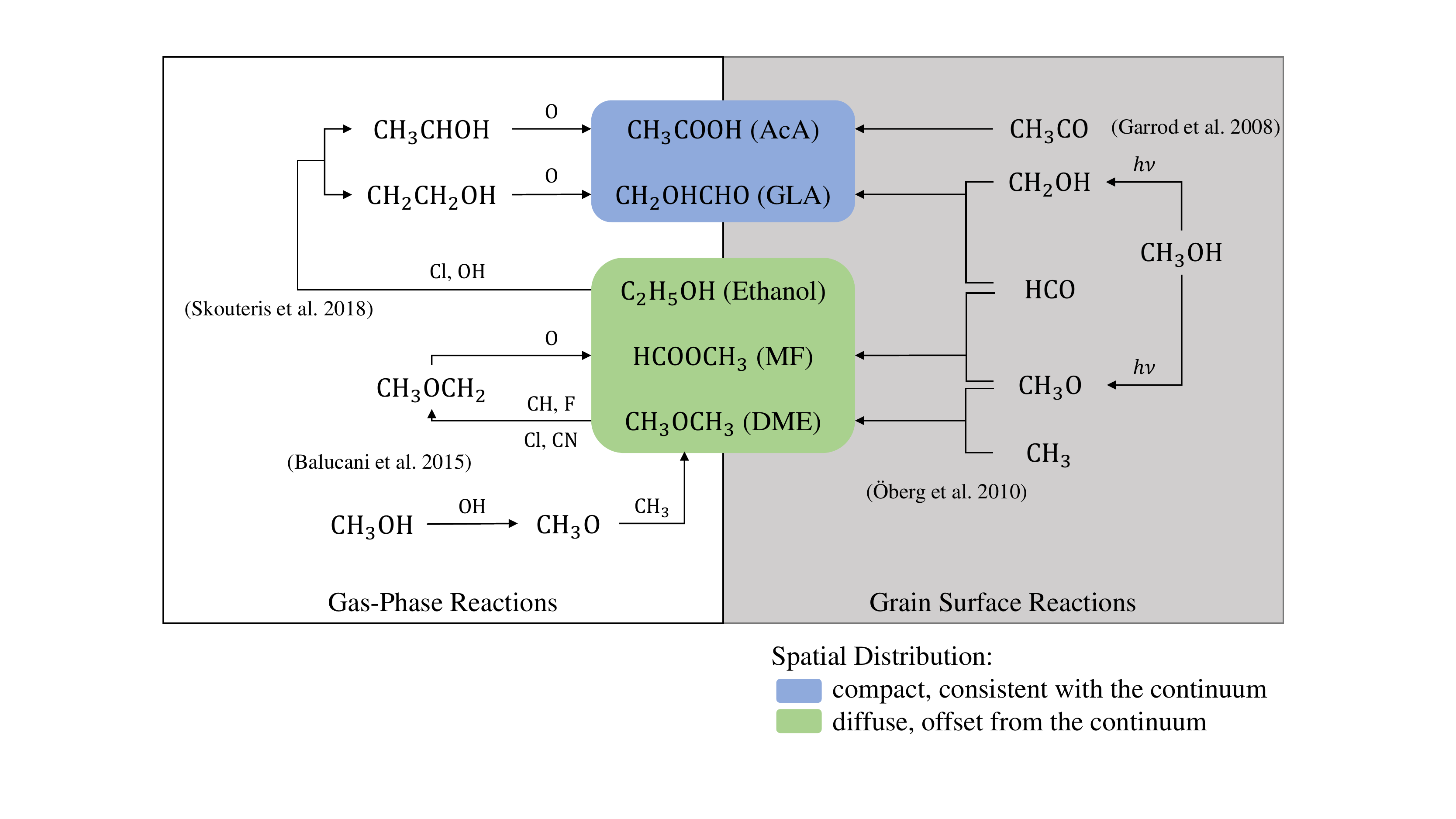}
        \caption{\footnotesize{Schematic diagram of the prevalent formation mechanisms related to the \ce{C2H4O2} isomers.}
        \label{fig:reactions}}
\end{figure*}

While the radical-radical reactions of MF and GLA only involve primary radicals, \citet{gwh2008} suggested that AcA is assembled from the \ce{OH} radical and the \ce{CH3CO} radical. Different from the dominant primary radicals, the \ce{CH3CO} radical, which is a secondary radical, is weakly produced and becomes mobile at higher temperatures than primary radicals due to its complexity \citep{gwh2008}. The involvement of the \ce{CH3CO} radical makes the AcA formation route different from its structural isomers. 

If the scenario of radical-radical reactions on grain surfaces \citep{gwh2008} is the main formation mechanism of the three isomers, it is reasonable to expect that the spatial distribution of MF and GLA would be similar to each other, since their precursors are \ce{HCO} and the isomeric radicals that both form from methanol photolysis \citep{lghw2011}, but different from that of AcA. However, the spatial separation of gaseous MF from its isomers and the spatial coincidence of gaseous GLA and AcA make this scenario insufficient for an overall explanation of the observational result. Additionally, there are other pathways involving other surface radicals to form GLA and AcA. GLA could also be formed from the reactions of the \ce{HCO} radical with the \ce{H2CO} and \ce{H} on grain surfaces \citep{bcvn2009}, which was suggested to be the most probable formation mechanism of GLA at low temperature \citep{wkv2012}. \citet{bk2007} also proposed that AcA could form via the association reaction of the \ce{HOCO} radical and the \ce{CH3} radical on grains. But the effectiveness of these proposed pathways is still not fully understood and requires further studies.

In addition to grain chemistry, it was suggested that MF could be also formed via gas-phase processes \citep{lghw2011, bct2015}. \citet{bct2015} outlined a possible mechanism for the synthesis of MF with gas-phase reactions in cold environments. Therefore, it is a natural outcome that molecules formed in the gas possess extended distributions \citep{hvsjl2001} but there remains the exception that non-thermal desorption can eject molecules formed on cold grain surfaces into a cold gas, as for the case of methanol formation \citep{wk2002}. Nevertheless, in the gas-phase model, dimethyl ether (\ce{CH3OCH3}, DME) works as the precursor of MF via the loss of a hydrogen atom to form the \ce{CH3OCH2} radical followed by reaction with atomic oxygen. In addition, an up-to-date scheme for the gas-phase synthesis of GLA and AcA was proposed, where GLA and AcA were both formed from ethanol (\ce{CH3CH2OH}) \citep{sbc2018}. With H-abstraction from ethanol, two different reactive radicals, \ce{CH2CH2OH} and \ce{CH3CHOH}, form and further react with atomic oxygen to produce GLA and AcA respectively. Based on this formation scheme involving ethanol, \citet{sbc2018} predicted a higher abundance of AcA than GLA, which agrees well with our findings. 

As a test of the different gas-phase formation pathways, the distributions of the chemical precursors could work as proxies to interpret the different morphology of these isomeric and related interstellar molecules. If there are either correlations or anti-correlations between the spatial distributions of the precursors and those of the corresponding products, that could be evidence to support the related chemistry. Therefore, we further imaged the chemical maps of the two proposed precursors, DME and ethanol, involved in the gas-phase reactions. We find that DME is co-spatial with MF, which is consistent with their distributions toward another star formation region, Orion KL \citep {bdb2013}. The correlation of MF and DME could also be explained with the common precursor, the surface \ce{CH3O} radical, in their grain-surface formation pathways \citep{obj2010}. We also suspect that ethanol, the precursor of GLA and AcA, should have a similar distribution to GLA and AcA. However, the distribution of ethanol is co-spatial to MF and DME proving that no relationship between the triplet of \ce{C2H4O2} and the proposed gas-phase precursors could be determined.

Different desorption behaviors of the three isomers could also contribute to the spatial separation between MF and the other isomers. In the warm-up model, desorption begins at different temperatures \citep{gwh2008}. Thus, temperature, which can be a substitute for time in the model, could be related to differing distributions. \citet{bpb2015} found via a Temperature Programmed Desorption (TPD) analysis that MF desorption starts at a lower temperature, $\sim$ 70 K, compared with the desorption of the other two isomers, $\sim$ 110 K. This different behavior during the thermal processing could account for the extended distribution of MF compared with GLA and AcA. A neglected drawback with the analysis presented in \citet{bpb2015}, however, is that they assumed identical grain-surface abundances of the isomers. To connect the effect of the desorption property to the morphological features, taking into account a chemical model, with which the abundances of the isomers could be constrained, seems to be required.

Future research is essential to understand the relative role of the gas-grain chemistry of these three isomers of \ce{C2H4O2}. By carrying out detailed model calculations, involving both grain-surface reactions and the newly proposed gas phase pathways of these three molecules \citep{bct2015, sbc2018}, we hope to constrain the formation mechanisms of the isomers. Observations of the \ce{C2H4O2} and other isomers toward a large sample of interstellar sources could help to constrain the relative importance of their formation pathways and explain the spatial distribution of different molecular isomers. 

\section{SUMMARY}\label{sec:sum}
In this paper, we report an intensive spectral and morphological analysis of the \ce{C2H4O2} isomers towards Sgr B2(N) with sensitive $ALMA$ Band 3 observations \citep{bmgk2016}.

We developed quantitative methods to identify weak and mostly uncontaminated transitions which are critical to automatically detect interstellar molecules within a source that has a dense spectrum and a high level of spectral confusion. From these methods, we identified the 26 least contaminated transitions for MF, 10 for GLA and 5 for AcA and further imaged the peak emission maps of these clean transitions. We stacked all of the consistent transition maps of each molecule to obtain their chemical maps, which show more reliable distributions of the molecules with higher $S/N$ ratios. 

From the high spatial resolution millimeter chemical maps of the three isomers, we find that they have distinct morphologies. Four emission cores of MF are resolved, in which two of them, \text{MF1 and MF2}, are connected to each other and located within the portion of the cloud with the low $V_\text{lsr}$ of $\sim 64\ \mathrm{km\ s^{-1}}$. For this velocity component, the distribution of MF is extended and offset from the continuum emission. The other two emission cores of MF, \text{MF3 and MF4}, with the high $V_\text{lsr}$, $\sim 73\ \mathrm{km\ s^{-1}}$, are separated by around 8.1$\arcsec$ from each other. GLA also displays two velocity components, while the $V_\text{lsr}$ of GLA1 is $\sim 64\ \mathrm{km\ s^{-1}}$ and of GLA2 is $\sim 73\ \mathrm{km\ s^{-1}}$. Only one velocity component of AcA, AcA1 at $\sim 64\ \mathrm{km\ s^{-1}}$, is resolved. In contrast to MF, the overall distribution of GLA and AcA are more compact and co-spatial with the continuum emission.

Because the distributions in the transition maps of each isomer are spatially consistent across a wide range of upper level energies, we rule out excitation conditions as the source of the distinct distributions of the \ce{C2H4O2} isomers. The dissimilar morphology of MF compared with GLA and AcA suggests that MF has a different formation mechanism than do GLA and AcA. Through a thorough investigation into the prevalent formation mechanisms involving both grain-surface and gas-phase routes, however, we find that the current existing chemical theory as summarized in Figure \ref{fig:reactions} can not explain the observed spatial distribution of the \ce{C2H4O2} isomers. In addition, different desorption behaviors of the three isomers could contribute to the spatial distributions of the \ce{C2H4O2} isomeric family. Therefore, more observations of the three isomers toward a larger sample of astronomical objects might help to constrain the morphological features and to probe the chemical formation mechanisms in these environments.

\acknowledgments
We acknowledge funding from the National Science Foundation for support of the astrochemistry program of E. H. Also special thanks go to K. H. Lam for helpful discussions on the quantification of the degree of consistency. We are indebted to B.A. McGuire for providing a Python program for simulating and visualizing molecular emission spectra, and to T. Hunter for providing codes to identify the continuum channels. A.M.B. was a Grote Reber Fellow for a portion of this project, and support for this work was provided by the NSF through the Grote Reber Fellowship Program administered by Associated Universities, Inc./National Radio Astronomy Observatory and the Virginia Space Grant Consortium. A.M.B. also acknowledges support from the Smithsonian Institution as a current Submillimeter Array (SMA) Fellow. This paper makes use of the following ALMA data: ADS/JAO.ALMA\#2011.0.00017.S, ADS/JAO.ALMA\#2012.1.00012.S. ALMA is a partnership of ESO (representing its member states), NSF (USA) and NINS (Japan), together with NRC (Canada), NSC and ASIAA (Taiwan), and KASI (Republic of Korea), in cooperation with the Republic of Chile. The Joint ALMA Observatory is operated by ESO, AUI/NRAO and NAOJ.

\vspace{5mm}
\software{CASA (https://casa.nrao.edu) \citep{shb2007}}

\newpage
\appendix

\section{Complementary table}
Tables \ref{MF-appendix-table}, \ref{GLA-appendix-table} and \ref{AcA-appendix-table} list the transitions of the \ce{C2H4O2} isomers that are detected above the 3$\sigma$ level toward Sgr B2(N). P and D factors are calculated based on the observed spectra toward LMH and the simulated spectra assuming an excited temperature of 190 K. Comments in these tables acknowledge the reference observations.

\startlongtable
\begin{deluxetable*}{llrrrccl}
    \tabletypesize{\scriptsize}
    \tablecaption{Transitions of MF with $S/N > 3$ covered by the EMoCA survey.\label{MF-appendix-table}}
    \tablehead{
        \\
        \colhead{Rest Frequency}	& \colhead{Transition}	& \colhead{$E_u$}	    & \colhead{$\log_{10}{\frac{A_{ul}}{\mathrm{s^{-1}}}}$}	&
        \colhead{$S_{ij}\mu^{2}$}	& \colhead{P Factor}	& \colhead{D Factor}	& Obs Ref\\
        \colhead{(MHz)}	            & ~                 	& \colhead{(K)} 	    & ~                             &
        \colhead{(Debye$^2$)}    	& (\%) 	                & (\%)                  &
    }
    \startdata
    84449.1690(100) & $7(2,6)-6(2,5)E$ & 19.00 & -5.0992 & 17.0 & 97.6 & 79.9 &  \\
84454.7540(100) & $7(2,6)-6(2,5)A$ & 18.98 & -5.0990 & 17.0 & 95.5 & 58.1 &  \\
85185.4660(100) & $7(5,3)-6(5,2)A\ v=1$ & 220.90 & -5.3602 & 9.1 & 97.4 & 63.4 &  \\
85186.0630(100) & $7(5,2)-6(5,1)A\ v=1$ & 220.90 & -5.3602 & 9.1 & " & " &  \\
85327.0290(100) & $7(4,4)-6(4,3)A\ v=1$ & 214.91 & -5.2200 & 12.5 & 79.2 & 6.4 &  \\
85360.7640(100) & $7(4,3)-6(4,2)A\ v=1$ & 214.91 & -5.2195 & 12.5 & 88.5 & 7.0 &  \\
85371.7300(100) & $7(3,5)-6(3,4)A\ v=1$ & 210.26 & -5.1362 & 15.1 & 79.9 & 11.8 &  \\
85506.2190(100) & $7(4,3)-6(4,2)E\ v=1$ & 214.56 & -5.2143 & 12.6 & 91.3 & 17.3 &  \\
85743.9760(100) & $7(4,4)-6(4,3)E\ v=1$ & 214.03 & -5.2119 & 12.5 & 87.7 & 54.0 &  \\
85761.8760(100) & $21(5,16)-21(4,17)E$ & 155.82 & -5.8778 & 7.8 & 76.6 & 9.3 &  \\
85773.3990(100) & $21(5,16)-21(4,17)A$ & 155.82 & -5.8778 & 7.8 & 83.3 & 4.7 &  \\
85780.6730(100) & $20(5,15)-20(4,16)E$ & 142.79 & -5.8864 & 7.2 & 73.6 & 24.5 &  \\
85785.3420(100) & $20(5,15)-20(4,16)A$ & 142.79 & -5.8865 & 7.2 & 93.9 & 2.8 &  \\
85919.2090(100) & $7(6,1)-6(6,0)E$ & 40.44 & -5.6139 & 4.9 & 71.1 & 7.2 &  \\
85926.5530(100) & $7(6,2)-6(6,1)E$ & 40.42 & -5.6137 & 4.9 & 94.6 & 8.6 &  \\
85927.2270(100) & $7(6,2)-6(6,1)A$ & 40.42 & -5.6136 & 4.9 & " & " &  \\
85927.2270(100) & $7(6,2)-6(6,1)A$ & 40.42 & -5.6136 & 4.9 & " & " &  \\
86021.1240(100) & $7(5,2)-6(5,1)E$ & 33.13 & -5.3461 & 9.1 & 89.7 & 21.9 & \citep{hvsjl2001} \\
86027.7230(100) & $7(5,3)-6(5,2)E$ & 33.12 & -5.3461 & 9.1 & 99.6 & 70.8 & \citep{hvsjl2001} \\
86029.4420(100) & $7(5,3)-6(5,2)A$ & 33.11 & -5.3459 & 9.1 & " & " & \citep{hvsjl2001} \\
86030.1860(100) & $7(5,2)-6(5,1)A$ & 33.11 & -5.3459 & 9.1 & " & " & \citep{hvsjl2001} \\
86034.0130(100) & $7(3,4)-6(3,3)E\ v=1$ & 209.82 & -5.1231 & 15.2 & 97.5 & 34.6 &  \\
86155.0780(100) & $7(3,4)-6(3,3)A\ v=1$ & 210.32 & -5.1242 & 15.1 & 92.2 & 12.8 &  \\
86172.7060(100) & $7(3,5)-6(3,4)E\ v=1$ & 209.41 & -5.1242 & 15.1 & 99.3 & 23.7 &  \\
86210.0570(100) & $7(4,4)-6(4,3)A$ & 27.15 & -5.2050 & 12.5 & 84.7 & 23.3 &  \\
86223.6550(100) & $7(4,3)-6(4,2)E$ & 27.17 & -5.2074 & 12.5 & 99.3 & 65.2 &  \\
86224.1600(100) & $7(4,4)-6(4,3)E$ & 27.16 & -5.2075 & 12.5 & " & " &  \\
86250.5520(100) & $7(4,3)-6(4,2)A$ & 27.15 & -5.2044 & 12.5 & 91.0 & 33.0 & \citep{hvsjl2001} \\
86265.7960(100) & $7(3,5)-6(3,4)A$ & 22.51 & -5.1210 & 15.2 & 99.7 & 63.2 & \citep{hvsjl2001} \\
86268.7390(100) & $7(3,5)-6(3,4)E$ & 22.53 & -5.1248 & 15.1 & 99.7 & 69.3 & \citep{hvsjl2001} \\
87143.2820(100) & $7(3,4)-6(3,3)E$ & 22.60 & -5.1116 & 15.1 & 95.1 & 51.1 &  \\
87160.8350(100) & $8(0,8)-7(1,7)E\ v=1$ & 207.04 & -5.8948 & 2.8 & 99.2 & 33.9 &  \\
87161.2850(100) & $7(3,4)-6(3,3)A$ & 22.58 & -5.1074 & 15.2 & " & " &  \\
87766.3820(100) & $8(0,8)-7(1,7)E$ & 20.08 & -5.8757 & 2.9 & 98.7 & 19.6 &  \\
87960.0490(100) & $7(1,6)-6(1,5)A\ v=1$ & 205.65 & -5.0223 & 18.0 & 93.9 & 11.5 &  \\
88053.9730(100) & $19(5,14)-19(4,15)E$ & 130.44 & -5.8713 & 6.6 & 99.4 & 31.3 &  \\
88054.4610(100) & $19(5,14)-19(4,15)A$ & 130.44 & -5.8716 & 6.6 & " & " &  \\
88220.7530(100) & $7(1,6)-6(1,5)E\ v=1$ & 204.91 & -5.0168 & 18.1 & 93.9 & 12.3 &  \\
88337.8140(100) & $22(5,17)-22(4,18)E$ & 169.52 & -5.8423 & 8.1 & 70.6 & 4.0 &  \\
88358.4870(100) & $22(5,17)-22(4,18)A$ & 169.52 & -5.8422 & 8.1 & 99.4 & 28.7 &  \\
88770.8680(100) & $8(1,8)-7(1,7)A\ v=1$ & 207.91 & -5.0025 & 20.8 & 97.9 & 32.1 &  \\
88843.1870(100) & $7(1,6)-6(1,5)E$ & 17.96 & -5.0081 & 18.0 & 89.1 & 25.0 &  \\
88851.6070(100) & $7(1,6)-6(1,5)A$ & 17.94 & -5.0079 & 18.0 & 95.0 & 35.8 &  \\
88998.3650(100) & $7(2,5)-6(2,4)A\ v=1$ & 207.37 & -5.0298 & 17.1 & 97.4 & 40.9 &  \\
89140.3750(100) & $7(2,5)-6(2,4)E\ v=1$ & 206.70 & -5.0277 & 17.1 & 96.2 & 6.2 &  \\
89314.6570(100) & $8(1,8)-7(1,7)E$ & 20.15 & -4.9931 & 20.8 & 99.2 & 15.2 &  \\
89316.6420(100) & $8(1,8)-7(1,7)A$ & 20.14 & -4.9929 & 20.8 & 100.0 & 8.1 &  \\
89731.6950(100) & $8(0,8)-7(0,7)A\ v=1$ & 207.83 & -4.9870 & 20.8 & 91.8 & 10.1 &  \\
89829.7040(100) & $8(0,8)-7(0,7)E\ v=1$ & 207.04 & -4.9832 & 20.9 & 98.4 & 21.9 &  \\
90145.7230(100) & $7(2,5)-6(2,4)E$ & 19.68 & -5.0114 & 17.1 & 99.8 & 25.4 & \citep{rsfls2003} \\
                &                  &       &         &      &      &      & \citep{rmr2006}\\
90156.4730(100) & $7(2,5)-6(2,4)A$ & 19.67 & -5.0111 & 17.1 & 98.7 & 28.2 & \citep{rsfls2003}  \\
                &                  &       &         &      &      &      & \citep{rmr2006}\\
90227.6590(100) & $8(0,8)-7(0,7)E$ & 20.08 & -4.9785 & 20.9 & 100.0 & 50.4 & \citep{rsfls2003} \\
                &                  &       &         &      &      &      & \citep{rmr2006}\\
90229.6240(100) & $8(0,8)-7(0,7)A$ & 20.06 & -4.9783 & 20.9 & 99.5 & 45.1 & \citep{rsfls2003}  \\
                &                  &       &         &      &      &      & \citep{rmr2006}\\
91775.9350(100) & $8(1,8)-7(0,7)E$ & 20.15 & -5.8125 & 2.9 & 99.5 & 29.1 &  \\
91776.8860(100) & $20(4,16)-20(3,17)E$ & 138.67 & -5.8587 & 6.3 & " & " &  \\
91777.2300(100) & $8(1,8)-7(0,7)A$ & 20.14 & -5.8128 & 2.9 & " & " &  \\
92073.1010(100) & $18(5,13)-18(4,14)A$ & 118.78 & -5.8392 & 5.9 & 87.8 & 10.7 &  \\
92074.0200(100) & $18(5,13)-18(4,14)E$ & 118.79 & -5.8391 & 5.9 & " & " &  \\
95242.0050(100) & $8(2,7)-7(2,6)A\ v=1$ & 211.31 & -4.9318 & 19.8 & 94.9 & 37.5 &  \\
95689.3990(100) & $8(2,7)-7(2,6)E\ v=1$ & 210.57 & -4.9247 & 19.8 & 83.2 & 18.4 &  \\
96070.7250(100) & $8(2,7)-7(2,6)E$ & 23.61 & -4.9191 & 19.8 & 97.7 & 97.6 &  \\
96076.8450(100) & $8(2,7)-7(2,6)A$ & 23.59 & -4.9189 & 19.8 & 99.4 & 51.8 &  \\
97199.1110(100) & $17(5,12)-17(4,13)E$ & 107.82 & -5.7969 & 5.2 & " & " &  \\
97199.2220(100) & $17(5,12)-17(4,13)A$ & 107.81 & -5.7970 & 5.2 & " & " &  \\
97304.7110(100) & $8(7,1)-7(7,0)A\ v=1$ & 241.56 & -5.5029 & 5.0 & 93.2 & 0.8 &  \\
97304.7110(100) & $8(7,1)-7(7,0)A\ v=1$ & 241.56 & -5.5029 & 5.0 & " & " &  \\
97350.4170(300) & $8(6,2)-7(6,1)A\ v=1$ & 232.90 & -5.2315 & 9.3 & 96.8 & 56.8 &  \\
97350.4170(300) & $8(6,2)-7(6,1)A\ v=1$ & 232.90 & -5.2315 & 9.3 & " & " &  \\
97395.3420(100) & $8(6,2)-7(6,1)E\ v=1$ & 232.82 & -5.2291 & 9.3 & 94.3 & 15.3 &  \\
97457.9670(100) & $8(5,4)-7(5,3)A\ v=1$ & 225.57 & -5.0864 & 12.9 & 89.1 & 17.6 &  \\
97460.3950(100) & $8(5,3)-7(5,2)A\ v=1$ & 225.57 & -5.0863 & 12.9 & 93.3 & 26.3 &  \\
97577.3030(100) & $8(5,3)-7(5,2)E\ v=1$ & 225.36 & -5.0824 & 13.0 & 72.9 & 43.2 &  \\
97597.1610(100) & $8(3,6)-7(3,5)A\ v=1$ & 214.94 & -4.9360 & 18.2 & 79.2 & 14.5 &  \\
97651.2700(300) & $10(4,7)-10(3,8)E$ & 43.17 & -5.9016 & 2.4 & 80.6 & 10.1 &  \\
97651.2700(300) & $10(4,7)-10(3,8)E$ & 43.17 & -5.9016 & 2.4 & " & " &  \\
97661.4010(100) & $8(4,5)-7(4,4)A\ v=1$ & 219.60 & -4.9937 & 15.9 & 93.4 & 24.8 &  \\
97738.7380(100) & $8(6,3)-7(6,2)E\ v=1$ & 232.09 & -5.2243 & 9.3 & " & " &  \\
97752.8850(100) & $8(4,4)-7(4,3)A\ v=1$ & 219.60 & -4.9924 & 15.9 & 80.4 & 58.4 &  \\
97885.6630(100) & $8(5,4)-7(5,3)E\ v=1$ & 224.73 & -5.0787 & 13.0 & 91.4 & 20.6 &  \\
97897.1180(100) & $8(4,4)-7(4,3)E\ v=1$ & 219.26 & -4.9875 & 16.0 & 99.3 & 42.5 &  \\
98176.2930(100) & $8(4,5)-7(4,4)E\ v=1$ & 218.74 & -4.9853 & 16.0 & 82.3 & 1.8 &  \\
98182.3360(100) & $8(7,1)-7(7,0)E$ & 53.78 & -5.4902 & 5.0 & 80.2 & 2.7 &  \\
98190.6580(100) & $8(7,2)-7(7,1)A$ & 53.76 & -5.4900 & 5.0 & 99.7 & 34.8 &  \\
98190.6580(100) & $8(7,2)-7(7,1)A$ & 53.76 & -5.4900 & 5.0 & " & " &  \\
98191.4600(100) & $8(7,2)-7(7,1)E$ & 53.76 & -5.4901 & 5.0 & " & " &  \\
98270.5010(100) & $8(6,2)-7(6,1)E$ & 45.15 & -5.2180 & 9.3 & 79.0 & 15.2 &  \\
98278.9210(100) & $8(6,3)-7(6,2)E$ & 45.13 & -5.2179 & 9.3 & 96.4 & 65.3 &  \\
98279.7620(100) & $8(6,3)-7(6,2)A$ & 45.13 & -5.2178 & 9.3 & " & " &  \\
98279.7620(100) & $8(6,3)-7(6,2)A$ & 45.13 & -5.2178 & 9.3 & " & " &  \\
98423.1650(100) & $9(0,9)-8(1,8)A\ v=1$ & 212.63 & -5.7011 & 3.4 & 91.4 & 36.1 &  \\
98424.2070(100) & $8(5,3)-7(5,2)E$ & 37.86 & -5.0722 & 13.0 & " & " &  \\
98431.8030(100) & $8(5,4)-7(5,3)E$ & 37.84 & -5.0721 & 13.0 & 98.9 & 38.8 &  \\
98432.7600(100) & $8(5,4)-7(5,3)A$ & 37.84 & -5.0720 & 13.0 & " & " &  \\
98435.8020(100) & $8(5,3)-7(5,2)A$ & 37.84 & -5.0720 & 13.0 & 99.7 & 33.3 &  \\
98606.8560(100) & $8(3,6)-7(3,5)E$ & 27.26 & -4.9216 & 18.2 & 82.3 & 9.4 &  \\
98611.1630(100) & $8(3,6)-7(3,5)A$ & 27.24 & -4.9209 & 18.3 & 97.9 & 4.0 &  \\
98682.4210(100) & $8(3,6)-7(3,5)E\ v=1$ & 214.15 & -4.9245 & 18.1 & 99.7 & 54.8 &  \\
98682.6150(100) & $8(4,5)-7(4,4)A$ & 31.89 & -4.9786 & 16.0 & " & " &  \\
98712.0010(100) & $8(4,5)-7(4,4)E$ & 31.90 & -4.9928 & 15.4 & 89.5 & 6.6 &  \\
98713.6550(100) & $26(8,18)-25(9,17)E$ & 250.95 & -6.7530 & 0.8 & " & " &  \\
98747.9060(100) & $8(4,4)-7(4,3)E$ & 31.91 & -4.9923 & 15.4 & 93.0 & 8.8 &  \\
98792.2890(100) & $8(4,4)-7(4,3)A$ & 31.89 & -4.9772 & 16.0 & 91.5 & 11.1 &  \\
98815.2900(100) & $8(3,5)-7(3,4)E\ v=1$ & 214.56 & -4.9193 & 18.2 & 95.6 & 36.2 &  \\
99089.5180(100) & $8(3,5)-7(3,4)A\ v=1$ & 215.08 & -4.9159 & 18.2 & 98.7 & 64.8 &  \\
99133.2720(100) & $9(0,9)-8(1,8)E$ & 24.91 & -5.6979 & 3.4 & 98.8 & 22.3 &  \\
99135.7620(100) & $9(0,9)-8(1,8)A$ & 24.89 & -5.6982 & 3.4 & 100.0 & 13.0 &  \\
99489.0570(100) & $9(1,9)-8(1,8)A\ v=1$ & 212.69 & -4.8499 & 23.4 & 99.1 & 39.4 &  \\
99575.5480(100) & $8(1,7)-7(1,6)A\ v=1$ & 210.43 & -4.8565 & 20.6 & 98.1 & 31.1 &  \\
99577.4190(100) & $9(1,9)-8(1,8)E\ v=1$ & 211.90 & -4.8461 & 23.6 & " & " &  \\
99869.1020(100) & $8(1,7)-7(1,6)E\ v=1$ & 209.71 & -4.8509 & 20.7 & 94.0 & 10.4 &  \\
100078.6080(100) & $9(1,9)-8(1,8)E$ & 24.96 & -4.8407 & 23.5 & 96.9 & 9.7 &  \\
100080.5420(100) & $9(1,9)-8(1,8)A$ & 24.94 & -4.8406 & 23.5 & 94.7 & 6.8 &  \\
100136.9130(100) & $9(0,9)-8(0,8)A\ v=1$ & 212.63 & -4.8407 & 23.5 & 73.1 & 13.2 &  \\
100226.6840(100) & $9(0,9)-8(0,8)E\ v=1$ & 211.85 & -4.8370 & 23.6 & 81.9 & 31.9 &  \\
100294.6040(100) & $8(3,5)-7(3,4)E$ & 27.41 & -4.8992 & 18.3 & 99.7 & 32.3 &  \\
100308.1790(100) & $8(3,5)-7(3,4)A$ & 27.40 & -4.8985 & 18.3 & 99.3 & 37.7 &  \\
100482.2410(100) & $8(1,7)-7(1,6)E$ & 22.78 & -4.8437 & 20.6 & 99.8 & 13.5 &  \\
100490.6820(100) & $8(1,7)-7(1,6)A$ & 22.76 & -4.8435 & 20.6 & 87.5 & 20.6 &  \\
100681.5450(100) & $9(0,9)-8(0,8)E$ & 24.91 & -4.8323 & 23.5 & 98.2 & 34.9 &  \\
100683.3680(100) & $9(0,9)-8(0,8)A$ & 24.89 & -4.8322 & 23.5 & " & " &  \\
101626.8840(100) & $9(1,9)-8(0,8)E$ & 24.96 & -5.6633 & 3.4 & 90.9 & 24.9 &  \\
101628.1490(100) & $9(1,9)-8(0,8)A$ & 24.94 & -5.6636 & 3.4 & " & " &  \\
102179.5100(100) & $8(2,6)-7(2,5)A\ v=1$ & 212.27 & -4.8370 & 19.9 & 96.5 & 8.8 &  \\
102503.1050(100) & $8(2,6)-7(2,5)E\ v=1$ & 211.62 & -4.8307 & 20.0 & 92.3 & 25.2 &  \\
102734.2120(100) & $16(5,11)-16(4,12)E$ & 97.52 & -5.7520 & 4.6 & 99.6 & 18.0 &  \\
102736.8690(100) & $16(5,11)-16(4,12)A$ & 97.51 & -5.7519 & 4.6 & 99.3 & 16.2 &  \\
103376.8420(100) & $24(6,18)-24(5,19)A$ & 203.82 & -5.6555 & 8.4 & 70.8 & 11.7 &  \\
103387.2000(100) & $24(6,18)-24(5,19)E$ & 203.82 & -5.6553 & 8.4 & 91.0 & 2.5 &  \\
103466.5720(100) & $8(2,6)-7(2,5)E$ & 24.65 & -4.8190 & 20.0 & 98.3 & 42.2 &  \\
103478.6630(100) & $8(2,6)-7(2,5)A$ & 24.63 & -4.8187 & 20.0 & 98.9 & 63.4 &  \\
106648.7540(100) & $9(2,8)-8(2,7)A\ v=1$ & 216.42 & -4.7760 & 22.5 & 95.6 & 8.8 &  \\
107022.1590(100) & $9(2,8)-8(2,7)E\ v=1$ & 215.71 & -4.7693 & 22.6 & 79.3 & 6.5 &  \\
107537.2580(100) & $9(2,8)-8(2,7)E$ & 28.77 & -4.7639 & 22.6 & 75.8 & 20.4 &  \citep{rsfls2003} \\
107543.7110(100) & $9(2,8)-8(2,7)A$ & 28.75 & -4.7637 & 22.6 & 98.3 & 5.9 &  \citep{rsfls2003}  \\
107590.3890(100) & $23(6,17)-23(5,18)A$ & 189.04 & -5.6251 & 7.7 & 91.8 & 1.0 &  \\
107604.3660(100) & $23(6,17)-23(5,18)E$ & 189.04 & -5.6248 & 7.7 & 93.9 & 1.1 &  \\
108045.9590(100) & $15(5,10)-15(4,11)E$ & 87.87 & -5.7107 & 4.1 & 94.6 & 25.3 &  \\
108050.9390(100) & $15(5,10)-15(4,11)A$ & 87.86 & -5.7099 & 4.1 & 88.1 & 25.4 &  \\
109501.2460(100) & $9(8,1)-8(8,0)A\ v=1$ & 256.83 & -5.3939 & 5.0 & 84.9 & 2.5 &  \\
109501.2460(100) & $9(8,1)-8(8,0)A\ v=1$ & 256.83 & -5.3939 & 5.0 & " & " &  \\
109502.6550(100) & $9(7,2)-8(7,1)E\ v=1$ & 246.89 & -5.1181 & 9.5 & " & " &  \\
109531.4130(100) & $9(7,2)-8(7,1)A\ v=1$ & 246.82 & -5.1190 & 9.4 & 99.5 & 34.9 &  \\
109531.4130(100) & $9(7,2)-8(7,1)A\ v=1$ & 246.82 & -5.1190 & 9.4 & " & " &  \\
109608.5400(300) & $9(6,3)-8(6,2)A\ v=1$ & 238.16 & -4.9703 & 13.3 & 64.0 & 11.3 &  \\
109608.5400(300) & $9(6,3)-8(6,2)A\ v=1$ & 238.16 & -4.9703 & 13.3 & " & " &  \\
109662.9440(100) & $9(6,3)-8(6,2)E\ v=1$ & 238.08 & -4.9679 & 13.3 & 90.5 & 66.7 &  \\
109763.7370(100) & $9(3,7)-8(3,6)A\ v=1$ & 220.21 & -4.7655 & 21.2 & 91.3 & 12.2 &  \\
109770.9950(100) & $9(5,5)-8(5,4)A\ v=1$ & 230.84 & -4.8736 & 16.5 & 89.0 & 11.5 &  \\
109778.8350(100) & $9(5,4)-8(5,3)A\ v=1$ & 230.84 & -4.8735 & 16.5 & 94.7 & 3.0 &  \\
109912.1680(100) & $9(5,4)-8(5,3)E\ v=1$ & 230.63 & -4.8697 & 16.6 & 80.5 & 3.2 &  \\
109924.8170(100) & $9(7,3)-8(7,2)E\ v=1$ & 246.10 & -5.1124 & 9.5 & 87.8 & 70.9 &  \\
110035.2690(100) & $9(4,6)-8(4,5)A\ v=1$ & 224.88 & -4.8060 & 19.2 & 97.6 & 7.2 &  \\
110050.3320(100) & $9(6,4)-8(6,3)E\ v=1$ & 237.38 & -4.9631 & 13.3 & 96.4 & 1.8 &  \\
110153.6520(100) & $10(1,10)-9(1,9)A\ v=1$ & 217.97 & -4.7139 & 26.1 & 94.6 & 14.8 &  \\
110364.5200(100) & $9(4,5)-8(4,4)E\ v=1$ & 224.55 & -4.7991 & 19.3 & 99.4 & 2.6 &  \\
110447.1800(100) & $9(8,1)-8(8,0)E$ & 69.04 & -5.3818 & 5.0 & 93.3 & 6.3 &  \\
110455.3720(100) & $9(8,2)-8(8,1)A$ & 69.03 & -5.3817 & 5.0 & 89.4 & 43.3 &  \\
110455.3720(100) & $9(8,2)-8(8,1)A$ & 69.03 & -5.3817 & 5.0 & " & " &  \\
110458.0140(100) & $9(8,2)-8(8,1)E$ & 69.02 & -5.3817 & 5.0 & " & " &  \\
110525.7410(100) & $9(7,2)-8(7,1)E$ & 59.09 & -5.1063 & 9.5 & 83.0 & 60.7 &  \\
110526.1900(100) & $7(2,6)-6(1,5)A\ v=1$ & 206.74 & -5.8286 & 1.4 & " & " &  \\
110535.1860(100) & $9(7,3)-8(7,2)A$ & 59.07 & -5.1061 & 9.5 & 91.3 & 67.5 &  \\
110535.1860(100) & $9(7,3)-8(7,2)A$ & 59.07 & -5.1061 & 9.5 & " & " &  \\
110536.0030(100) & $9(7,3)-8(7,2)E$ & 59.07 & -5.1061 & 9.5 & " & " &  \\
110571.6320(100) & $10(0,10)-9(0,9)A\ v=1$ & 217.94 & -4.7086 & 26.1 & 93.4 & 28.8 &  \\
110652.8130(100) & $9(6,3)-8(6,2)E$ & 50.46 & -4.9568 & 13.3 & 98.5 & 30.4 &  \\
110655.3100(100) & $10(0,10)-9(0,9)E\ v=1$ & 217.16 & -4.7051 & 26.3 & 98.6 & 16.3 &  \\
110662.3150(100) & $9(6,4)-8(6,3)E$ & 50.45 & -4.9567 & 13.3 & 77.1 & 19.0 &  \\
110663.2730(100) & $9(6,4)-8(6,3)A$ & 50.44 & -4.9566 & 13.3 & " & " &  \\
110663.4290(100) & $9(6,3)-8(6,2)A$ & 50.44 & -4.9566 & 13.3 & " & " &  \\
110684.1230(100) & $9(4,6)-8(4,5)E\ v=1$ & 224.06 & -4.7970 & 19.2 & 92.6 & 2.9 &  \\
110776.4990(100) & $9(1,8)-8(1,7)A\ v=1$ & 215.75 & -4.7150 & 23.1 & 96.9 & 68.0 &  \\
110788.6640(100) & $10(1,10)-9(1,9)E$ & 30.27 & -4.7051 & 26.2 & 96.6 & 62.5 &  \\
110790.5260(100) & $10(1,10)-9(1,9)A$ & 30.26 & -4.7049 & 26.2 & " & " &  \\
110873.9550(100) & $9(5,4)-8(5,3)E$ & 43.18 & -4.8594 & 16.6 & 96.8 & 57.9 &  \\
110879.7660(100) & $9(3,7)-8(3,6)E$ & 32.58 & -4.7510 & 21.2 & 97.1 & 89.6 &  \\
110880.4470(100) & $9(5,5)-8(5,4)A$ & 43.16 & -4.8592 & 16.6 & " & " &  \\
110882.3310(100) & $9(5,5)-8(5,4)E$ & 43.16 & -4.8594 & 16.6 & " & " &  \\
110887.0920(100) & $9(3,7)-8(3,6)A$ & 32.57 & -4.7507 & 21.3 & 99.8 & 52.7 &  \\
110890.2560(100) & $9(5,4)-8(5,3)A$ & 43.16 & -4.8590 & 16.6 & 99.4 & 39.2 &  \\
111005.6720(100) & $9(3,7)-8(3,6)E\ v=1$ & 219.48 & -4.7583 & 20.8 & 70.7 & 13.4 &  \\
111094.1240(100) & $9(1,8)-8(1,7)E\ v=1$ & 215.04 & -4.7094 & 23.2 & 71.8 & 14.7 &  \\
111169.9030(100) & $10(0,10)-9(0,9)E$ & 30.25 & -4.7002 & 26.2 & 99.6 & 82.7 &  \\
111171.6340(100) & $10(0,10)-9(0,9)A$ & 30.23 & -4.7001 & 26.2 & " & " &  \\
111195.9620(100) & $9(4,6)-8(4,5)A$ & 37.22 & -4.7909 & 19.2 & 93.9 & 55.9 &  \\
111223.4910(100) & $9(4,6)-8(4,5)E$ & 37.23 & -4.8145 & 18.2 & 98.7 & 53.0 &  \\
111224.9860(100) & $14(3,11)-13(4,10)E\ v=1$ & 256.70 & -6.3187 & 0.9 & " & " &  \\
111408.4120(100) & $9(4,5)-8(4,4)E$ & 37.26 & -4.8123 & 18.2 & 94.9 & 86.1 & \citep{rmr2006} \\
111453.3000(100) & $9(4,5)-8(4,4)A$ & 37.24 & -4.7878 & 19.2 & 99.9 & 75.0 & \citep{rsfls2003} \\
111674.1310(100) & $9(1,8)-8(1,7)E$ & 28.14 & -4.7037 & 23.2 & 99.6 & 78.7 & \citep{rmr2006} \\
111682.1890(100) & $9(1,8)-8(1,7)A$ & 28.12 & -4.7035 & 23.2 & 99.6 & 58.9 & \citep{rmr2006} \\
111734.0020(100) & $10(1,10)-9(0,9)E$ & 30.27 & -5.5269 & 3.8 & 95.9 & 36.0 &  \\
111735.3070(100) & $10(1,10)-9(0,9)A$ & 30.26 & -5.5271 & 3.8 & " & " &  \\
112011.9660(100) & $9(3,6)-8(3,5)E\ v=1$ & 219.94 & -4.7426 & 21.0 & " & " &  \\
112306.9410(100) & $9(3,6)-8(3,5)A\ v=1$ & 220.47 & -4.7352 & 21.2 & 96.7 & 18.8 &  \\
113743.1070(100) & $9(3,6)-8(3,5)E$ & 32.87 & -4.7171 & 21.3 & 98.1 & 14.4 &  \\
113756.6100(100) & $9(3,6)-8(3,5)A$ & 32.86 & -4.7168 & 21.3 & 99.1 & 30.5&
    \enddata
    \tablecomments{Pertinent parameters of the detected transitions of MF taken from JPL catalog. The transitions with ditto marks in the column of P and D factors are blended with the transitions in the previous row.}
\end{deluxetable*}

\startlongtable
\begin{deluxetable*}{llrrrccl}
    \tabletypesize{\scriptsize}
    \tablecaption{Transitions of GLA with $S/N > 3$ covered by the EMoCA survey.\label{GLA-appendix-table}}
    \tablehead{
        \\
        \colhead{Rest Frequency}	& \colhead{Transition}	& \colhead{$E_u$}	    & \colhead{$\log_{10}{\frac{A_{ul}}{\mathrm{s^{-1}}}}$}	&
        \colhead{$S_{ij}\mu^{2}$}	& \colhead{P Factor}	& \colhead{D Factor}	& Obs Ref\\
        \colhead{(MHz)}	            & ~                 	& \colhead{(K)} 	    & ~                             &
        \colhead{(Debye$^2$)}    	& (\%) 	                & (\%)                  &
    }
    \startdata
    85288.3467(47) & $8(4,4)-8(3,5)$ & 29.74 & -5.0287 & 22.0 & 73.0 & 63.8 &   \\
85476.6666(92) & $11(1,10)-11(0,11)$ & 37.91 & -5.3021 & 15.8 & 97.9 & 41.4 &   \\
85782.2440(55) & $8(1,8)-7(0,7)$ & 18.87 & -4.8264 & 34.5 & 73.6 & 31.2 &\citep{hawpz2006}   \\
86600.5997(67) & $17(5,12)-17(4,13)$ & 101.14 & -4.8624 & 63.6 & 91.0 & 86.6 &   \\
87764.5007(47) & $5(2,4)-4(1,3)$ & 10.69 & -5.0665 & 12.0 & 84.5 & 27.8 &   \\
87812.2080(47) & $6(4,2)-6(3,3)$ & 21.38 & -5.0745 & 13.9 & 94.3 & 8.9 &   \\
88425.7474(47) & $7(4,4)-7(3,5)$ & 25.27 & -5.0198 & 17.8 & 99.0 & 20.7 &   \\
88463.7084(47) & $6(4,3)-6(3,4)$ & 21.38 & -5.0662 & 13.9 & 90.7 & 15.9 &   \\
88530.4001(46) & $8(4,5)-8(3,6)$ & 29.73 & -4.9874 & 21.7 & 94.9 & 68.9 &   \\
88691.2605(68) & $12(3,10)-12(2,11)$ & 48.77 & -5.0203 & 29.4 & 92.2 & 58.9 &   \\
88892.4467(45) & $9(4,6)-9(3,7)$ & 34.76 & -4.9607 & 25.4 & 85.0 & 14.1 &   \\
89616.3982(132) & $20(4,16)-20(3,17)$ & 130.44 & -4.8374 & 71.2 & 98.4 & 29.8 &   \\
89644.1477(45) & $10(4,7)-10(3,8)$ & 40.35 & -4.9352 & 29.1 & 89.7 & 2.2 &   \\
89832.2856(115) & $14(2,12)-14(1,13)$ & 62.93 & -5.0296 & 32.1 & 91.8 & 39.5 &   \\
89868.6318(54) & $9(1,8)-8(2,7)$ & 26.42 & -5.0389 & 20.6 & 94.0 & 16.8 &\citep{hawpz2006}   \\
90187.0697(109) & $23(5,18)-23(4,19)$ & 172.97 & -4.7716 & 93.1 & 95.7 & 28.2 &   \\
90507.2610(133) & $17(3,14)-17(2,15)$ & 93.78 & -4.9064 & 50.3 & 98.1 & 80.4 &   \\
90591.3442(78) & $11(2,10)-11(1,11)$ & 38.17 & -5.2329 & 15.5 & 86.6 & 1.8 &   \\
90922.2144(46) & $11(4,8)-11(3,9)$ & 46.51 & -4.9075 & 32.5 & 58.2 & 22.1 &   \\
91712.9604(71) & $16(5,11)-16(4,12)$ & 91.41 & -4.8169 & 56.0 & 74.2 & 7.9 &   \\
92738.2867(52) & $25(6,19)-25(5,20)$ & 206.18 & -4.7044 & 108.5 & 85.7 & 27.2 &   \\
92853.9487(49) & $12(4,9)-12(3,10)$ & 53.23 & -4.8762 & 35.7 & 97.2 & 79.2 &   \\
93048.5007(46) & $24(6,18)-24(5,19)$ & 191.55 & -4.7088 & 102.2 & 94.2 & 27.1 &   \\
93052.6745(56) & $9(0,9)-8(1,8)$ & 23.34 & -4.7062 & 39.8 & 95.7 & 44.1 &\citep{hlj2000}   \\
               &                 &       &         &      &      &      &\citep{hawpz2006}  \\
95070.0983(60) & $9(1,9)-8(0,8)$ & 23.37 & -4.6763 & 40.0 & 78.9 & 10.5 &\citep{hawpz2006}   \\
95175.1800(76) & $26(6,20)-26(5,21)$ & 221.45 & -4.6731 & 112.1 & 84.4 & 31.5 &   \\
95300.4082(75) & $13(3,11)-13(2,12)$ & 55.93 & -4.9488 & 30.2 & 84.9 & 26.7 &   \\
95443.0896(53) & $6(2,5)-5(1,4)$ & 13.97 & -4.9612 & 14.0 & 99.5 & 1.2 &   \\
95544.7538(53) & $13(4,10)-13(3,11)$ & 60.52 & -4.8397 & 38.5 & 82.6 & 18.2 &   \\
95741.7260(94) & $12(1,11)-12(0,12)$ & 44.35 & -5.1911 & 15.8 & 79.9 & 4.6 &   \\
95756.2369(57) & $23(6,17)-23(5,18)$ & 177.57 & -4.6897 & 94.0 & 83.9 & 58.9 &   \\
96078.4307(75) & $11(2,9)-10(3,8)$ & 40.66 & -5.1981 & 14.1 & 75.8 & 13.6 &   \\
96870.7828(70) & $15(5,10)-15(4,11)$ & 82.31 & -4.7720 & 49.5 & 89.6 & 67.1 &   \\
97919.5791(49) & $3(3,1)-2(2,0)$ & 8.80 & -4.6806 & 13.4 & 7.1 & 49.1 &   \\
98070.5113(49) & $3(3,0)-2(2,1)$ & 8.80 & -4.6793 & 13.3 & 87.8 & 12.2 &   \\
99068.4727(59) & $14(4,11)-14(3,12)$ & 68.36 & -4.7976 & 40.8 & 71.4 & 20.6 &   \\
99109.9990(83) & $12(2,11)-12(1,12)$ & 44.52 & -5.1491 & 15.7 & 99.6 & 41.8 &   \\
99346.4488(133) & $24(5,19)-24(4,20)$ & 187.08 & -4.6741 & 90.9 & 93.1 & 11.8 &   \\
100323.8733(73) & $22(6,16)-22(5,17)$ & 164.25 & -4.6532 & 85.1 & 92.8 & 2.2 &   \\
100506.6546(103) & $27(6,21)-27(5,22)$ & 237.35 & -4.6155 & 112.8 & 92.7 & 21.8 &   \\
101116.3124(145) & $21(4,17)-21(3,18)$ & 142.69 & -4.7161 & 68.7 & 99.6 & 39.5 &   \\
101232.1720(115) & $15(2,13)-15(1,14)$ & 71.31 & -4.9076 & 31.8 & 79.8 & 21.7 &   \\
101527.8515(65) & $14(5,9)-14(4,10)$ & 73.82 & -4.7334 & 44.0 & 93.9 & 15.0 &   \\
102549.7240(60) & $7(2,6)-6(1,5)$ & 17.78 & -4.8570 & 16.6 & 76.3 & 8.1 &\citep{hawpz2006}   \\
102572.9391(82) & $14(3,12)-14(2,13)$ & 63.60 & -4.8764 & 30.7 & 99.0 & 62.2 &   \\
102614.3429(137) & $18(3,15)-18(2,16)$ & 104.10 & -4.7784 & 49.0 & 97.9 & 64.4 &   \\
103391.2854(62) & $10(0,10)-9(1,9)$ & 28.34 & -4.5552 & 45.5 & 93.5 & 26.3 &\citep{hlj2000}   \\
                &                   &       &         &      &      &      &\citep{hawpz2006}  \\
103461.3106(66) & $15(4,12)-15(3,13)$ & 76.75 & -4.7501 & 42.8 & 95.0 & 77.7 &   \\
103667.9094(57) & $10(1,9)-9(2,8)$ & 31.93 & -4.7910 & 26.2 & 91.3 & 58.6 &\citep{hlj2000}   \\
                &                   &       &         &      &      &      &\citep{hawpz2006}  \\
104587.7032(64) & $10(1,10)-9(0,9)$ & 28.36 & -4.5394 & 45.5 & 88.2 & 6.9 &\citep{hawpz2006}   \\
105355.6639(59) & $13(5,8)-13(4,9)$ & 65.93 & -4.7051 & 39.1 & 99.2 & 16.0 &   \\
106067.4262(84) & $21(6,15)-21(5,16)$ & 151.60 & -4.6068 & 76.6 & 87.7 & 21.6 &   \\
106710.9951(60) & $29(7,22)-29(6,23)$ & 275.99 & -4.5255 & 124.4 & 87.1 & 22.4 &   \\
107380.1425(74) & $30(7,23)-30(6,24)$ & 293.59 & -4.5135 & 129.8 & 81.9 & 21.5 &   \\
107663.7353(115) & $19(5,14)-18(6,13)$ & 122.53 & -5.3224 & 12.8 & 88.9 & 13.9 &   \\
107874.8614(87) & $13(2,12)-13(1,13)$ & 51.36 & -5.0703 & 15.7 & 99.4 & 13.3 &   \\
108261.4557(56) & $12(5,7)-12(4,8)$ & 58.63 & -4.6885 & 34.7 & 94.7 & 40.1 &   \\
108719.6747(73) & $16(4,13)-16(3,14)$ & 85.69 & -4.6979 & 44.2 & 94.6 & 3.8 &   \\
108824.4021(62) & $28(7,21)-28(6,22)$ & 259.05 & -4.5133 & 116.5 & 98.2 & 59.3 &   \\
109114.1722(53) & $4(3,2)-3(2,1)$ & 11.02 & -4.6411 & 13.6 & 79.0 & 18.3 &   \\
109280.0092(66) & $8(2,7)-7(1,6)$ & 22.11 & -4.7522 & 19.8 & 87.7 & 60.2 &\citep{hawpz2006}   \\
109877.1408(54) & $4(3,1)-3(2,2)$ & 11.02 & -4.6354 & 13.5 & 77.9 & 1.9 &   \\
110330.6028(55) & $11(5,6)-11(4,7)$ & 51.91 & -4.6836 & 30.5 & 99.2 & 3.3 &   \\
110402.5367(87) & $15(3,13)-15(2,14)$ & 71.79 & -4.8048 & 31.0 & 99.4 & 80.3 &   \\
110608.5089(148) & $25(5,20)-25(4,21)$ & 201.73 & -4.5659 & 88.0 & 84.5 & 3.9 &   \\
110845.1604(102) & $14(3,11)-13(4,10)$ & 65.84 & -5.1366 & 13.4 & 74.9 & 2.8 &   \\
111098.2681(97) & $31(7,24)-31(6,25)$ & 311.83 & -4.4755 & 132.1 & 97.6 & 53.3 &   \\
111734.8569(55) & $10(5,5)-10(4,6)$ & 45.76 & -4.6897 & 26.4 & 98.5 & 47.0 &   \\
112177.7128(51) & $13(5,9)-13(4,10)$ & 65.90 & -4.6338 & 38.2 & 75.4 & 27.1 &   \\
112210.4422(52) & $12(5,8)-12(4,9)$ & 58.62 & -4.6471 & 34.3 & 92.2 & 70.7 &   \\
112247.9894(87) & $20(6,14)-20(5,15)$ & 139.58 & -4.5575 & 69.0 & 81.7 & 25.9 &   \\
112341.6773(114) & $16(2,14)-16(1,15)$ & 80.14 & -4.8011 & 31.6 & 94.4 & 22.8 &   \\
112444.3079(53) & $11(5,7)-11(4,8)$ & 51.90 & -4.6614 & 30.3 & 86.6 & 31.1 &   \\
112472.7946(50) & $14(5,10)-14(4,11)$ & 73.76 & -4.6197 & 42.0 & 98.6 & 98.4 &   \\
112656.0535(56) & $9(5,4)-9(4,5)$ & 40.19 & -4.7071 & 22.4 & 69.6 & 8.7 &   \\
112773.4455(55) & $10(5,6)-10(4,7)$ & 45.76 & -4.6788 & 26.4 & 81.6 & 9.2 &   \\
113326.8996(77) & $12(2,10)-11(3,9)$ & 47.58 & -4.9081 & 18.2 & 88.5 & 14.4 &\citep{hawpz2006}   \\
113429.3652(57) & $8(5,4)-8(4,5)$ & 35.18 & -4.7360 & 18.4 & 95.7 & 5.9 &   \\
113569.5317(66) & $11(0,11)-10(1,10)$ & 33.81 & -4.4222 & 51.0 & 75.3 & 42.3 &\citep{hawpz2006}   \\
113616.7028(57) & $7(5,2)-7(4,3)$ & 30.73 & -4.7891 & 14.3 & 83.5 & 5.3 &   \\
113679.1292(57) & $7(5,3)-7(4,4)$ & 30.73 & -4.7884 & 14.3 & 95.8 & 7.9 &   \\
113694.9880(150) & $22(4,18)-22(3,19)$ & 155.43 & -4.5962 & 66.7 & 95.6 & 10.1 &   \\
114264.4356(67) & $11(1,11)-10(0,10)$ & 33.82 & -4.4139 & 51.1 & 95.8 & 29.3& \citep{hawpz2006}
    \enddata
    \tablecomments{Pertinent parameters of the detected transitions of GLA taken from CDMS catalog. The transitions with ditto marks in the column of P and D factors are blended with the transitions in the previous row.}
\end{deluxetable*}

\startlongtable
\begin{deluxetable*}{llrrrccl}
    \tabletypesize{\scriptsize}
    \tablecaption{Transitions of AcA with $S/N > 3$ covered by the EMoCA survey.\label{AcA-appendix-table}}
    \tablehead{
        \\
        \colhead{Rest Frequency}	& \colhead{Transition}	& \colhead{$E_u$}	    & \colhead{$\log_{10}{\frac{A_{ul}}{\mathrm{s^{-1}}}}$}	&
        \colhead{$S_{ij}\mu^{2}$}	& \colhead{P Factor}	& \colhead{D Factor}	& Obs Ref\\
        \colhead{(MHz)}	            & ~                 	& \colhead{(K)} 	    & ~                             &
        \colhead{(Debye$^2$)}    	& (\%) 	                & (\%)                  &
    }
    \startdata
    84857.3100(10) & $10(2,8)-10(2,9)+- $ & 39.18 & -6.4449 & 1.1 & 89.4 & 62.3  &   \\
84857.3200(10) & $10(2,8)-10(1,9)+- $ & 39.18 & -5.7818 & 4.9 & " & "  &   \\
84857.8300(10) & $10(3,8)-10(2,9)+- $ & 39.18 & -5.7818 & 4.9 & " & "  &   \\
84857.8300(10) & $10(3,8)-10(2,9)+- $ & 39.18 & -5.7818 & 4.9 & " & "  &   \\
85399.0300(10) & $19(10,9)-19(9,10)+- $ & 167.72 & -5.3809 & 22.4 & 89.8 & 50.3  &   \\
85981.5900(10) & $19(-10,9)-19(-9,10) $ & 167.84 & -5.3722 & 22.4 & 81.1 & 22.7  &   \\
86473.4400(20) & $23(15,9)-23(14,10)+- $ & 251.63 & -5.3148 & 30.2 & 95.1 & 4.7  &   \\
86569.9100(10) & $20(12,9)-20(11,10)-+ $ & 187.34 & -5.3465 & 24.4 & 80.8 & 86.5  &   \\
86879.0100(20) & $23(13,10)-23(12,11)-+ $ & 247.08 & -5.3194 & 29.5 & 96.5 & 6.0  &   \\
87631.1000(40) & $15(-11,4)-15(-9,6) $ & 113.06 & -7.0292 & 0.4 & 95.5 & 5.6  &   \\
87632.9100(10) & $19(11,9)-19(10,10) $ & 167.92 & -5.3486 & 22.3 & " & "  &   \\
87759.3100(30) & $14(-14,0)-14(-13,1) $ & 112.51 & -5.9766 & 3.9 & 91.1 & 14.8  &   \\
87761.2200(10) & $19(11,9)-19(10,10)+- $ & 167.84 & -5.3475 & 22.3 & " & "  &   \\
87822.7200(10) & $4(-4,0)-3(-3,0) $ & 10.91 & -5.2581 & 6.3 & 84.7 & 3.6  &   \\
87823.6800(10) & $6(2,4)-5(3,3)++ $ & 16.93 & -5.4130 & 6.4 & " & "  &   \\
89516.9600(10) & $7(-1,6)-6(2,5) $ & 19.59 & -5.1941 & 11.5 & 93.2 & 9.2  &   \\
89518.9500(10) & $7(2,6)-6(2,5) $ & 19.59 & -5.6395 & 4.1 & " & "  &   \\
89531.1900(10) & $7(-1,6)-6(-1,5) $ & 19.59 & -5.6393 & 4.1 & 88.9 & 55.7  &   \\
89533.1800(10) & $7(2,6)-6(-1,5) $ & 19.59 & -5.1939 & 11.5 & " & "  &   \\
89653.2800(10) & $7(1,6)-6(2,5)-- $ & 19.12 & -5.1974 & 11.3 & 92.7 & 3.4  &   \\
89677.4100(10) & $7(2,6)-6(1,5)-- $ & 19.12 & -5.1971 & 11.3 & 90.3 & 3.2  &   \\
90203.4400(10) & $8(*,8)-7(1,7) $ & 20.83 & -4.9646 & 21.6 & 99.0 & 23.3  &\citep{msml1997}   \\
               &                  &       &         &      &      &       &\citep{rsfls2003}   \\
90203.4600(10) & $8(*,8)-7(0,7) $ & 20.83 & -4.9646 & 21.6 & " & "  &\citep{msml1997}   \\
               &                  &       &         &      &   &    &\citep{rsfls2003}   \\
90246.2400(10) & $8(*,8)-7(1,7)++ $ & 20.33 & -4.9642 & 21.6 & 96.8 & 47.1  &\citep{msml1997}   \\
               &                  &         &         &      &      &       &\citep{rsfls2003}  \\
90246.2700(10) & $8(*,8)-7(0,7)++ $ & 20.33 & -4.9642 & 21.6 & " & "  &\citep{msml1997}   \\
               &                  &         &         &      &   &    &\citep{rsfls2003}  \\
93760.4300(10) & $12(-3,9)-12(*,10) $ & 58.24 & -5.4784 & 8.7 & 66.9 & 16.1  &   \\
93760.6900(10) & $12(4,9)-12(3,10) $ & 58.24 & -5.5568 & 7.2 & " & "  &   \\
93760.6900(10) & $12(4,9)-12(3,10) $ & 58.24 & -5.5568 & 7.2 & " & "  &   \\
94499.3200(10) & $12(3,9)-12(3,10)-+ $ & 57.87 & -6.2182 & 1.5 & 76.9 & 33.9  &   \\
94499.3300(10) & $12(3,9)-12(2,10)-+ $ & 57.87 & -5.5502 & 7.2 & " & "  &   \\
94499.8200(10) & $12(4,9)-12(3,10)-+ $ & 57.87 & -5.5502 & 7.2 & " & "  &   \\
94499.8300(10) & $12(4,9)-12(2,10)-+ $ & 57.87 & -6.2182 & 1.5 & " & "  &   \\
95199.7500(10) & $22(13,10)-22(12,11) $ & 225.12 & -5.2227 & 26.8 & 75.9 & 1.8  &   \\
95243.0100(10) & $22(13,10)-22(12,11)+- $ & 225.11 & -5.2236 & 26.7 & 84.1 & 18.5  &   \\
95568.4100(10) & $21(-11,10)-21(-10,11) $ & 203.65 & -5.2349 & 24.6 & 72.9 & 23.6  &   \\
96694.7700(10) & $21(12,10)-21(11,11) $ & 203.71 & -5.2204 & 24.6 & 98.5 & 19.2  &   \\
96707.2300(10) & $21(12,10)-21(11,11)-+ $ & 203.68 & -5.2216 & 24.5 & 80.5 & 17.0  &   \\
97347.3100(20) & $25(14,11)-25(13,12)+- $ & 290.34 & -5.1772 & 31.6 & 82.6 & 37.2  &   \\
97508.8900(10) & $20(10,10)-20(9,11)+- $ & 183.14 & -5.2308 & 22.3 & 72.8 & 22.7  &   \\
97685.6600(10) & $20(-10,10)-20(-9,11) $ & 183.22 & -5.2269 & 22.4 & 70.2 & 20.6  &   \\
98134.2400(10) & $20(11,10)-20(10,11) $ & 183.25 & -5.2211 & 22.4 & 96.5 & 4.3  &   \\
98194.4400(10) & $20(11,10)-20(10,11)+- $ & 183.18 & -5.2221 & 22.3 & 96.0 & 13.6  &   \\
99262.3700(10) & $19(-9,10)-19(-8,11) $ & 163.72 & -5.2288 & 20.2 & 98.6 & 27.4  &   \\
99341.5700(10) & $7(-2,5)-6(3,4) $ & 22.19 & -5.1689 & 8.9 & 80.3 & 18.1  &   \\
99424.7200(10) & $7(3,5)-6(3,4) $ & 22.20 & -5.5518 & 3.7 & 77.5 & 67.2  &   \\
99425.9200(10) & $19(10,10)-19(9,11) $ & 163.73 & -5.2266 & 20.2 & " & "  &   \\
99487.5000(10) & $7(2,5)-6(3,4)++ $ & 21.74 & -5.1743 & 8.8 & 86.7 & 15.5  &   \\
99579.4600(10) & $19(10,10)-19(9,11)-+ $ & 163.63 & -5.2268 & 20.1 & 65.9 & 12.3  &   \\
99862.2800(10) & $7(3,5)-6(-2,4) $ & 22.19 & -5.1601 & 8.9 & 79.9 & 13.5  &   \\
100168.7100(10) & $8(-1,7)-7(2,6) $ & 24.40 & -5.0256 & 13.7 & 100.0 & 10.4  &   \\
100168.9700(10) & $8(2,7)-7(2,6) $ & 24.40 & -5.4811 & 4.8 & " & "  &   \\
100170.7000(10) & $8(-1,7)-7(-1,6) $ & 24.40 & -5.4810 & 4.8 & 99.8 & 18.4  &   \\
100170.9600(10) & $8(2,7)-7(-1,6) $ & 24.40 & -5.0256 & 13.7 & " & "  &   \\
100202.9200(10) & $7(3,5)-6(2,4)++ $ & 21.74 & -5.1625 & 8.8 & 92.4 & 13.2  &   \\
100306.5300(10) & $8(1,7)-7(2,6)-- $ & 23.94 & -5.0295 & 13.5 & 87.4 & 23.4  &   \\
100306.9800(10) & $8(2,7)-7(2,6)-- $ & 23.94 & -5.4650 & 5.0 & " & "  &   \\
100309.1100(40) & $14(11,4)-14(9,5)+- $ & 100.77 & -6.8014 & 0.4 & 92.0 & 15.3  &   \\
100309.6900(10) & $8(1,7)-7(1,6)-- $ & 23.94 & -5.4650 & 5.0 & " & "  &   \\
100310.1300(10) & $8(2,7)-7(1,6)-- $ & 23.94 & -5.0295 & 13.5 & " & "  &   \\
100481.8000(10) & $18(-8,10)-18(-7,11) $ & 145.13 & -5.2386 & 18.1 & 96.9 & 2.1  &   \\
100536.1900(10) & $18(9,10)-18(8,11) $ & 145.15 & -5.2378 & 18.1 & 73.8 & 40.3  &   \\
100855.4300(10) & $9(*,9)-8(*,8) $ & 25.67 & -4.5117 & 49.0 & 94.7 & 70.7  &\citep{msml1997}   \\
               &                  &         &         &      &      &       &\citep{rsfls2003}  \\
100897.4500(10) & $9(*,9)-8(*,8)++ $ & 25.17 & -4.5113 & 49.0 & 92.0 & 40.8  &\citep{msml1997}   \\
               &                  &         &         &      &      &       &\citep{rsfls2003}  \\
101448.7200(10) & $17(-7,10)-17(7,11) $ & 127.51 & -6.0030 & 2.9 & 65.9 & 10.4  &   \\
101449.4900(10) & $17(-7,10)-17(-6,11) $ & 127.51 & -5.2563 & 16.0 & " & "  &   \\
101465.8700(10) & $17(8,10)-17(7,11) $ & 127.51 & -5.2561 & 16.0 & 80.3 & 49.9  &   \\
101466.6400(10) & $17(8,10)-17(-6,11) $ & 127.51 & -6.0028 & 2.9 & " & "  &   \\
102227.0100(10) & $16(-6,10)-16(6,11) $ & 110.83 & -6.0173 & 2.5 & 90.0 & 2.1  &   \\
102227.1700(10) & $16(-6,10)-16(-5,11) $ & 110.83 & -5.2827 & 13.8 & " & "  &   \\
102231.5800(10) & $16(7,10)-16(6,11) $ & 110.83 & -5.2827 & 13.8 & 73.9 & 4.6  &   \\
102231.7500(10) & $16(7,10)-16(-5,11) $ & 110.83 & -6.0173 & 2.5 & " & "  &   \\
102728.0300(10) & $16(6,10)-16(6,11)+- $ & 110.62 & -5.9798 & 2.7 & 93.9 & 23.0  &   \\
102728.3900(10) & $16(6,10)-16(5,11)+- $ & 110.62 & -5.2801 & 13.7 & " & "  &   \\
102736.4900(10) & $16(7,10)-16(6,11)+- $ & 110.62 & -5.2800 & 13.7 & 97.0 & 13.6  &   \\
102736.8400(10) & $16(7,10)-16(5,11)+- $ & 110.62 & -5.9796 & 2.7 & " & "  &   \\
102853.7400(10) & $15(-5,10)-15(5,11) $ & 95.10 & -6.0463 & 2.2 & 94.8 & 54.1  &   \\
102853.7700(10) & $15(-5,10)-15(-4,11) $ & 95.10 & -5.3206 & 11.7 & " & "  &   \\
102854.8100(10) & $15(6,10)-15(5,11) $ & 95.10 & -5.3206 & 11.7 & " & "  &   \\
102854.8400(10) & $15(6,10)-15(-4,11) $ & 95.10 & -6.0463 & 2.2 & " & "  &   \\
103356.4500(10) & $14(-4,10)-14(4,11) $ & 80.35 & -6.0910 & 1.8 & 94.6 & 21.3  &   \\
103356.4600(10) & $14(-4,10)-14(-3,11) $ & 80.35 & -5.3743 & 9.5 & " & "  &   \\
103356.6700(10) & $14(5,10)-14(4,11) $ & 80.35 & -5.3743 & 9.5 & " & "  &   \\
103356.6800(10) & $14(5,10)-14(-3,11) $ & 80.35 & -6.0910 & 1.8 & " & "  &   \\
104077.7400(10) & $14(4,10)-14(4,11)+- $ & 80.06 & -6.0455 & 2.0 & 77.5 & 8.2  &   \\
104077.7500(10) & $14(4,10)-14(3,11)+- $ & 80.06 & -5.3703 & 9.4 & " & "  &   \\
104078.1700(10) & $14(5,10)-14(4,11)+- $ & 80.06 & -5.3703 & 9.4 & " & "  &   \\
104078.1900(10) & $14(5,10)-14(3,11)+- $ & 80.06 & -6.0455 & 2.0 & " & "  &   \\
104078.5800(10) & $24(14,11)-24(13,12)-+ $ & 266.33 & -5.1099 & 29.0 & " & "  &   \\
104574.9600(10) & $13(3,10)-13(*,11)-+ $ & 66.21 & -5.3636 & 8.8 & 92.2 & 5.9  &   \\
104575.0300(10) & $13(4,10)-13(*,11)-+ $ & 66.21 & -5.3636 & 8.8 & " & "  &   \\
107339.7000(10) & $22(12,11)-22(11,12)-+ $ & 220.54 & -5.1055 & 24.5 & 69.8 & 4.0  &   \\
107353.9100(10) & $22(12,11)-22(11,12) $ & 220.55 & -5.1025 & 24.7 & 98.8 & 11.9  &   \\
108601.4600(10) & $21(-10,11)-21(-9,12) $ & 199.07 & -5.1076 & 22.5 & 90.1 & 82.6  &   \\
108621.1100(10) & $21(10,11)-21(9,12)+- $ & 199.03 & -5.1107 & 22.3 & 96.2 & 43.9  &   \\
108709.3300(10) & $21(11,11)-21(10,12) $ & 199.08 & -5.1065 & 22.5 & 88.2 & 1.3  &   \\
108797.3900(10) & $21(11,11)-21(10,12)+- $ & 199.04 & -5.1086 & 22.3 & 76.9 & 6.8  &   \\
108912.9900(10) & $5(5,1)-4(4,1) $ & 15.80 & -4.9257 & 8.7 & 75.7 & 27.9  &   \\
109835.4100(10) & $20(-9,11)-20(9,12) $ & 178.54 & -5.8830 & 3.5 & 59.3 & 1.6  &   \\
109837.5600(10) & $20(-9,11)-20(-8,12) $ & 178.54 & -5.1158 & 20.4 & " & "  &   \\
109873.5600(10) & $20(10,11)-20(9,12) $ & 178.54 & -5.1154 & 20.4 & 94.2 & 0.5  &   \\
109875.7100(10) & $20(10,11)-20(-8,12) $ & 178.54 & -5.8826 & 3.5 & " & "  &   \\
110079.5700(10) & $20(10,11)-20(9,12)-+ $ & 178.47 & -5.1168 & 20.2 & 95.3 & 4.2  &   \\
110145.1600(10) & $8(-2,6)-7(3,5) $ & 27.47 & -4.9909 & 11.2 & 92.7 & 4.1  &   \\
110179.7400(10) & $5(5,1)-4(4,0)++ $ & 15.62 & -4.9042 & 8.8 & 75.8 & 6.7  &   \\
110358.1900(10) & $8(2,6)-7(3,5)++ $ & 27.04 & -4.9951 & 11.0 & 88.8 & 5.1  &   \\
110393.5000(10) & $5(-5,0)-4(-4,0) $ & 16.20 & -4.9056 & 8.7 & 64.7 & 9.8  &   \\
110499.9800(10) & $8(3,6)-7(2,5)++ $ & 27.04 & -4.9934 & 11.0 & 88.6 & 81.4  &   \\
110817.2500(10) & $9(-1,8)-8(2,7) $ & 29.72 & -4.8768 & 15.9 & 73.4 & 40.0  &   \\
110817.2800(10) & $9(2,8)-8(2,7) $ & 29.72 & -5.3410 & 5.5 & " & "  &   \\
110817.5000(10) & $9(-1,8)-8(-1,7) $ & 29.72 & -5.3410 & 5.5 & " & "  &   \\
110817.5400(10) & $9(2,8)-8(-1,7) $ & 29.72 & -4.8768 & 15.9 & " & "  &   \\
110954.1100(10) & $9(1,8)-8(2,7)-- $ & 29.27 & -4.8820 & 15.7 & 99.0 & 39.3  &   \\
110954.1700(10) & $9(2,8)-8(2,7)-- $ & 29.27 & -5.3223 & 5.7 & " & "  &   \\
110954.5500(10) & $9(1,8)-8(1,7)-- $ & 29.27 & -5.3223 & 5.7 & " & "  &   \\
110954.6100(10) & $9(2,8)-8(1,7)-- $ & 29.27 & -4.8820 & 15.7 & " & "  &   \\
111507.2800(10) & $10(*,10)-9(*,9) $ & 31.02 & -4.3757 & 54.8 & 99.4 & 86.7  &\citep{rsfls2003}  \\
111548.5300(10) & $10(*,10)-9(*,9)++ $ & 30.53 & -4.3753 & 54.8 & 99.5 & 38.4  &\citep{rsfls2003}   \\
112357.4100(10) & $17(-6,11)-17(6,12) $ & 122.64 & -5.9181 & 2.6 & 83.0 & 25.5  &   \\
112357.4400(10) & $17(-6,11)-17(-5,12) $ & 122.64 & -5.1808 & 14.0 & " & "  &   \\
112358.1900(10) & $17(7,11)-17(6,12) $ & 122.64 & -5.1808 & 14.0 & " & "  &   \\
112358.2200(10) & $17(7,11)-17(-5,12) $ & 122.64 & -5.9181 & 2.6 & " & "  &   \\
112917.6300(10) & $16(-5,11)-16(*,12) $ & 105.92 & -5.1475 & 14.0 & 86.4 & 19.8  &   \\
112917.8000(10) & $16(6,11)-16(*,12) $ & 105.92 & -5.1475 & 14.0 & " & "  &   \\
112925.8100(10) & $17(6,11)-17(6,12)+- $ & 122.46 & -5.8710 & 2.8 & 99.6 & 9.4  &   \\
112925.8700(10) & $17(6,11)-17(5,12)+- $ & 122.46 & -5.1802 & 13.8 & " & "  &   \\
112927.3800(10) & $17(7,11)-17(6,12)+- $ & 122.46 & -5.1802 & 13.8 & " & "  &   \\
112927.4400(10) & $17(7,11)-17(5,12)+- $ & 122.46 & -5.8710 & 2.8 & " & "  &   \\
113374.0500(10) & $15(-4,11)-15(*,12) $ & 90.17 & -5.2031 & 11.4 & 96.8 & 2.9  &   \\
113374.0800(10) & $15(5,11)-15(*,12) $ & 90.17 & -5.2031 & 11.4 & " & "  &   \\
113600.7200(10) & $16(5,11)-16(5,12)-+ $ & 105.69 & -5.8990 & 2.4 & 96.0 & 8.8  &   \\
113600.7300(10) & $16(5,11)-16(4,12)-+ $ & 105.69 & -5.2201 & 11.7 & " & "  &   \\
113601.0800(10) & $16(6,11)-16(5,12)-+ $ & 105.69 & -5.2201 & 11.7 & " & "  &   \\
113601.0900(10) & $16(6,11)-16(4,12)-+ $ & 105.69 & -5.8990 & 2.4 & " & "  &   \\
113742.2700(20) & $14(*,11)-14(*,12) $ & 75.38 & -4.9837 & 17.6 & 99.0 & 3.1  &   \\
114166.3000(10) & $15(4,11)-15(*,12)+- $ & 89.89 & -5.1921 & 11.5 & 96.1 & 8.0  &   \\
114166.3700(10) & $15(5,11)-15(*,12)+- $ & 89.89 & -5.1921 & 11.5 & " & "
    \enddata
    \tablecomments{Pertinent parameters of the detected transitions of AcA taken from SLAIM catalog. The transitions with ditto marks in the column of P and D factors are blended with the transitions in the previous row.}
\end{deluxetable*}


\begin{thebibliography}{}
\footnotesize
%
\bibitem[Balucani et al.(2015)]{bct2015} Balucani, N., Ceccarelli, C., \& Taquet, V.\ 2015, \mnras, 449, L16 
%
\bibitem[Belloche et al.(2014)]{2014Sci...345.1584B} Belloche, A., Garrod, R.~T., M{\"u}ller, H.~S.~P., \& Menten, K.~M.\ 2014, Science, 345, 1584 
%
\bibitem[Belloche et al.(2008)]{bmcmso2008} Belloche, A., Menten, K. M., Comito, C., M\"uller, H. S. P., Schilke, P., Ott, J., Thorwirth, S. \& Hieret, C. 2008, \aap, 482, 179
%
\bibitem[Belloche et al.(2016)]{bmgk2016} Belloche, A., M\"uller, H. S. P., Garrod, R. T. \& Menten, K. M. 2016, \aap, 587, A91
%
\bibitem[Belloche et al.(2013)]{bmmsc2013} Belloche, A., M\"uller, H. S. P., Menten, K. M,; Schilke, P. \& Comito, C. 2013, \aap, 559, 47
%
\bibitem[Beltr{\'a}n et al.(2009)]{bcvn2009} Beltr{\'a}n, M.~T., Codella, C., Viti, S., Neri, R., \& Cesaroni, R.\ 2009, \apjl, 690, L93 
%
\bibitem[Bennett \& Kaiser(2007)]{bk2007} Bennett, C.~J., \& Kaiser, R.~I.\ 2007, \apj, 660, 1289 
%
\bibitem[Bisschop et al.(2007)]{bjvd2007} Bisschop, S.~E., J{\o}rgensen, J.~K., van Dishoeck, E.~F., \& de Wachter, E.~B.~M.\ 2007, \aap, 465, 913 
%
\bibitem[Bonfand et al.(2017)]{bbmgm2017} Bonfand, M., Belloche, A., Menten, K. M., Garrod,R. T. \& M\"uller, H. S. P. 2017, \aap, 604, A60
%
\bibitem[Brouillet et al.(2013)]{bdb2013} Brouillet, N., Despois, D., Baudry, A., et al.\ 2013, \aap, 550, A46 
%
\bibitem[Brouillet et al.(2015)]{bdl2015} Brouillet, N., Despois, D., Lu, X.-H., et al.\ 2015, \aap, 576, A129 
%
\bibitem[Brown et al.(1975)]{bcggrw1975} Brown, R. D., Crofts, J. G., Godfrey, P. D., Gardner, F. F., Robinson, B. J. \& Whiteoak, J. B. 1975, \apj, 197, 29
%
\bibitem[Burke et al.(2015)]{bpb2015} Burke, D.~J., Puletti, F., Brown, W.~A., et al.\ 2015, \mnras, 447, 1444 
%
\bibitem[Butler et al.(2001)]{blp2001} Butler, R. A. H., De Lucia, F. C., Petkie, D. T., Doug T., Møllendal, H., Horn, A. \& Herbst, E. 2001, \apjs, 134, 319 
%
\bibitem[Calcutt et al.(2014)]{cvc2014} Calcutt, H., Viti, S., Codella, C., et al.\ 2014, \mnras, 443, 3157 
%
\bibitem[Carroll et al.(2010)]{cdw2010} Carroll, P. B., Drouin, B. J., \& Widicus Weaver, S. L. 2010, \apj, 723, 845
%
\bibitem[Carroll et al.(2015)]{cmb2015} Carroll, P.~B., McGuire, B.~A., Blake, G.~A., et al.\ 2015, \apj, 799, 15 
%
\bibitem[Carvajal et al.(2007)]{cwdk2007} Carvajal, M., Willaert, F., Demaison, J., \& Kleiner, I.\ 2007, Journal of Molecular Spectroscopy, 246, 158
%
\bibitem[Chuang et al.(2016)]{cfiv2016} Chuang, K.-J., Fedoseev, G., Ioppolo, S., van Dishoeck, E.~F., \& Linnartz, H. 2016, \mnras, 455, 170
%
\bibitem[Condon \& Ransom (2016)]{cr2016} Condon, J. J. \& Ransom, S. M.\ 2016, Essential Radio Astronomy (2nd ed.; Princeton: Princeton Univ. Press)
%
\bibitem[Demyk et al.(2008)]{dwc2008} Demyk, K., Wlodarczak, G., \& Carvajal, M.\ 2008, \aap, 489, 589 
%
\bibitem[Garrod \& Herbst(2006)]{gh2006} Garrod, R.~T., \& Herbst, E.\ 2006, \aap, 457, 927 
%
\bibitem[Garrod et al.(2008)]{gwh2008} Garrod, R.~T., Widicus Weaver, S.~L., \& Herbst, E.\ 2008, \apj, 682, 283 
%
\bibitem[Graninger et al.(2015)]{goqk2015} Graninger, D., {\"O}berg, K.~I., Qi, C., \& Kastner, J.\ 2015, \apjl, 807, L15 
%
\bibitem[Halfen et al.(2006)]{hawpz2006} Halfen, D. T., Apponi, A. J., Woolf, N., Polt, R. \& Ziurys, L. M. 2006, \apj, 639, 237
%
\bibitem[Halfen et al.(2015)]{hiz2015} Halfen, D.~T., Ilyushin, V.~V., \& Ziurys, L.~M.\ 2015, \apjl, 812, L5 
%
\bibitem[Herbst \& van Dishoeck(2009)]{hd2009} Herbst, E., \& van Dishoeck, E.~F.\ 2009, \araa, 47, 427 
%
\bibitem[Hollis et al.(2004)]{hjlr2004} Hollis, J. M., Jewell, P. R., Lovas, F. J. \& Remijan, A. 2004, \apj, 613, 45
%
\bibitem[Hollis et al.(2004)]{hjlrm2004} Hollis, J.~M., Jewell, P.~R., Lovas, F.~J., Remijan, A., \& M{\o}llendal, H.\ 2004, \apjl, 610, L21 
%
\bibitem[Hollis et al.(2000)]{hlj2000} Hollis, J. M., Lovas, F. J. \& Jewell, P. R. 2000, \apj, 540, L107
%
\bibitem[Hollis et al.(2001)]{hvsjl2001} Hollis, J. M., Vogel, S. N., Snyder, L. E., Jewell, P. R. \& Lovas, F. J. 2001, \apj, 554, 81
%
\bibitem[Ilyushin et al.(2001)]{iadp2001} Ilyushin, V. V., Alekseev, E. A., Dyubko, S. F., Podnos, S. V., Kleiner, I. Margulès, L., Wlodarczak, G., Demaison, J., Cosléou, J., Maté, B., Karyakin, E. N., Golubiatnikov, G. Yu., Fraser, G. T., Suenram, R. D., \& Hougen J. T. 2001, Journal of Molecular Spectroscopy, 205, 286 
%
\bibitem[Ilyushin et al.(2013)]{ielsd2013} Ilyushin, V. V., Endres, C. P., Lewen, F., Schlemmer, S., \& Drouin, B.~J. 2013, Journal of Molecular Spectroscopy, 290, 31
%
\bibitem[Karton \& Talbi (2014)]{kt2014} Karton, A.\& Talbi, D. 2014, Chemical Physics, 436, 22
%
\bibitem[Kobayashi et al.(2007)]{kott2007} Kobayashi, K., Ogata, K., Tsunekawa, S., \& Takano, S.\ 2007, \apjl, 657, L17 
%
\bibitem[Laas et al.(2011)]{lghw2011} Laas, J.~C., Garrod, R. T., Herbst, E., \& Widicus Weaver, S. L. 2011, \apj, 728, 71 
%
\bibitem[Lattelais et al.(2011)]{lbm2011} Lattelais, M., Bertin, M., Mokrane, H., et al. 2011, \aap, 532, A12
%
\bibitem[Lattelais et al.(2009)]{lpec2009} Lattelais, M., Pauzat, F., Ellinger, Y., \& Ceccarelli, C. 2009, \apjl,
696, L133
%
\bibitem[Lattelais et al.(2010)]{lpec2010} Lattelais, M., Pauzat, F., Ellinger, Y., \& Ceccarelli, C. 2010, \aap, 519, A30
%
\bibitem[Linnartz et al.(2015)]{lif2015} Linnartz, H., Ioppolo, S., \& Fedoseev, G. 2015, Int. Rev. Phys. Chem. 34, 205
%
\bibitem[Liu et al.(2002)]{lgrs2002} Liu, S.-Y., Girart, J. M., Remijan, A. J., \& Snyder, L. E. 2002, \apj, 576, 255
%
\bibitem[Liu et al.(2001)]{lms2001} Liu, S.-Y., Mehringer, D. M., \& Snyder, L. E. 2001, \apj, 552, 654
%
\bibitem[Loomis et al.(2015)]{lmsjbrr2015} Loomis, R. A., McGuire, B. A., Shingledecker, C., Johnson, C. H., Blair, S., Robertson, A. \& Remijan, A. J. 2015, \apj, 799, 34
%
\bibitem[Mangum \& Shirley(2015)]{ms2015} Mangum, J.~G., \& Shirley, Y.~L.\ 2015, \pasp, 127, 266
%
\bibitem[McGuire et al.(2016)]{mcl2016} McGuire, B. A., Carroll, P. B., Loomis, R. A., et al. 2016, Science, 352, 1449
%
\bibitem[McGuire et al.(2017)]{gswb2017} McGuire, B. A., Shingledecker, C. N., Willis, E. R., Burkhardt, A. M., El-Abd, S., Motiyenko, R. A., Brogan, C. L., Hunter, T. R., Margulès, L., Guillemin, J.-C., Garrod, R. T., Herbst, E. \& Remijan, A. J. 2017, \apjl, 851, L46 
%
\bibitem[McMullin et al.(2007)]{mws2007} McMullin, J.~P., Waters, B., Schiebel, D., Young, W., \& Golap, K.\ 2007, Astronomical Data Analysis Software and Systems XVI, 376, 127
%
\bibitem[Mehringer et al.(1997)]{msml1997} Mehringer, D. M., Snyder, L. E., Miao, Y. \& Lovas, F. J. 1997, \apj, 480, 71
%
\bibitem[Mottl et al.(2007)]{mgkm2007} Mottlad, M. J., Glazerad, B. T., Kaiserbd, R. I. \& Meechcd, K. J. 2007, ChEG 67, 253
%
\bibitem[M{\"u}ller et al.(2005)]{mssw2005} M{\"u}ller, H. S. P., Schl{\"o}der, F., Stutzki, J., \& Winnewisser, G. 2005, Journal of Molecular Structure, 742, 215
%
\bibitem[{\"O}berg et al.(2010)]{obj2010} {\"O}berg, K.~I., Bottinelli, S., J{\o}rgensen, J.~K., \& van Dishoeck, E.~F.\ 2010, \apj, 716, 825 
%
\bibitem[\"Oberg et al.(2015)]{ogfca2015} \"Oberg, K. I., Guzmán, V. V., Furuya, K., Qi, C., Aikawa, Y., Andrews, S. M., Loomis, R. \& Wilner, D. J. 2015, \nat, 520, 198
%
\bibitem[Oesterling et al.(1999)]{oads1999} Oesterling, L. C., Albert, S., De Lucia, F. C., Sastry, K. V. L. N., \& Herbst, E. 1999, \apj, 521, 255
%
\bibitem[Ogata et al.(2004)]{oott2004} Ogata, K., Odashima, H., Takagi, K., \& Tsunekawa, S. 2004, Journal of Molecular Spectroscopy, 225, 14
%
\bibitem[Pickett et al.(1998)]{ppcd1998} Pickett, H. M., Poynter, R. L., Cohen, E. A., Delitsky, M. L., Pearson, J. C. \& Muller, H. S. P. 1998, J. Quant. Spectrosc. \& Rad. Transfer, 60, 883
%
\bibitem[Puletti et al.(2010)]{pmmc2010} Puletti, F., Malloci, G., Mulas, G., \& Cecchi-Pestellini, C.\ 2010, \mnras, 402, 1667 
%
\bibitem[Qin et al.(2011)]{qsrcl2011} Qin, S.-L., Schilke, P., Rolffs, R., Comito, C., Lis, D. C. \& Zhang, Q. 2011, \aap, 530, 9
%
\bibitem[Remijan et al.(2015)]{raj2015} Remijan, A. J., Adams, M. \& Warmels, R. 2015, ALMA Cycle 3 Technical Handbook Version 1.0, ALMA
%
\bibitem[Remijan et al.(2005)]{rhlpj2005} Remijan, A. J., Hollis, J. M., Lovas, F. J., Plusquellic, D. F. \& Jewell, P. R. 2005, \apj, 632, 333
%
\bibitem[Remijan et al.(2003)]{rsfls2003} Remijan, A. J., Snyder, L. E., Friedel, D. N., Liu, S.-Y. \& Shah, R. Y. 2003, \apj, 590, 314
%
\bibitem[Remijan et al.(2014)]{rsm2014} Remijan, A. J., Snyder, L. E., McGuire, B. A., et al.\ 2014, \apj, 783, 77
%
\bibitem[Remijan et al.(2002)]{rslmk2002} Remijan, A., Snyder, L. E., Liu, S.-Y., Mehringer, D., \& Kuan, Y.-J. 2002, \apj, 576, 264 
%
\bibitem[Requena-Torres et al.(2006)]{rmr2006} Requena-Torres, M.~A., Mart{\'{\i}}n-Pintado, J., Rodr{\'{\i}}guez-Franco, A., et al.\ 2006, \aap, 455, 971
%
\bibitem[Sakai et al.(2015)]{skh2015} Sakai, Y., Kobayashi, K., \& Hirota, T.\ 2015, \apj, 803, 97 
%
\bibitem[S{\'a}nchez-Monge et al.(2017)]{sss2017} S{\'a}nchez-Monge, {\'A}., Schilke, P., Schmiedeke, A., et al.\ 2017, \aap, 604, A6
%
\bibitem[Shaw et al.(2007)]{shb2007} Shaw, R.~A., Hill, F., \& Bell, D.~J.\ 2007, Astronomical Data Analysis Software and Systems XVI, 376,
%
\bibitem[Skouteris et al.(2018)]{sbc2018} Skouteris, D., Balucani, N., Ceccarelli, C., et al.\ 2018, \apj, 854, 135 
%
\bibitem[Snyder, Kuan \& Miao(1994)]{skm1994} Snyder, L. E., Kuan, Y.-J., \& Miao, Y 1994, in The Structure and Content of Molecular Clouds, ed. T. L. Wilson \& K. J. Johnson (Berlin: Springer), 87
%
\bibitem[Snyder et al.(2005)]{slhf2005} Snyder, L. E., Lovas, F. J., Hollis, J. M., Friedel, D. N., Jewell, P. R., Remijan, A., Ilyushin, V. V., Alekseev, E. A. \& Dyubko, S. F. 2005, \apj, 619, 914 
%
\bibitem[Townes \& Schawlow (1975)]{ts1975} Townes, C. H., \& Schawlow, A. L.\ 1975, Microwave Spectroscopy (New York: Dover)
%
\bibitem[Turner(1991)]{tur1991} Turner, B. E. 1991, \apjs, 76, 617
%
\bibitem[W\"achtersha\"user(2000)]{w2000} W\"achtersha\"user G., 2000, Science, 289, 1307
%
\bibitem[Watanabe \& Kouchi(2002)]{wk2002} Watanabe, N., \& Kouchi, A.\ 2002, \apjl, 571, L173 
%
\bibitem[Widicus Weaver et al.(2005)]{wbd2005} Widicus Weaver, S.~L., Butler, R.~A.~H., Drouin, B.~J., et al.\ 2005, \apjs, 158, 188
%
\bibitem[Woods et al.(2012)]{wkv2012} Woods, P.~M., Kelly, G., Viti, S., et al.\ 2012, \apj, 750, 19 

\end{thebibliography}
\end{document}